\documentclass[12pt]{article}
\pdfoutput=1
\usepackage{dcolumn}
\usepackage{bm}
\usepackage{graphicx}
\usepackage{amssymb,amsmath}
\usepackage{multirow}
\usepackage{afterpage}
\usepackage{cite,color,url}
\usepackage[colorlinks=true
,urlcolor=magenta
,anchorcolor=blue
,citecolor=blue
,filecolor=blue
,linkcolor=blue
,menucolor=blue
,linktocpage=true
,pdfproducer=medialab
,pdfa=true
]{hyperref}

\usepackage{tikz-feynman}
\usetikzlibrary{trees}
\usetikzlibrary{decorations.pathmorphing}
\usetikzlibrary{decorations.markings}

\usepackage{array,booktabs}
\usepackage{subcaption}

\usepackage{slashed}
\usepackage{epsfig,psfrag,rotating,soul}
\usepackage{rotfloat}
\usepackage[font={small}]{caption}
\usepackage{colortbl}

\usepackage{array,booktabs}
\usepackage{subcaption}
\usepackage{contour}

\evensidemargin \oddsidemargin
\allowdisplaybreaks

\let\OLDthebibliography\thebibliography
\renewcommand\thebibliography[1]{
	\OLDthebibliography{#1}
	\setlength{\parskip}{0pt}
	\setlength{\itemsep}{0pt plus 0.3ex}
}

\newcommand{\beq}{\begin{equation}}
	\newcommand{\eeq}{\end{equation}}

\usepackage{xspace}

\newcommand{\ba}{\begin{array}}
	\newcommand{\ea}{\end{array}}
\newcommand{\bd}{\begin{displaymath}}
	\newcommand{\ed}{\end{displaymath}}
\newcommand{\besub}{\begin{subequations}}
	\newcommand{\eesub}{\end{subequations}}
\newcommand{\bea}{\begin{eqnarray}}
	\newcommand{\eea}{\end{eqnarray}}

\newcommand{\sla}[1]{/\!\!\!#1}

\def\q2 {q^2}

\def\bt{\begin{table}}
	\def\et{\end{table}}

\definecolor{mygray}{gray}{0.85} 
\definecolor{myblue}{cmyk}{0.65, 0.37, 0.0, 0.19}

\begin{document}
	\thispagestyle{empty}
	
	\def\thefootnote{\fnsymbol{footnote}}
	
	\vspace*{1cm}
	
	\begin{center}
		
		\begin{Large}
			\textbf{\textsc{Exploring Optimal Sensitivity of Lepton Flavor Violating Effective Couplings at the $e^+e^-$ Colliders}}
		\end{Large}
		
		\vspace{0.7cm}
		
		{
			Sahabub Jahedi\footnote{{\tt \href{mailto:sahabub@iitg.ac.in}{sahabub@iitg.ac.in}}~(Corresponding author)}%
			~and Abhik Sarkar\footnote{{\tt \href{mailto:sarkar.abhik@iitg.ac.in}{sarkar.abhik@iitg.ac.in}}}
		}
		
		\vspace*{.2cm}
		
		{
			Department of Physics, Indian Institute of Technology Guwahati, Assam 781039, India \\ 
		}
		
	\end{center}
	
\vspace{0.1cm}

\begin{abstract}
	We analyze lepton flavor violation (LFV) using the Standard Model Effective Field Theory (SMEFT) framework at the future lepton colliders. Our focus is on the associated production of tau lepton with electron/muon at the electron-positron ($e^+e^-$) colliders, related to four-Fermi SMEFT effective operators. In accordance with the upper limits on effective couplings from lepton flavor violating tau decays, we conduct a cut-based analysis to achieve sufficient signal significance. We utilize the optimal observable technique (OOT) to estimate the optimal sensitivity of the effective couplings. The impact of electron beam polarization and the interplay of signal and background in enhancing the optimal sensitivity of the effective couplings are discussed in detail. We find that the sensitivity of flavor-violating effective couplings is enhanced by order of one for 3 TeV center-of-mass (CM) energy and 1000 $\rm{fb}^{-1}$ integrated luminosity at the $e^+e^-$ colliders.
\end{abstract}
\def\thefootnote{\arabic{footnote}}
\setcounter{page}{0}
\setcounter{footnote}{0}

\newpage

\section{Introduction}
\label{sec:intro}
In the Standard Model (SM), lepton flavor violation (LFV) is absent at tree level due to the alignment of flavor eigenstates with corresponding mass eigenstates for leptons. The observation of neutrino oscillations \cite{Super-Kamiokande:1998kpq,SNO:2002tuh} indicate that neutrinos possess finite but tiny mass, leading to the breakdown of lepton flavor conservation. If neutrinos are the sole origin of LFV emerging from loop-level processes, the magnitude of this phenomenon is minuscule, rendering it practically undetectable in foreseeable experiments. Hence, the detection of the LFV in future experiments serves as compelling evidence for the existence of New Physics (NP). 

A comprehensive array of experiments has been conducted across different scales to explore LFV in charged leptons. However, none of these experiments have yielded substantial evidence supporting LFV. Consequently, upper limits have been established for LFV branching ratios (BRs) in decay processes, as well as LFV couplings in the decays of charged leptons. The SINDRUM experiment puts an upper bound on BR($\mu \to e \gamma$) at $10^{-12}$ \cite{SINDRUM:1987nra}. In case of $\tau$-$\mu$ and $\tau$-$e$ conversions, the Belle experiment puts bound on BR($\tau \to \mu \gamma$) $< 4.2 \times 10^{-8}$ and BR($\tau \to e \gamma$) $< 5.6 \times 10^{-8}$ \cite{Belle:2021ysv}. On the other hand, the MEG experiment provides an upper bound on BR($\mu \to e \gamma$) at $4.2 \times 10^{-13}$ \cite{MEG:2016leq}.  In the scenario of three body decays, the Belle experiment sets an upper bound on BR($\tau \to 3e$) $<2.7 \times 10^{-8}$ and BR($\tau \to 3\mu$) $<4.4 \times 10^{-8}$ \cite{Hayasaka:2010np}. All the limits are determined at 90\% confidence level (C.L.). LFV via neutral bosons decays is also investigated at different collider experiments. The CMS experiment at the Large Hadron Collider (LHC) has established that the BRs of the Higgs boson in the $e \mu$, $e \tau$, and $\tau \mu$ channels are constrained to be less than $3.5 \times 10^{-4}$ \cite{CMS:2016cvq}, $6.1 \times 10^{-3}$, and $2.5 \times 10^{-3}$ \cite{CMS:2017con}, respectively, at 95\% C.L. The LFV decays of $Z$ boson have undergone measurement in the $e \mu$, $e \tau$, and $\mu \tau$ channels at the LEP by the OPAL and DELPHI collaborations \cite{OPAL:1995grn,DELPHI:1996iox}. Nevertheless, the latest investigations of these decays at the LHC by the ATLAS collaboration have surpassed the preceding limits set by LEP \cite{ATLAS:2014vur,ATLAS:2020zlz,ATLAS:2021bdj}. LEP has further examined LFV $2 \to 2$ processes and imposed restrictions on the cross-sections associated with these processes.  The OPAL analysis conducted at LEP presents confidence level boundaries on the cross-sections for $e^+e^- \to \tau \mu,~\tau e$, and $\mu e$ processes at various CM energies, as detailed in Table \ref{tab:x.sec}.

\begin{table}[htb!]
	\centering
	{\renewcommand{\arraystretch}{1.4}%
		\begin{tabular}{|c|c|c|c|}
			\hline
			\multirow{1.5}*{CM energy ($\sqrt{s}$)} &
			\multicolumn{3}{c|}{Cross-section (fb)} \\
			\cline{2-4}
			(GeV)& $\tau \mu$ & $\tau e$ & $\mu e$\\
			\hline
			189 & 115 & 95 & 58   \\
			$192 \le \sqrt{s} \le 196$  & 116 & 144 & 162  \\
			$200 \le \sqrt{s} \le 209$  & 64 &  78 & 22\\
			\hline
	\end{tabular}}
	\caption{Upper bound on different flavor violating cross-sections from the OPAL experiment \cite{OPAL:2001qhh}.}
	\label{tab:x.sec}
\end{table}

In this analysis, we investigate the production of muon-tau ($\mu \tau$) and electron-tau ($e \tau$) pairs at the Compact Linear Collider (CLIC) \cite{Aicheler:2018arh}, a proposed electron-positron collider. The absence of the QCD effects in the initial electron-positron beams proves invaluable for estimating potential NP against a significantly cleaner background. Additionally, the availability of partially polarized beams offers a distinct advantage in suppressing the SM background, thereby facilitating the dominance of the NP signal over the SM background. We consider the Standard Model Effective Field Theory (SMEFT) framework \cite{Buchmuller:1985jz,Grzadkowski:2010es,Lehman:2014jma,Bhattacharya:2015vja,Liao:2016qyd,Li:2020gnx} to probe the LFV through four-Fermi effective operators, which we discuss in the next section. LFV under the SMEFT framework has been studied at hadron colliders \cite{Han:2010sa,Davidson:2012wn,Cai:2015poa,Cai:2018cog,Angelescu:2020uug} and lepton colliders \cite{Murakami:2014tna,Cho:2018mro,Etesami:2021hex,Altmannshofer:2023tsa}. Due to heightened background contamination in the leptonic decay modes of $\tau$ lepton, our study centers on the hadronic decay modes of the $\tau$ lepton that manifest as light jets at high-energy colliders. Despite the predominance of the hadronic modes in $\tau$ lepton decays, distinguishing $\tau$ jets from other hadronic activities proves challenging due to formidable background interference at both hadron and hadron-electron colliders. Hence, exploring such signals in the cleaner environments of lepton colliders offers a more favorable avenue for investigation.

We perform the optimal observable technique (OOT) \cite{Atwood:1991ka,Davier:1992nw,Diehl:1993br,Gunion:1996vv} to estimate the optimal sensitivity of dimension-6 effective couplings through the signal process $e^+e^- \to \ell \tau$ ($\ell = e, \mu$). The OOT has been utilized in constraining top-quark couplings \cite{Grzadkowski:1996pc,Grzadkowski:1997cj,Grzadkowski:1998bh,Grzadkowski:1999kx,Grzadkowski:2000nx,Bhattacharya:2023mjr} and Higgs couplings \cite{Hagiwara:2000tk,Dutta:2008bh} at the $e^+e^-$ colliders. Its application extends to the examination of top quark interactions at $\gamma \gamma$ colliders \cite{Grzadkowski:2004iw,Grzadkowski:2005ye}, the measurement of top-Yukawa couplings at the LHC \cite{Gunion:1998hm}, muon colliders \cite{Hioki:2007jc}, and $e \gamma$ colliders \cite{Cao:2006pu}. Recent studies of the OOT includes the investigation of $Z$ couplings of heavy charged fermions at the $e^+e^-$ colliders \cite{Bhattacharya:2021ltd,Bhattacharya:2023zln,Jahedi:2024koa}, explored neutral triple gauge couplings \cite{Jahedi:2022duc,Jahedi:2023myu}, and investigated NP effects in flavor physics scenarios \cite{Bhattacharya:2015ida,Calcuttawala:2017usw,Calcuttawala:2018wgo,Bhattacharya:2023beo}.

Our paper is organized as follows: In Section~\ref{sec:pheno}, we point out the relevant dimension-6 effective operators pertinent to our study and evaluate the upper bound on NP couplings from flavor violating $\tau$ lepton decays. We describe the collider simulation in Section~\ref{sec:col}. A brief overview of the OOT and optimal sensitivity are discussed in Section~\ref{sec:oot}. Finally, in Section~\ref{sec:con}, we present summary and conclusion of our study.

\section{LFV via dimension-6 SMEFT}
\label{sec:pheno}

The lack of signals in direct searches for new particle production at the LHC implies a gap in energy between the electroweak scale and the potential scale where the NP responsible for inducing LFV may manifest. This leads us to work under the SMEFT framework (referred in the introduction), which involves introducing a series of higher-dimensional operators along with the SM Lagrangian.  The general definition of SMEFT Lagrangian is given by
\beq
\mathcal{L}^{\rm EFT}=\mathcal{L}^{\rm SM}+\sum_{i}\sum_d\frac{C_i}{\Lambda^{d-4}}\mathcal{O}^d_i,
\eeq
where $\Lambda$ is the scale of NP, $C_i$'s are Wilson coefficients (WCs) through which the effects of NP are understood. $\mathcal{O}^d_i$'s are the $d$-dimensional operators constructed from SM fields and respect SM gauge symmetry. Flavor-violating dilepton production at the lepton colliders is primarily governed by three classes of SMEFT operators, presented in Table~\ref{tab:ops.class}.
\begin{center}
	\begin{table}[htb!]
		\centering
		\begin{tabular}{|c|c|c|} 
			\hline
			Higgs-current & Dipole & Four-Fermi \\
			operators & operators & operators \\
			\hline
			$(\mathcal{O}_{\varphi \ell}^{(1)})_{ij}: (H^\dagger i \overset{\leftrightarrow}{D_\mu} H)(\bar \ell_i \gamma^\mu \ell_j)$ & $(\mathcal{O}_{eW})_{ij}: (\bar \ell_i \sigma^{\mu\nu} e_j) \tau^I H W_{\mu\nu}^I$ & $(\mathcal{O}_{\ell \ell})_{ijkl}: (\bar \ell_i \gamma^\mu \ell_j) (\bar \ell_k \gamma^\mu \ell_l)$ \\ 
			$(\mathcal{O}_{\varphi \ell}^{(3)})_{ij}: (H^\dagger i \overset{\leftrightarrow}{D^I_\mu} H)(\bar \ell_i \gamma^\mu \tau^I \ell_j)$ & $(\mathcal{O}_{eB})_{ij}: (\bar \ell_i \sigma^{\mu\nu} e_j) H B_{\mu\nu}$ & $(\mathcal{O}_{ee})_{ijkl}: (\bar e_i \gamma^\mu e_j) (\bar e_k \gamma^\mu e_l)$ \\ 
			$(\mathcal{O}_{\varphi e})_{ij}: (H^\dagger i \overset{\leftrightarrow}{D_\mu} H)(\bar e_i \gamma^\mu e_j)$ & & $(\mathcal{O}_{\ell e})_{ijkl}: (\bar \ell_i \gamma^\mu \ell_j) (\bar e_k \gamma^\mu e_l)$\\
			\hline
		\end{tabular}
		\caption{Three classes of dimension-6 operators contributing to the flavor-violating dilepton production at the lepton colliders \cite{Buchmuller:1985jz,Grzadkowski:2010es}.}
		\label{tab:ops.class}
	\end{table}
\end{center}
\noindent
In operators expressions, $\ell$ and $e$ are the $SU(2)_L$ lepton doublets and iso-spin singlets, $H$ is the Higgs doublet, and $W_{\mu\nu}^I$, $B_{\mu\nu}$ are the field strength tensors of the $SU(2)_L$ and $U(1)_Y$ gauge group. The Higgs currents are written as
\begin{equation}
	H^\dagger i \overset{\leftrightarrow}{D_\mu} H = i \left( H^\dagger (D_\mu H) - (D_\mu H^\dagger) H \right) ~,~~
	H^\dagger i \overset{\leftrightarrow}{D^I_\mu} H = i \left( H^\dagger \tau^I (D_\mu H) - (D_\mu H^\dagger) \tau^I H \right) ~,
\end{equation}
where $\tau^I$ are the Pauli matrices.

Experimental constraints derived from the muon decay process $\mu \to 3e$ at the SINDRUM experiment \cite{SINDRUM:1987nra}, $\mu-e$ conversion \cite{SINDRUMII:2006dvw}, and $\mu \to e \gamma$ transition \cite{MEG:2016leq} have firmly restricted flavor violation between the first and second generations of leptons. However, the constraints pertaining to flavor violation between electrons/muons and tau leptons appear less stringent. Given these observations, we examine the $ee \ell \tau$ couplings through $\ell \tau$ production at future $e^{+}e^{-}$ colliders. 

\subsection{$\ell \tau (\ell=\mu,e)$ production at the $e^+e^-$ collider}
\label{sec:x.sec}
\begin{figure}[htb!]
	\centering
	\includegraphics[width = 0.4 \textwidth]{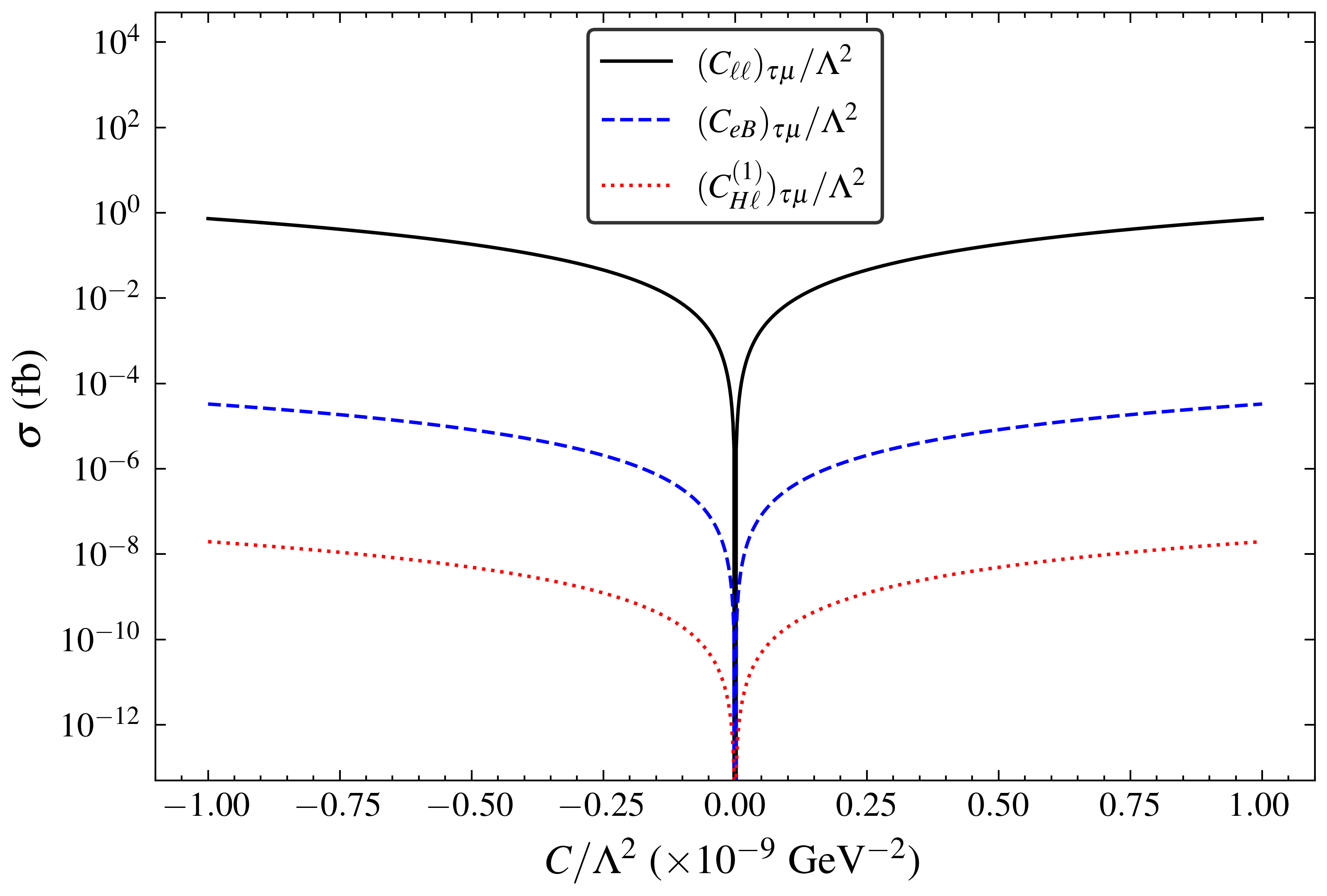}
	\includegraphics[width = 0.4 \textwidth]{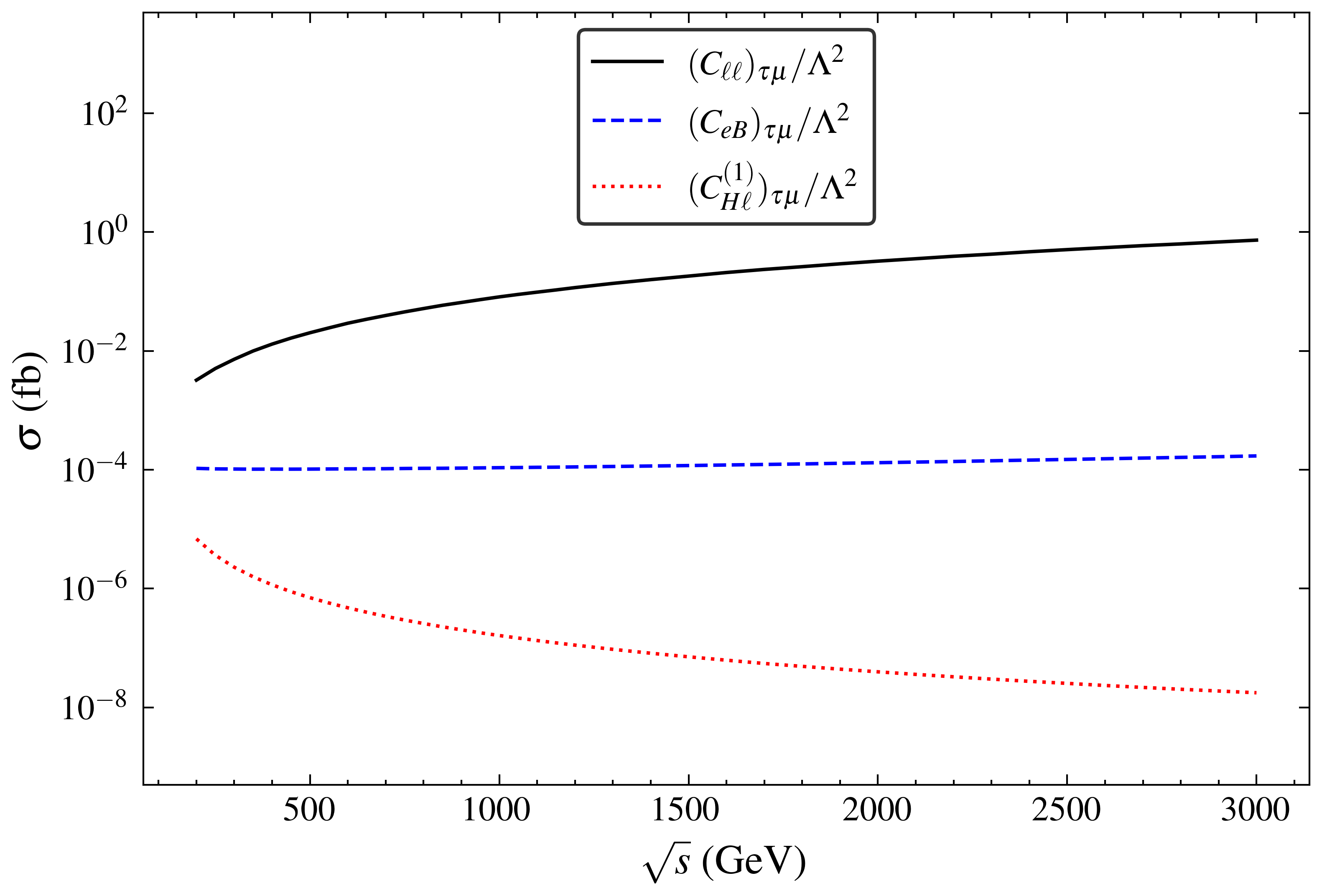}
	\caption{Left: Variation of cross-section for the process $e^{+} e^{-} \rightarrow \mu^{\pm} \tau^{\mp}$ with the change in WCs ($C/ \Lambda^{2}$) at $\sqrt{s} = 3$ TeV. Right: Variation of the cross-section for the same process with change in CM energy, $\sqrt{s}$ ($C / \Lambda^{2}$ set to $1.0 \times 10^{-9}$ GeV$^{-2}$).}
	\label{fig:lfv-sens}
\end{figure}

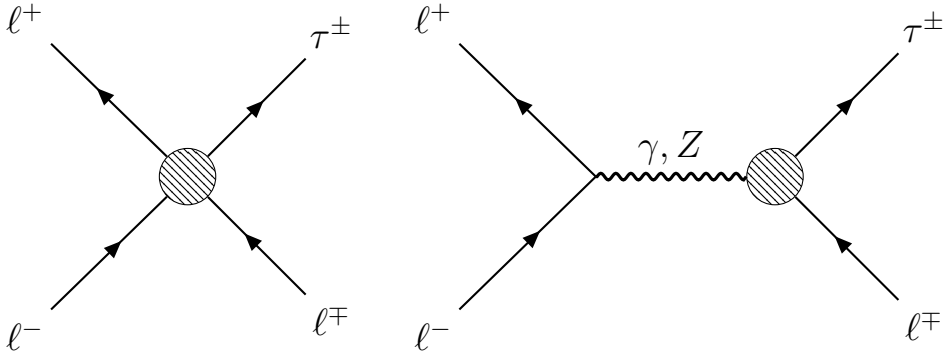
\begin{figure}[htb!]
	\centering
	{\begin{tikzpicture}[baseline={(current bounding box.center),style={scale=0.7,transform shape}}]
			\begin{feynman}
				\vertex (a);
				\vertex [above left=2.5cm of a] (b) {\large $\ell^{+}$};
				\vertex [below left=2.5cm of a] (c) {\large $\ell^{-}$};
				\vertex[blob]  (d)  {\contour{black}{}};
				\vertex [above right=2.7cm of d] (e) {\large $\tau^{\pm}$};
				\vertex [below right=2.7cm of d] (f) {\large $\ell^{\mp}$};
				\diagram{
					(b) -- [anti fermion, thick] (d) -- [ anti fermion, thick] (c);
					(d);
					(d) -- [fermion, thick] (e);
					(d) -- [ anti fermion, thick] (f);
				};
			\end{feynman}
	\end{tikzpicture}} \quad
	{\begin{tikzpicture}[baseline={(current bounding box.center),style={scale=0.7,transform shape}}]
			\begin{feynman}
				\vertex (a);
				\vertex [above left=2.5cm of a] (b) {\large $\ell^{+}$};
				\vertex [below left=2.5cm of a] (c) {\large $\ell^{-}$};
				\vertex[blob] [ right=2cm of a] (d)  {\contour{black}{}};
				\vertex [above right=2.8cm of d] (e) {\large $\tau^{\pm}$};
				\vertex [below right=2.8cm of d] (f) {\large $\ell^{\mp}$};
				\diagram{
					(b) -- [anti fermion, thick] (a) -- [ anti fermion, thick] (c);
					(a) -- [ very thick, boson, edge label={\large $\gamma,Z$}] (d);
					(d) -- [fermion, thick] (e);
					(d) -- [ anti fermion, thick] (f);
				};
			\end{feynman}
	\end{tikzpicture}}	
	\caption{Feynman diagrams that induce $\ell^{\mp} \tau^{\pm}$ production at the lepton colliders; left: effective four-Fermi contribution, 
		right: dipole and Higgs-current contributions. }
	\label{fig:tcprod}
\end{figure}

Among the various classes of operators listed in Table~\ref{tab:ops.class} to produce $\tau \ell$, the four-Fermi operators result in contact interactions, as illustrated in left diagram of Fig.~\ref{fig:tcprod}. Meanwhile, the Dipole and Higgs-current operators contribute to the $Z \ell \tau$ and $\gamma \ell \tau$ vertices, respectively, as shown in right diagram Fig.~\ref{fig:tcprod}. Considering that the couplings are of same order, at a fixed CM energy, the contribution of four-Fermi operators to $\tau \ell$ production dominates over the Dipole/Higgs-current operators evidently due to the absence of $s$-channel suppression in case of this class of operators, and the dominance is amplified as we tend towards higher CM energies, as shown in Fig.~\ref{fig:lfv-sens}.  The contribution of Higgs-current operators to the $\tau \ell$ production drops rapidly with increase in CM energy. Concerning the Dipole operators, the cross-section remains nearly constant throughout the range of $\sqrt{s}$, but for a similar value of WC, the cross-section pertaining to four-Fermi operators dominate over the Dipole operators by $\mathcal{O}(100)$ at $\sqrt{s}=3$ TeV. It should be noted that owing to their sensitivity, four-Fermi operators are more strongly constrained from LFV measurements in comparison to the other two classes. However, it is clear from Fig. \ref{fig:tcprod} that only four-Fermi-type operators can address tree level decoupled NP concerned with $e^+e^- \to \ell \tau$ production process at the lepton colliders. For our analysis, we restrict ourselves to four-Fermi operators only. Since, we study the processes at the $e^{+} e^{-}$ collider, to further simplify the notation, we drop the $ee$ indices associated with these operators and different index combinations contributing to same vertex are naturally assumed to be equal, detailed in Eq. (\ref{eqn:index}) for $\mu \tau$ production. Same applies for $e \tau$ production as well.

\begin{figure}[htb!]
	\centering
	\includegraphics[height=6cm, width=5.4cm]{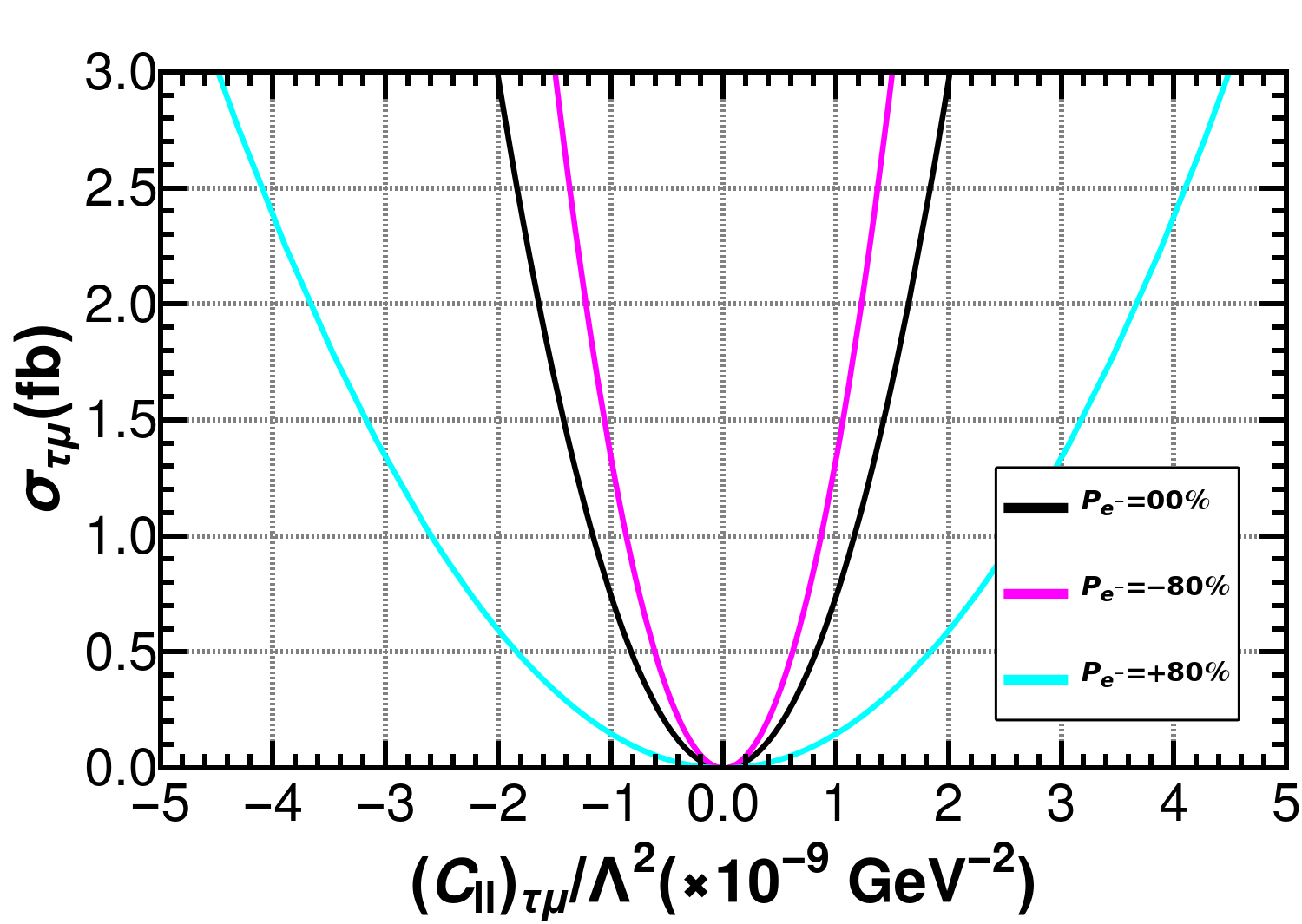}~
	\includegraphics[height=6cm, width=5.4cm]{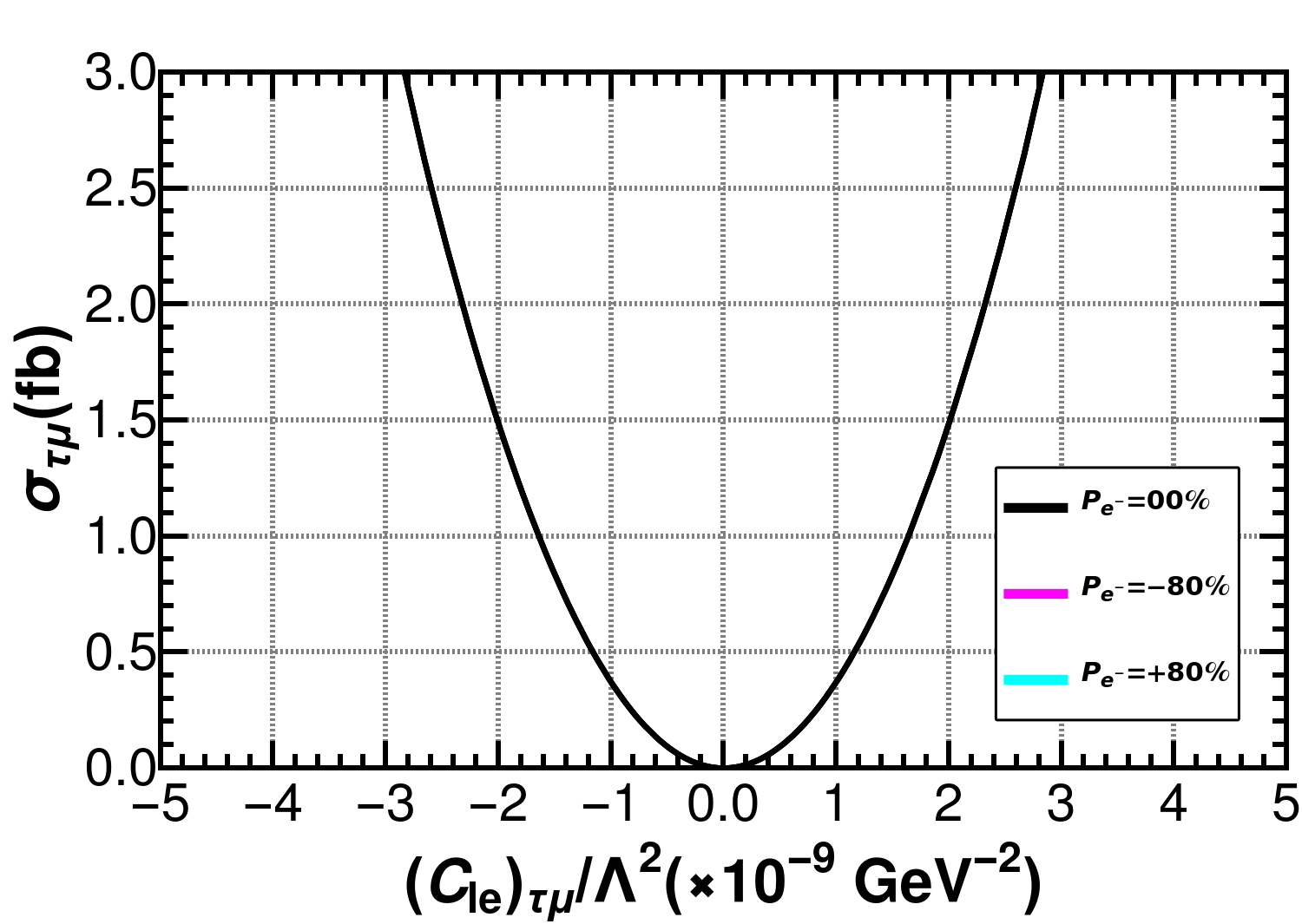}~
	\includegraphics[height=6cm, width=5.4cm]{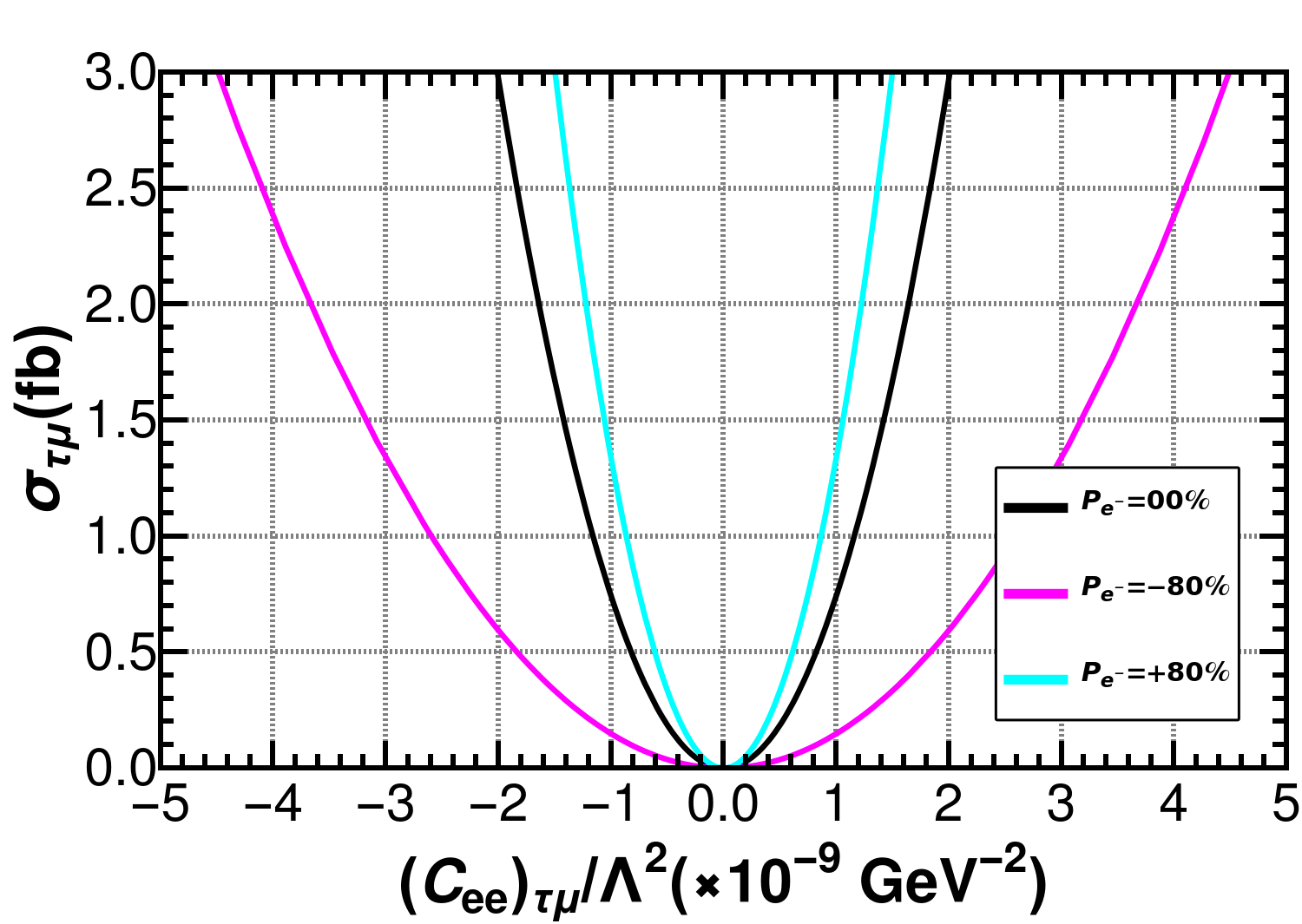}
	\caption{Variation of $\mu \tau$ cross-section with various flavor violating effective couplings for different choices of beam polarization. Left: $(C_{\ell \ell})_{\tau \mu}/\Lambda^2$, middle: $(C_{\ell e})_{\tau \mu}/\Lambda^2$, right: $(C_{e e})_{\tau \mu}/\Lambda^2$.}
	\label{fig:xsec}
\end{figure}

\begin{equation} \label{eqn:index}
	\begin{split}
		\left(C_{\ell \ell / \ell e / ee} \right)_{\tau \mu} &= \left(C_{\ell \ell / \ell e / ee} \right)_{\mu \tau ee} = \left(C_{\ell \ell / \ell e / ee} \right)_{ee \mu \tau} \\ &= \left(C_{\ell \ell / \ell e / ee} \right)_{e \tau \mu e} = \left(C_{\ell \ell / \ell e / ee} \right)_{\mu ee \tau}, \\
		\left(C_{\ell \ell / \ell e / ee} \right)_{\tau e} &= \left(C_{\ell \ell / \ell e / ee} \right)_{e \tau ee} = 
		\left(C_{\ell \ell / \ell e / ee} \right)_{eee \tau}, \\
		\left(C_{\ell \ell / \ell e / ee} \right)_{\mu e} &= \left(C_{\ell \ell / \ell e / ee} \right)_{e \mu ee} = 
		\left(C_{\ell \ell / \ell e / ee} \right)_{eee \mu}. \\
	\end{split}
\end{equation}
Considering the $\mu \tau$ production at the $e^+e^-$ colliders, the four-Fermi operators, following simplification and Fierz transformation, can be expressed as
\begin{align}
	\begin{split}
		(C_V^{LL})_{\tau\mu} & \frac{1}{\Lambda^2} (\bar e \gamma_\alpha P_L e)(\bar \mu \gamma^{\alpha} P_L \tau) ,~~~~~~~~  (C_V^{RR})_{\tau\mu}  \frac{1}{\Lambda^2} (\bar e \gamma_\alpha P_R e)(\bar \mu \gamma^{\alpha} P_R \tau) , \\
		(C_V^{LR})_{\tau\mu} & \frac{1}{\Lambda^2} (\bar e \gamma_\alpha P_L e)(\bar \mu \gamma^{\alpha} P_R \tau) ,~~~~~~~~ (C_V^{RL})_{\tau\mu}  \frac{1}{\Lambda^2} (\bar e \gamma_\alpha P_R e)(\bar \mu \gamma^{\alpha} P_L \tau) , \\
		(C_S^{LR})_{\tau\mu} & \frac{1}{\Lambda^2} (\bar e P_L e)(\bar \mu P_R \tau) ,~~~~~~~~~~~~~~ (C_S^{RL})_{\tau\mu} \frac{1}{\Lambda^2} (\bar e P_R e)(\bar \mu P_L \tau),
	\end{split}
\end{align}
where the WCs are defined as follows:
\begin{equation}
	\begin{split}
		(C_{V}^{LL})_{\tau\mu} &=~~~ (C_{\ell \ell})_{ee\mu\tau} ~+ (C_{\ell \ell})_{\mu\tau ee} ~+ (C_{\ell \ell})_{e\tau \mu e} ~+ (C_{\ell \ell})_{\mu ee \tau} ~=~~~ 4(C_{\ell \ell})_{\tau\mu}, \\
		(C_{V}^{RR})_{\tau\mu} &=~~~ (C_{ee})_{ee\mu\tau} ~+ (C_{ee})_{\mu\tau ee} ~+ (C_{ee})_{e\tau \mu e} ~+ (C_{ee})_{\mu ee \tau} ~=~~~ 4(C_{ee})_{\tau\mu}, \\
		(C_{V}^{LR})_{\tau\mu} &= (C_{\ell e})_{ee\mu\tau} = (C_{\ell e})_{\tau\mu},~~~~~~~~~~(C_{V}^{RL})_{\tau\mu} = (C_{\ell e})_{\mu\tau ee} = (C_{\ell e})_{\tau\mu}, \\
		(C_{S}^{LR})_{\tau\mu} &= -2(C_{\ell e})_{\mu ee \tau} = -2(C_{\ell e})_{\tau\mu},~~(C_{S}^{RL})_{\tau\mu} = -2(C_{\ell e})_{e \tau \mu e} = -2(C_{\ell e})_{\tau\mu}. \\
	\end{split}
\end{equation}
The helicity amplitudes, $\mathcal{M}(\lambda_{e^-},\lambda_{e^+};\lambda_{\mu^-},\lambda_{\tau^+})$ for the process $e^+e^- \to \mu^{+} \tau^{-}$ induced by the four-Fermi operators are given by\footnote{These amplitudes are calculated in the massless limit of initial and final particles.}
\begin{align}
	\begin{split}
		\mathcal{M}(+\lambda,-\lambda;+\lambda^{'},-\lambda^{'})=&-\frac{s}{\Lambda^2}\left[(C^{LL}_{V})_{\tau\mu} \delta_{\lambda,-1}\delta_{\lambda^{'},-1} + (C^{RR}_{V})_{\tau\mu} \delta_{\lambda,1}\delta_{\lambda^{'},1}\right](1+  \cos \theta)\\
		&+\frac{s}{\Lambda^2}\left[(C^{LR}_{V})_{\tau\mu} \delta_{\lambda,-1}\delta_{\lambda^{'},1} + (C^{RL}_{V})_{\tau\mu} \delta_{\lambda,1}\delta_{\lambda^{'},-1}\right](1-\cos \theta),\\
		\mathcal{M}(+\lambda,-\lambda;+\lambda^{'},+\lambda^{'})=&\; 0,\\
		\mathcal{M}(+\lambda,+\lambda;+\lambda^{'},-\lambda^{'})=&\; 0,\\
		\mathcal{M}(+\lambda,+\lambda;+\lambda^{'},+\lambda^{'})=& \; \frac{s}{\Lambda^2} \left[ (C^{LR}_S)_{\tau\mu} \delta_{\lambda, -1} \delta_{\lambda', -1} +   (C^{RL}_S)_{\tau\mu} \delta_{\lambda, 1} \delta_{\lambda', 1} \right],\\
	\end{split}
\end{align}
where $\theta$ is the scattering angle in CM frame. 
$\lambda,\lambda^{'}=-1(+1)$ denotes the left(right)-handed helicity of initial beam particle. The differential cross-section with partial initial beam polarization $(-1 \le P_{e^{\pm}} \le+1)$ is written as
\begin{equation}\label{eq:pp}
	\begin{split}
		\frac{d\sigma(P_{e^-},P_{e^+})}{d\phi}=& \frac{(1-P_{e^-})(1-P_{e^+})}{4}\left(\frac{d\sigma}{d\phi}\right)_{\text{LL}}+\frac{(1-P_{e^-})(1+P_{e^+})}{4}\left(\frac{d\sigma}{d\phi}\right)_{\text{LR}}\\
		+&\frac{(1+P_{e^-})(1-P_{e^+})}{4}\left(\frac{d\sigma}{d\phi}\right)_{\text{RL}}+\frac{(1+P_{e^-})(1+P_{e^+})}{4}\left(\frac{d\sigma}{d\phi}\right)_{\text{RR}},\\
		=&g_i f_i(\phi),
	\end{split}
\end{equation}
where $P_{e^{-}(e^+)}$ is the electron(positron) beam polarization and $\phi$ is the phase-space co-ordinate.
\begin{align}
	\begin{split}
		g_1=& (1 - P_{e^-}) \left(\frac{1}{\Lambda^{4}}\right) \big[ (1 + P_{e^+})(C_{V}^{LR})_{\tau\mu}^2 + (1 - P_{e^+}) \big\{(C_V^{LL})_{\tau\mu}^2 + (C_S^{LR})_{\tau\mu}^2 \big\} \big]\\
		&+ (1 + P_{e^-})\left(\frac{1}{\Lambda^{4}}\right) \big[ (1 - P_{e^+})(C_V^{RL})_{\tau\mu}^2 + (1 + P_{e^+})\left\{(C_V^{RR})_{\tau\mu}^2  +  (C_S^{RL})_{\tau\mu}^2 \right\} \big],\\
		\vspace{50cm}
		g_2=&(1 - P_{e^-})(1 + P_{e^+}) \left(\frac{1}{\Lambda^{4}}\right) \big[(C_V^{LL})_{\tau\mu}^2 - (C_V^{LR})_{\tau\mu}^2\big] \\ &+ (1 + P_{e^-})(1 - P_{e^+}) \left(\frac{1}{\Lambda^{4}}\right) \big[(C_V^{RL})_{\tau\mu}^2 - (C_V^{RR})_{\tau\mu}^2\big],\\
		g_3=&(1 - P_{e^-})(1 + P_{e^+}) \left(\frac{1}{\Lambda^{4}}\right) \big[(C_V^{LR})_{\tau\mu}^2 + (C_V^{LL})_{\tau\mu}^2\big] \\ &+ (1 + P_{e^-})(1 - P_{e^+}) \left(\frac{1}{\Lambda^{4}}\right) \big[(C_V^{RL})_{\tau\mu}^2 + (C_V^{RR})_{\tau\mu}^2\big],
	\end{split}
	\label{eq:gi}
\end{align}
and
\begin{align}
	\begin{split}
		f_1(\theta)&=\frac{s}{256\pi^2},\\
		f_2(\theta)&=\frac{s}{128\pi^2}\cos \theta,\\
		f_3(\theta)&=\frac{s}{256\pi^2}\cos^2 \theta.
	\end{split}
	\label{eq:fi}
\end{align}
This above decomposition (Eqs.~\eqref{eq:gi}-\eqref{eq:fi}) is required to determine the optimal covariance matrix as we discuss in Section~\ref{sec:oot}. The total cross-section is evaluated as
\beq
\sigma(P_{e^-},P_{e^+})=\frac{s}{96 \pi} (3 g_1 +g_2).
\eeq
We show that variation of $\mu \tau$ cross-section with three different flavor-violating dimension-6 effective couplings for various choices of beam polarization in Fig.~\ref{fig:xsec}. For the $(\mathcal{O}_{\ell \ell})_{\tau \mu}$ operator, both currents are left-handed. Thus, using a left-polarized electron beam is advantageous for enhancing the cross-section compared to an unpolarized beam (left plot of Fig.~\ref{fig:xsec}). Conversely, for the $(\mathcal{O}_{ee})_{\tau \mu}$ operator, both currents are right-handed, making a right-handed electron beam more beneficial for increasing the total cross-section (right plot of Fig.~\ref{fig:xsec}). In the case of the $(\mathcal{O}_{\ell e})_{\tau \mu}$ operator, one current is left-handed and the other is right-handed. Therefore, regardless of the beam polarization, this operator contributes in the same manner (middle plot of Fig.~\ref{fig:xsec}).

\subsection{Constraints from LFV processes}
As discussed in previous sections, the LFV processes are constrained from a wide array of low-energy experiments. In this section, we translate these experimental bounds to the LFV operators. The four-Fermi operators contributing to $\ell^{'} \to \ell \gamma$ processes (via leptonic loop) vanish when all the diagrams corresponding to this process are taken into account. Hence, bounds on the four-Fermi operators are insensitive to $\ell^{'} \to \ell \gamma$ branching. The four-Fermi operators are mostly constrained from flavor-violating three body decays of $\mu$ and $\tau$ {\it i.e.} $\ell^{'} \to 3\ell$. These decay modes have been studied explicitly in all possible channels and the most recent bounds are quoted in Table~\ref{tab:lfv}. The BRs are parametrized in terms of EFT coefficients in Eq. \ref{eq:lfv1}. Since the phase space of three body decays are universal, the parametrization is done for the ratio of LFV $\ell^{'} \to 3\ell$ decay to lepton flavor-conserving (LFC) $\ell^{'} \to \ell \nu^{'} \overline{\nu}$ decay. This removes common constant parameters and the parametrization is simplified. The branchings $\mathcal{B}(\tau^{-} \rightarrow \mu^{-} \nu_{\tau} \overline{\nu}_{\mu})$ and $\mathcal{B}(\mu^{-} \rightarrow e^{-} \nu_{\mu} \overline{\nu}_{e})$ are taken to be 0.174 and 1, respectively (based on combined fits by PDG \cite{Workman:2022ynf}).
\begin{figure}[htb!]
	\centering
	{\begin{tikzpicture}[baseline={(current bounding box.center),style={scale=0.7,transform shape}}]
			\begin{feynman}
				\vertex [blob] (a) {\contour{black}{}};
				\vertex [left = 2.5cm of a] (b) {\large $\ell^{'}$};
				\vertex [right = 2.5cm of a] (c) {\large $\ell$};
				\vertex [above right = 2.5cm and 2.5cm of a] (d) {\large $\gamma$};
				\diagram{
					(b) -- [fermion, thick] (a);
					(a) -- [fermion, thick] (c);
					(a) -- [boson, thick] (d);
				};
			\end{feynman}
	\end{tikzpicture}} \quad
	{\begin{tikzpicture}[baseline={(current bounding box.center),style={scale=0.7,transform shape}}]
			\begin{feynman}
				\vertex [blob] (a) {\contour{black}{}};
				\vertex [left = 2.5cm of a] (b) {\large $\ell^{'}$};
				\vertex [right = 2.5cm of a] (e) {\large $\ell$};
				\vertex [above right = 2.5cm and 2.5cm of a] (d) {\large $\ell$};
				\vertex [above right = 1.25cm and 2.5cm of a] (c) {\large $\ell$};
				\diagram{
					(b) -- [fermion, thick] (a);
					(c) -- [fermion, thick] (a);
					(a) -- [fermion, thick] (d);
					(a) -- [fermion, thick] (e);
				};
			\end{feynman}
	\end{tikzpicture}}	
	\caption{Feynman diagrams of LFV decay modes: (a) $\ell^{'} \to \ell \gamma$ processes. (b) $\ell^{'} \to 3 \ell$ processes.}
	\label{fig:lfvd}
\end{figure}
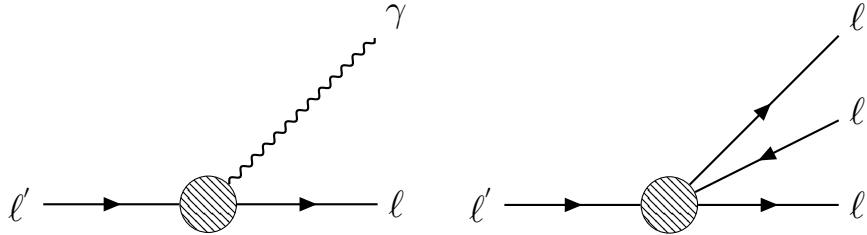 
\begin{equation} \label{eq:lfv1}
	\begin{split}
		\dfrac{\mathcal{B}(\mu^{-} \rightarrow e^{-} e^{+} e^{-})}{\mathcal{B}(\mu^{-} \rightarrow e^{-} \nu_{\mu} \overline{\nu}_{e})} & \sim \frac{v^{4}}{\Lambda^4} \left\{4 \left|(C_{\ell \ell})_{\mu e}\right|^{2} + 4 \left|(C_{ee})_{\mu e}\right|^{2} + \left|(C_{\ell e})_{\mu e}\right|^{2} \right\}, \\
		\dfrac{\mathcal{B}(\tau^{-} \rightarrow e^{-} e^{+} e^{-})}{\mathcal{B}(\tau^{-} \rightarrow \mu^{-} \nu_{\tau} \overline{\nu}_{\mu})} & \sim \frac{v^{4}}{\Lambda^4} \left\{4 \left|(C_{\ell \ell})_{\tau e}\right|^{2} + 4 \left|(C_{ee})_{\tau e}\right|^{2} + \left|(C_{\ell e})_{\tau e}\right|^{2} \right\}, \\
		\dfrac{\mathcal{B}(\tau^{-} \rightarrow \mu^{-} e^{+} e^{-})}{\mathcal{B}(\tau^{-} \rightarrow \mu^{-} \nu_{\tau} \overline{\nu}_{\mu})} & \sim \frac{v^{4}}{\Lambda^4} \left\{4 \left|(C_{\ell \ell})_{\tau \mu}\right|^{2} + 4 \left|(C_{ee})_{\tau \mu}\right|^{2} + \left|(C_{\ell e})_{\tau \mu}\right|^{2} \right\}. \\
	\end{split}
\end{equation}

\begin{table}[h!]
	\centering
	{\renewcommand{\arraystretch}{1.2}
		\begin{tabular}{|c|c|c|}
			\hline
			Observable & Upper Bounds & Bounds on EFT coefficients (GeV$^{-2}$) \\ \hline
			& & $\left|(C_{\ell \ell})_{\mu e}\right|/\Lambda^2 < 8.26 \times 10^{-12}$ \\
			$\mathcal{B} (\mu^{-} \rightarrow e^{-} e^{+} e^{-})$ & $< 1.0 \times 10^{-12}$ \cite{SINDRUM:1987nra} & $\left|(C_{e e})_{\mu e}\right|/\Lambda^2 < 8.26 \times 10^{-12}$ \\
			& & $\left|(C_{\ell e})_{\mu e}\right|\Lambda^2 < 1.65 \times 10^{-11}$ \\ \hline
			& & $\left|(C_{\ell \ell})_{\tau e}\right|/\Lambda^2 < 3.26 \times 10^{-9}$ \\
			$\mathcal{B} (\tau^{-} \rightarrow e^{-} e^{+} e^{-})$ & $< 2.7 \times 10^{-8}$ \cite{Hayasaka:2010np} & $\left|(C_{e e})_{\tau e}\right|/\Lambda^2 < 3.26 \times 10^{-9}$ \\
			& & $\left|(C_{\ell e})_{\tau e}\right|/\Lambda^2 < 6.51 \times 10^{-9}$ \\ \hline
			& & $\left|(C_{\ell \ell})_{\tau \mu}\right|/\Lambda^2 < 2.66 \times 10^{-9}$ \\
			$\mathcal{B} (\tau^{-} \rightarrow \mu^{-} e^{+} e^{-})$ & $< 1.8 \times 10^{-8}$ \cite{Hayasaka:2010np} & $\left|(C_{e e})_{\tau \mu}\right|/\Lambda^2 < 2.66 \times 10^{-9}$ \\
			& & $\left|(C_{\ell \ell})_{\tau \mu}\right|/\Lambda^2 < 5.32 \times 10^{-9}$ \\
			\hline
	\end{tabular}}
	\caption{Flavor bounds from lepton number violating observables and processes. $\mathcal{B}$ refers to branching ratio ($\Gamma_{i}/\Gamma$).}
	\label{tab:lfv}
\end{table}
The bounds on operators contributing to $e^{+} e^{-} \to \mu e$ production, as presented in Table~\ref{tab:lfv}, are very stringent, therefore, it is very unlikely\footnote{Considering the allowed upper bound on, say, $(C_{\ell \ell})_{\mu e} = 8.26 \times 10^{-12}$, the cross-section of $e^{+} e^{-} \to \mu e$ is $< 5 \times 10^{-5}$ fb (without any cuts), i.e. even $\mathfrak{L}_{\text{int}} = $ 10 ab$^{-1}$ won't yield even a single event ($< 0.5$ events).} that these operators will be probed in this particular channel even at CM energy as high as 3 TeV. Future high energy muon collider \cite{Black:2022cth} could be a possibility to probe LFV through this process with satisfactory statistics. We would like to point out that the renomalization group (RG) evolution on four-Fermi effective couplings are small, given the upper bound from LFV decays (see Appendix~\ref{sec:rge}). 

\section{Collider simulation}
\label{sec:col}
In this section, we study the sensitivity of the flavor violating four-Fermi effective operators at the CLIC with 3 TeV CM energy via $\ell \tau$ production. Since these operators are not flavor universal, we probe $(\mathcal{O}_{\ell \ell})_{\tau e}$, $(\mathcal{O}_{e e})_{\tau e}$ and $(\mathcal{O}_{\ell e})_{\tau e}$ operators with $e^{\mp}\tau^{\pm}$ production and $(\mathcal{O}_{\ell \ell})_{\tau \mu}$, $(\mathcal{O}_{e e})_{\tau \mu}$ and $(\mathcal{O}_{\ell e})_{\tau \mu}$ operators with $\mu^{\mp}\tau^{\pm}$ production. The tau leptons are heavier than electrons and muons and hence, decay before leaving an observable track in the detectors. So the $\tau$ signal has to be studied in one of its decay modes. The dominant modes are the hadronic modes where $\tau$ lepton decays to $\nu_{\tau}$ and quarks, which in turn form mesons and baryons. Such decay modes manifest as light jets of hadronic particles in the detector. At hadron-hadron and hadron-lepton colliders, radiative QCD jets are ubiquitous and light QCD jets ($u, d, c, g$ jets) adeptly mimic the $\tau$ jets and discrimination of these two classes becomes a daunting task. However, lepton colliders are mostly immune to the hadronic activities and hence, $\tau$ jet tagging is much easier. The other significant decay modes of $\tau$ lepton are the leptonic modes where $\tau$ lepton decays to leptons and neutrinos. Even though these modes are very clean due to the better detectability of leptons, distinction from dilepton processes and leptonic decays of weak bosons is difficult owing to the fact that both processes have missing particles. For our signal, we will restrict ourselves to the hadronic modes. Hence, the signal process in our case is $e^{+} e^{-} \rightarrow \ell \tau_{h}$ (+ missing energy, $\sla{E}$, from the neutrino in $\tau$ decay). The dominant SM backgrounds come from $e^{+} e^{-} \rightarrow W^{+} W^{-}$, $e^{+} e^{-} \rightarrow \tau^{+} \tau^{-}$ and $e^{+} e^{-} \rightarrow \nu \overline{\nu} Z$.

The signal model is implemented in \texttt{FeynRules2.3} \cite{Alloul:2013bka}. The signal and background events are generated in \texttt{MG5\_aMC} \cite{Alwall:2011uj}. The generated MC events are fed into \texttt{Pythia8} \cite{Sjostrand:2014zea} for parton showering (ISR, FSR, and hadronization effects). The showered events are further fed into \texttt{Delphes3} \cite{deFavereau:2013fsa}, where the detector resolution and efficiency factors are taken into account. The electron and muon efficiencies in different kinematic regions are tabulated in Table~\ref{tab:eff}. The jet reconstruction task is done using \texttt{FastJet3} \cite{Cacciari:2011ma}. The hadronic $\tau$ ($\tau$ jet) tagging efficiency is taken as 0.6 and the misstagging efficiency of light jets as $\tau$ is 0.01 (as per \texttt{Delphes3} default card).

\begin{table}[h!]
	\centering
	\begin{tabular}{|c|c|c|c|}
		\hline
		Electrons & Efficiency & Muons & Efficiency \\ \hline
		$p_{T} < 10.0$ GeV & 0.00 & $p_{T} < 10.0$ GeV & 0.00 \\
		$p_{T} > 10.0$ GeV, $|\eta| \in [0.0, 1.5]$  & 0.95 & $p_{T} > 10.0$ GeV, $|\eta| \in [0.0, 1.5]$ & 0.95 \\
		$p_{T} > 10.0$ GeV, $|\eta| \in (1.5, 2.5]$ & 0.85 & $p_{T} > 10.0$ GeV, $|\eta| \in (1.5, 2.4]$ & 0.95 \\
		$|\eta| > 2.5 $ & 0.00 & $|\eta| > 2.4 $ & 0.00 \\ \hline
	\end{tabular}
	\caption{Efficiency of electron and muon detection for different kinematic regions.}
	\label{tab:eff}
\end{table}
\subsection{Cut based analysis}
\label{sec:x.sec1}
\begin{figure}[h!]
	\centering
	\includegraphics[width = 0.4 \textwidth]{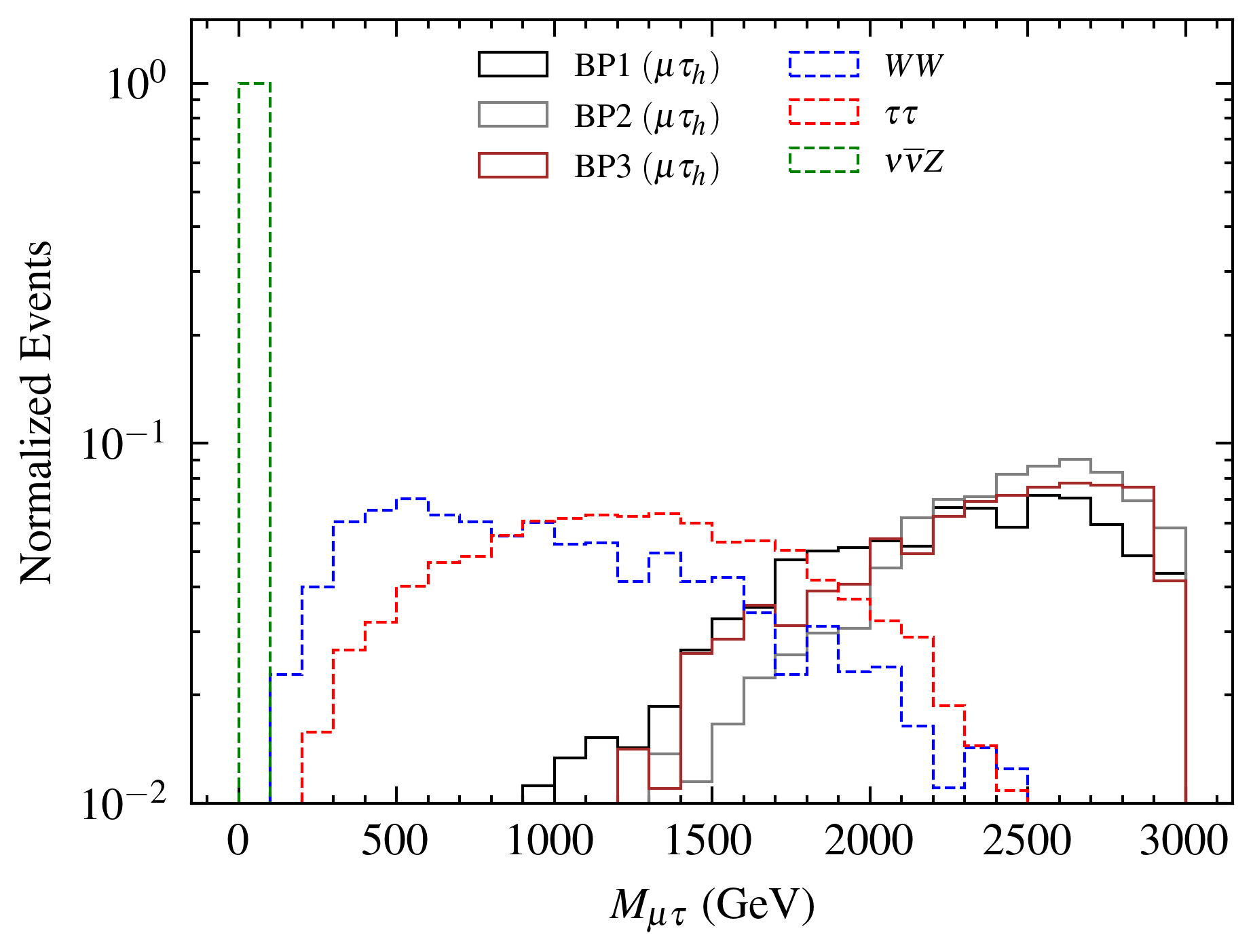}
	\includegraphics[width = 0.4 \textwidth]{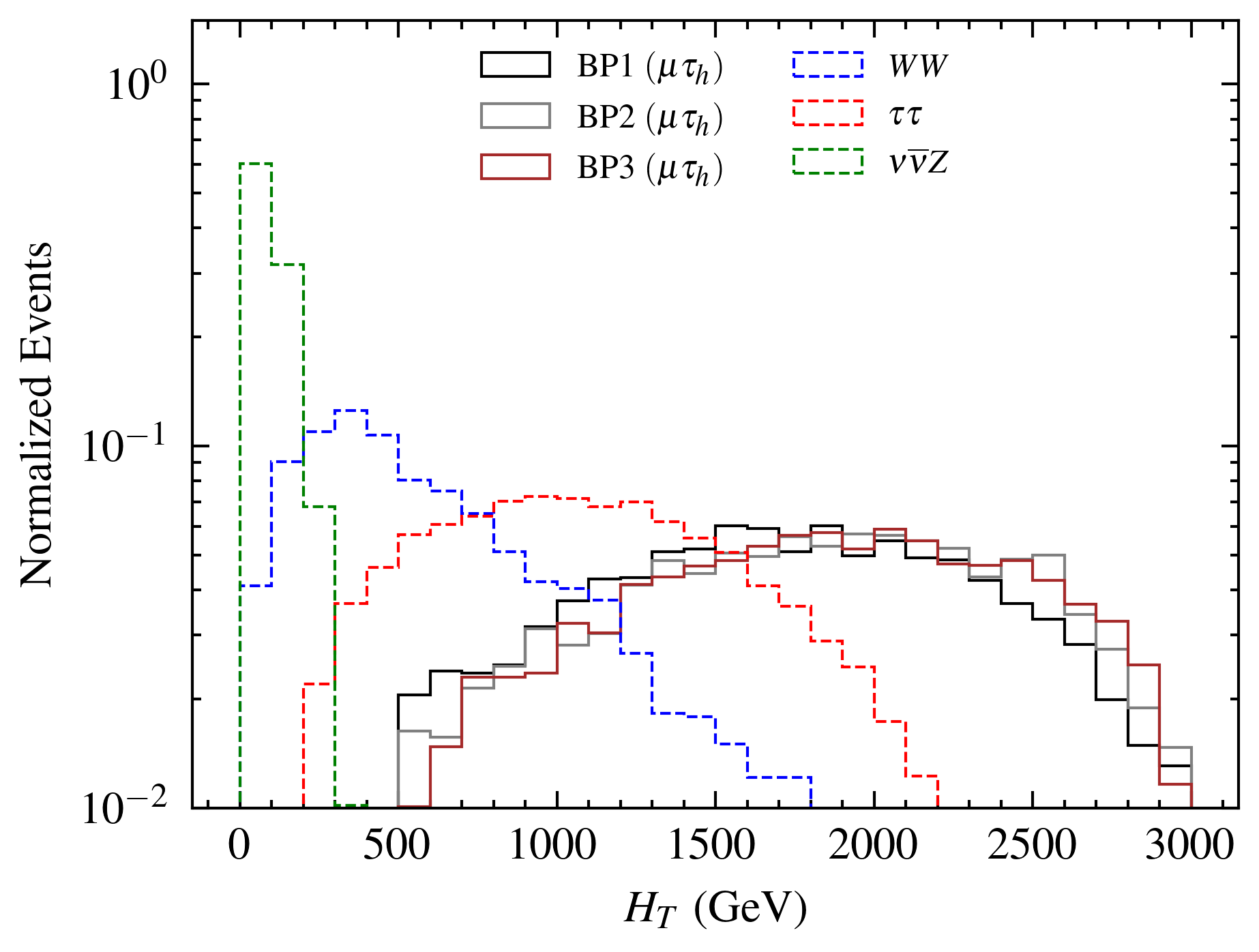}
	\caption{Kinematic distributions corresponding to signal and main background processes for $e^{+} e^{-} \rightarrow \mu \tau_h$ production at CLIC 3 TeV.}
	\label{fig:dist-mt}
\end{figure}

\begin{figure}[h!]
	\centering
	\includegraphics[width = 0.4 \textwidth]{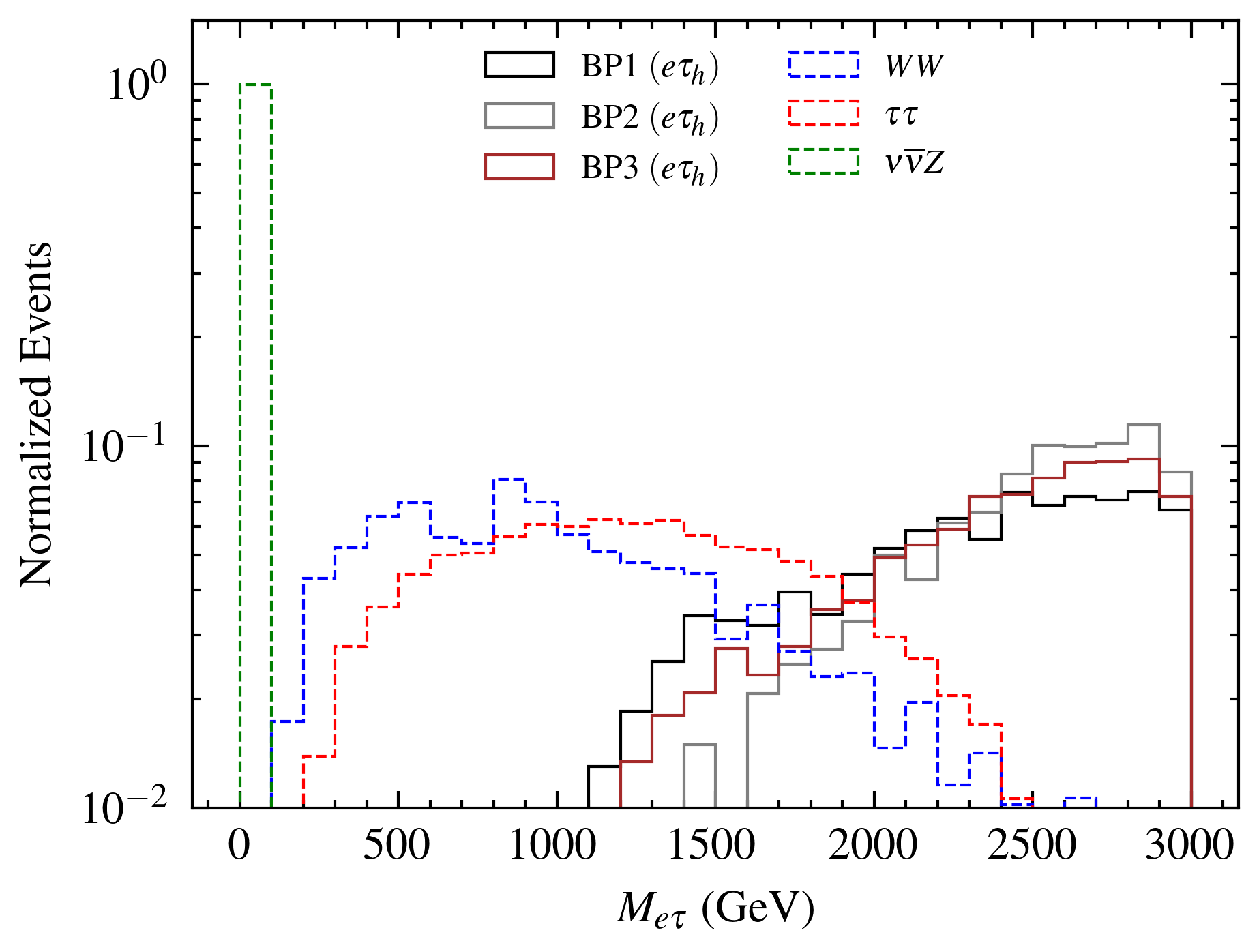}
	\includegraphics[width = 0.4 \textwidth]{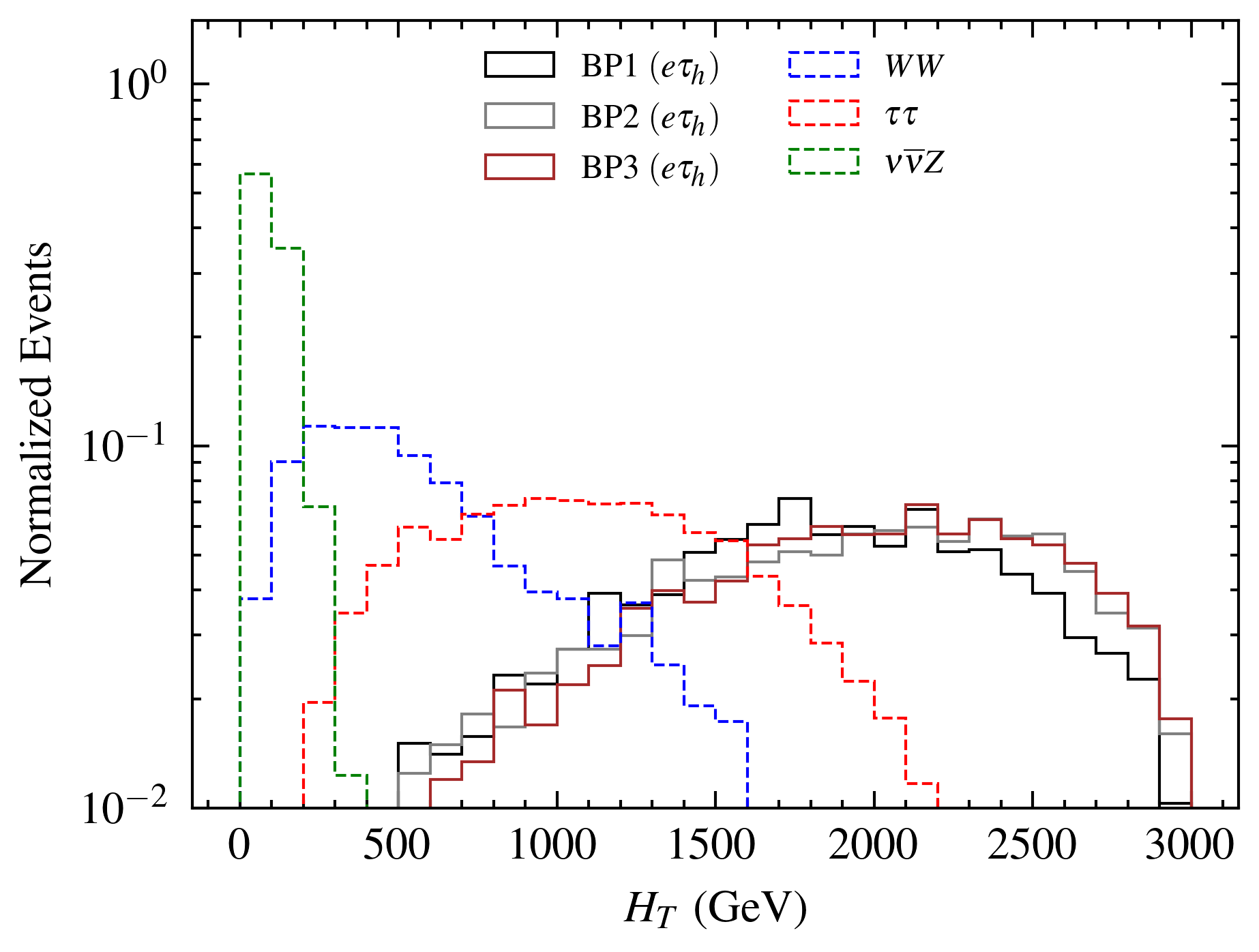}
	\caption{Kinematic distributions corresponding to signal and main background processes for $e^{+} e^{-} \rightarrow e \tau_h$ production at CLIC 3 TeV.}
	\label{fig:dist-et}
\end{figure}

The total cross-sections at $\sqrt{s} = 3$ TeV for different polarization settings are tabulated in Table~\ref{tab:BPs1} for $\ell \tau$ and background processes at production level. The signal cross-sections are noted for three benchmarks of EFT coefficients, adhering to the allowed bounds on these operators as stated in Table~\ref{tab:lfv}. As discussed previously, for signal process, the operators $(\mathcal{O}_{\ell \ell})_{\tau \ell}$ and $(\mathcal{O}_{e e})_{\tau \ell}$ have fixed chirality, they are very sensitive to the polarization settings, $P_{e^{-}} = \pm 80 \%$. The operator $(\mathcal{O}_{\ell e})_{\tau \ell e e}$ has mixed chirality and is unaffected by the different polarization tuning. The dominant backgrounds $WW$ and $\nu \overline{\nu} Z$ are left chiral owing to the gauge structure of the SM, hence, polarization choice of $P_{e^{-}} = +80\%$ significantly reduce the background cross-section and $P_{e^{-}} = -80\%$ choice enhance the SM background. The invariant mass, $M_{\mu \tau}/M_{e \tau}$ and $H_{T}$ distributions for the signal benchmarks and the major backgrounds are plotted in Figs.~\ref{fig:dist-mt} and \ref{fig:dist-et}. The invariant mass is defined as:
\begin{equation}
	M_{\ell \tau} = \sqrt{\left(p_{\ell} + p_{\tau_{h}}\right)^{2}},
\end{equation}
where, $p_{\ell}$ and $p_{\tau_{h}}$ are the 4-momenta of the lepton and  $\tau$ jet, respectively. The $H_{T}$ variable is defined as:
\begin{equation}
	\label{eq:ht}
	H_{T} = \sum_{visible} p_{T}.
\end{equation}
This is essentially the scalar sum of $p_{T}$ of visible particles. There are different ways in which $H_{T}$ is defined in the collider literature \cite{Schwartz:2017hep}, but we will resort to the definition in Eq.~\eqref{eq:ht}. Additional distributions are shown in Figs.~\ref{fig:dist-mt1} and \ref{fig:dist-et1} of Appendix \ref{sec:ca}.

We perform a cut and count analysis based on the distributions. The cutflows are detailed in Table~\ref{tab:cutflow1}. We apply three sequential cuts as itemized below. Prior to $\mathcal{C}_{0}$, we apply detector resolution and efficiency criteria.
\begin{itemize}
	\item $\mathcal{C}_{0}$: $N_{\mu / e} = 1$, $N_{\tau_{h}} = 1$.
	\item $\mathcal{C}_{1}$: $M_{\mu \tau}/M_{e \tau} > 2$ TeV.
	\item $\mathcal{C}_{2}$: $H_{T} >$ 1.5 TeV.
\end{itemize}
Here, $N_{\mu/e}$ is the number of muons/electrons. For $\mu \tau$, $N_{e}$ is set to 0 and for $e \tau$, $N_{\mu}$ is set 0. $N_{\tau_{h}}$ refers to the number of tau tagged jets. No additional jet is allowed. $M_{\mu \tau}/M_{e \tau}$ refers to the invariant mass of $\mu/e$ and $\tau$ jet. The signal process arises from a contact interaction with minimal branching, hence the invariant mass distribution is expected to peak close to the CM energy of the process, segregating it from the backgrounds which essentially peak at lower values. The invariant mass of $\nu \overline{\nu} Z$ is expected to peak entirely around the Z pole, and gets entirely wiped out by the invariant mass cut. The $H_{T}$ distribution of the signal is shifted towards the higher end of the distribution and that of the background is peaked towards the lower end, owing to the fact that the backgrounds mimicking $\ell \tau_{h}$ final state is usually accompanied by huge swarm of invisible particles, reducing the energies of the visible particles. Also, it should be noted that the effect of the $H_{T}$ cut can also be replicated by a missing energy ($\slashed{E}$) cut instead, due to the same reason. After employing all the kinematical cuts on collider variables, we estimate the efficiency factor ($\epsilon$) which is crucial to estimate the optimal sensitivity of NP couplings as we discuss next. The $\epsilon$ is defined as $\epsilon=\sigma^{\text{sig}}/\sigma^{\text{prod}}$, where $\sigma_{\text{prod}}$ is the production cross-section and $\sigma^{\text{sig}}$ is the signal cross-section for the chosen final state after implementing all the cuts along the branching ratios.The signal efficiency ($\epsilon_s$) for three benchmark points (BPs) is as follows: for BP1, $\epsilon_{s}$ is 0.157; for BP2, $\epsilon_{s}$ is 0.190; and for BP3, $\epsilon_{s}$ is 0.182. The background efficiency ($\epsilon_b$) for the dominant SM background is 0.01.

\begin{table}[htb!]
	\centering
	{\renewcommand{\arraystretch}{1.2}
		\begin{tabular}{|c|c|c|c|}
			\hline
			$\left\{(C_{\ell \ell})_{\tau \ell} / \Lambda^{2}, (C_{e e})_{\tau \ell} / \Lambda^{2}, (C_{\ell e})_{\tau \ell} / \Lambda^{2} \right\}$ &
			\multicolumn{3}{c|}{Cross-section (fb)} \\
			\cline{2-4}
			($\times 10^{-9}$ GeV$^{-2}$) & $P_{e^{-}}= 0\%$ & $P_{e^{-}}= +80\%$ & $P_{e^{-}}= -80\%$ \\
			\hline
			BP1: $\{1.0,0.0,0.0\}$ & 0.74 & 0.15 & 1.33  \\
			BP2: $\{0.0,1.0,0.0\}$  & 0.74 & 1.33 & 0.15 \\
			BP3: $\{0.0,0.0,1.0\}$  & 0.37 & 0.37 & 0.37 \\
			\hline
			Backgrounds & $P_{e^{-}}= 0\%$ & $P_{e^{-}}= +80\%$ & $P_{e^{-}}= -80\%$ \\
			\hline
			$W^{+} W^{-}$ & 453.7 & 91.89 & 814.2 \\
			$\tau^{+} \tau^{-}$ & 12.23 & 11.70 & 12.76 \\
			$\nu \overline{\nu} Z$ & 2090 & 419.6 & 3751 \\
			\hline
	\end{tabular}}
	\caption{Total cross-section of $e^{+} e^{-} \rightarrow \ell \tau$ ($\ell = e, \mu$) for different four-Fermi couplings as well as SM backgrounds for different choices of beam polarization combination with $\sqrt{s}=3$ TeV.}
	\label{tab:BPs1}
\end{table}

\begin{table}[h!]
	\centering
	{\renewcommand{\arraystretch}{1.2}
		\begin{tabular}{|c|c|c|c|c|c|c|c|c|c|}
			\hline
			\multirow{2}*{Processes} & \multicolumn{3}{c|}{$\mathcal{C}_{0}:$ Selection cuts} & \multicolumn{3}{c|}{$\mathcal{C}_{1}: M_{\mu \tau} > 2$ TeV} & \multicolumn{3}{c|}{$\mathcal{C}_{2}: H_{T} > 1.5$ TeV} \\
			\cline{2-10}
			& $P_{0}$ & $P_{+}$ & $P_{-}$ & $P_{0}$ & $P_{+}$ & $P_{-}$ & $P_{0}$ & $P_{+}$ & $P_{-}$ \\
			\hline
			BP1 $(\mu \tau_h)$ & 0.238 & 0.048 & 0.427 & 0.157 & 0.032 & 0.281 & 0.116 & 0.024 & 0.208 \\
			BP2 $(\mu \tau_h)$ & 0.232 & 0.415 & 0.047 & 0.190 & 0.342 & 0.038 & 0.141 & 0.253 & 0.028 \\
			BP3 $(\mu \tau_h)$ & 0.118 & 0.118 & 0.118 & 0.087 & 0.087 & 0.087 & 0.067 & 0.067 & 0.067 \\
			\hline
			Background & 5.979 & 2.275 & 9.663 & 0.312 & 0.194 & 0.430 & 0.162 & 0.123 & 0.201 \\
			\hline
	\end{tabular}}
	\vspace{0.25cm}
	
	{\renewcommand{\arraystretch}{1.2}
		\begin{tabular}{|c|c|c|c|c|c|c|c|c|c|}
			\hline
			\multirow{2}*{Processes} & \multicolumn{3}{c|}{$\mathcal{C}_{0}:$ Selection cuts} & \multicolumn{3}{c|}{$\mathcal{C}_{1}: M_{e \tau} > 2$ TeV} & \multicolumn{3}{c|}{$\mathcal{C}_{2}: H_{T} > 1.5$ TeV} \\
			\cline{2-10}
			& $P_{0}$ & $P_{+}$ & $P_{-}$ & $P_{0}$ & $P_{+}$ & $P_{-}$ & $P_{0}$ & $P_{+}$ & $P_{-}$ \\
			\hline
			BP1 $(e \tau_h)$ & 0.216 & 0.044 & 0.387 & 0.146 & 0.030 & 0.262 & 0.114 & 0.023 & 0.206 \\
			BP2 $(e \tau_h)$ & 0.208 & 0.372 & 0.042 & 0.174 & 0.313 & 0.035 & 0.133 & 0.239 & 0.027 \\
			BP3 $(e \tau_h)$ & 0.105 & 0.105 & 0.105 & 0.080 & 0.080 & 0.080 & 0.066 & 0.066 & 0.066 \\
			\hline
			Background & 5.134 & 1.957 & 8.295 & 0.252 & 0.161 & 0.342 & 0.133 & 0.104 & 0.162 \\
			\hline
	\end{tabular}}
	\caption{Cutflow cross-sections (in fb) corresponding to signal and background for different beam polarization choices at the CLIC with $\sqrt{s} = 3$ TeV. Here, $P_{0} \to P_{e^{-}}= 0\%$, $P_{+} \to P_{e^{-}}= +80\%$ and $P_{-} \to P_{e^{-}}= -80\%$.}
	\label{tab:cutflow1}
\end{table}

\subsection{Signal significance}
\begin{figure}[h!]
	\centering
	\includegraphics[width = 0.4 \textwidth, height = 0.37 \textwidth]{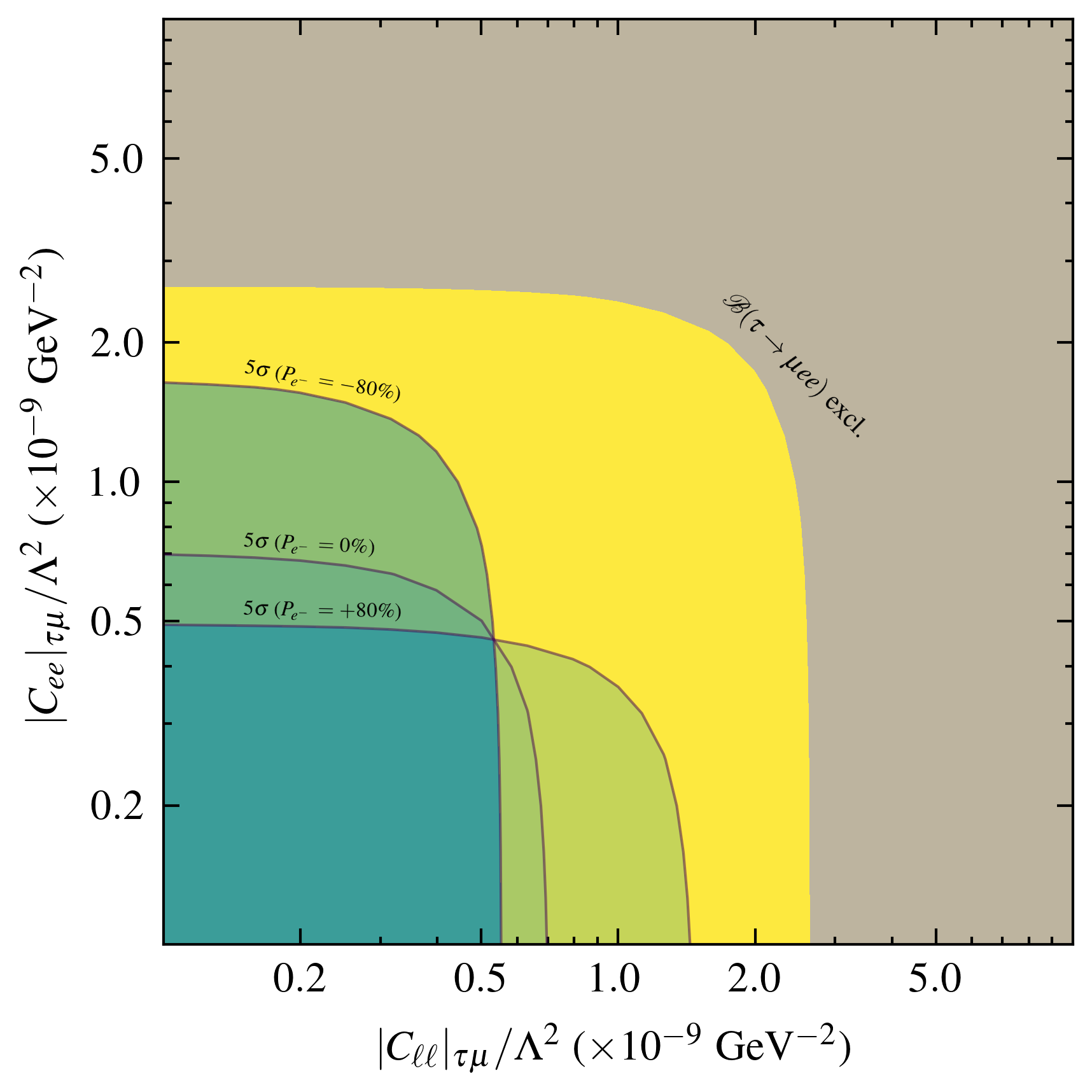}
	\includegraphics[width = 0.4 \textwidth, height = 0.37 \textwidth]{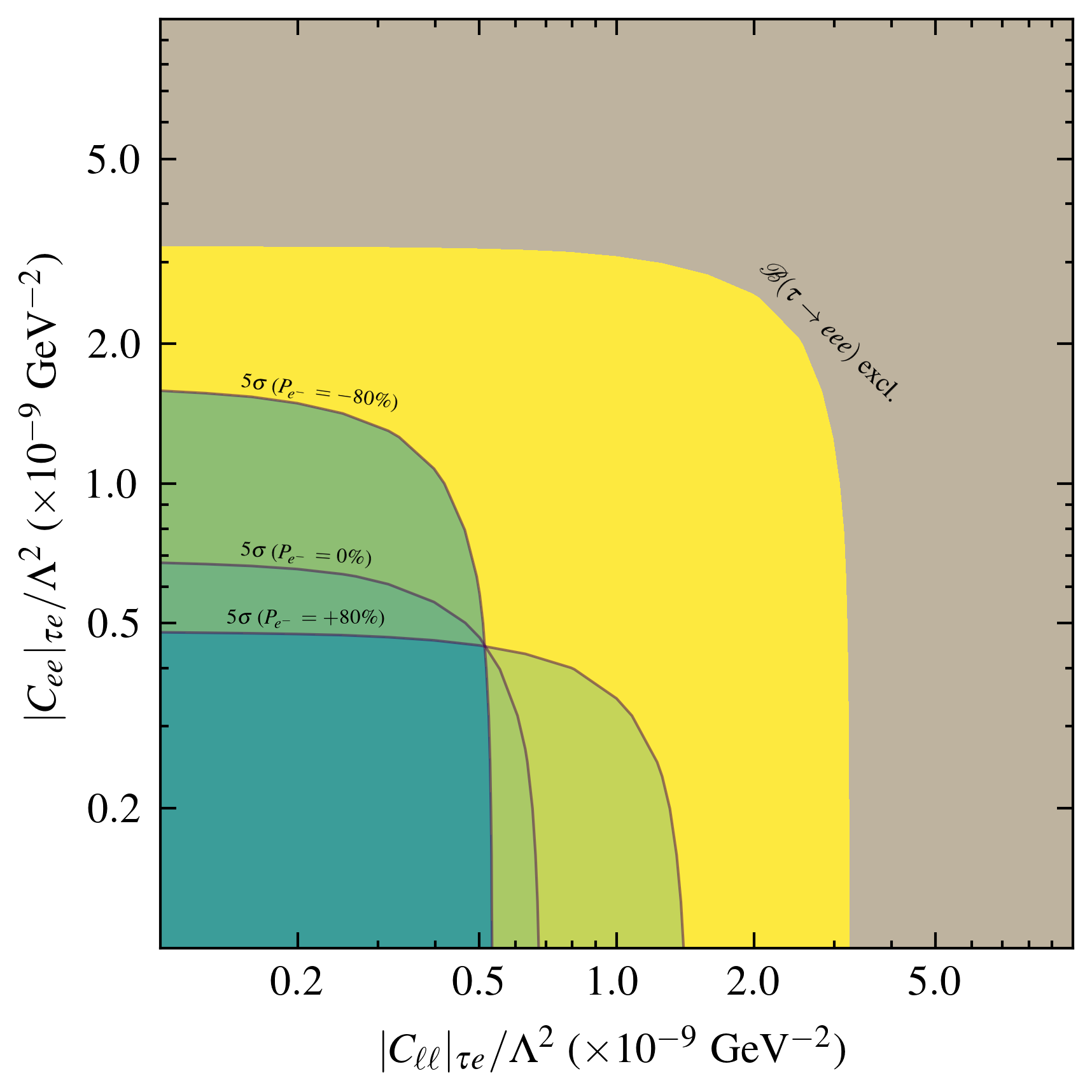} \\
	\includegraphics[width = 0.4 \textwidth, height = 0.37 \textwidth]{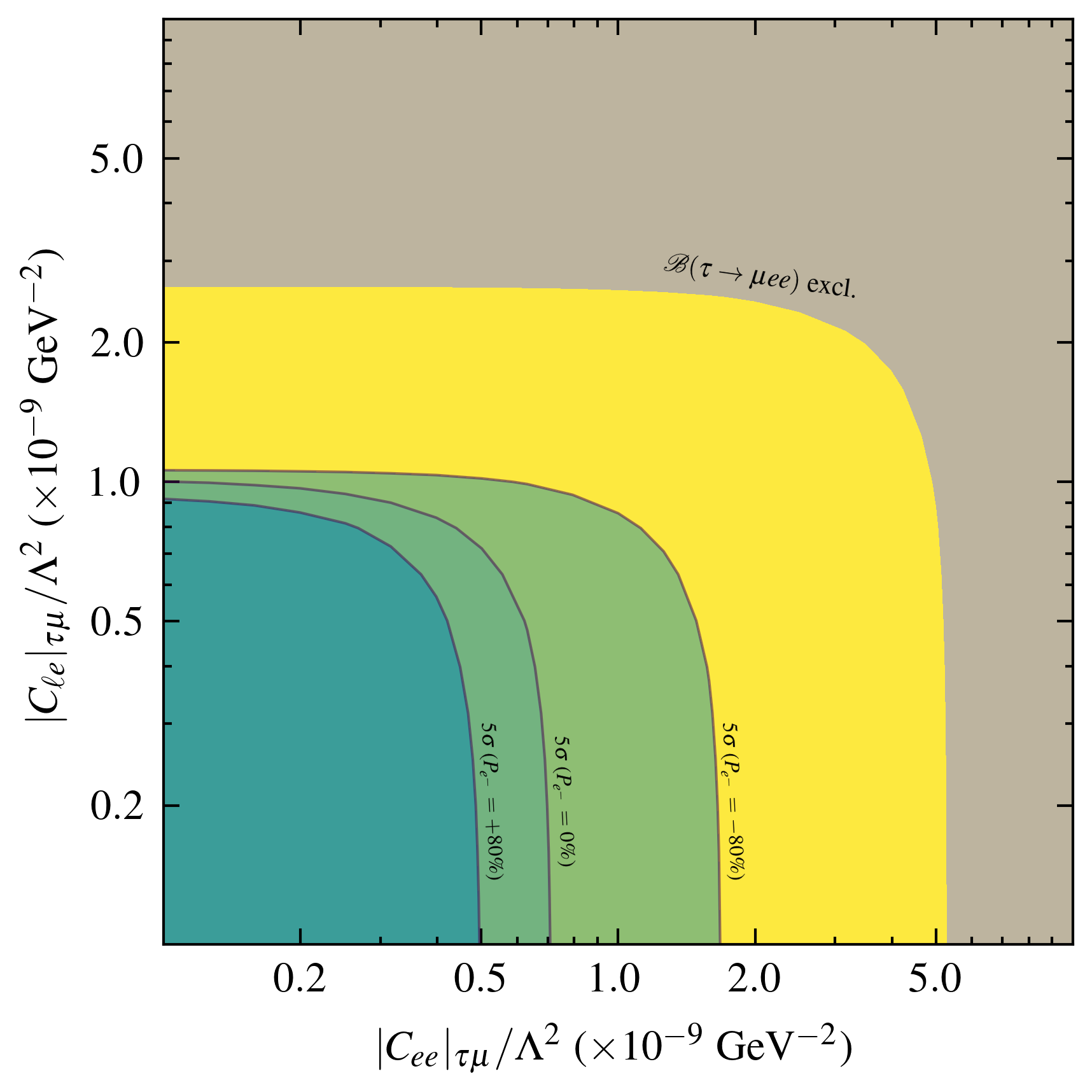}
	\includegraphics[width = 0.4 \textwidth, height = 0.37 \textwidth]{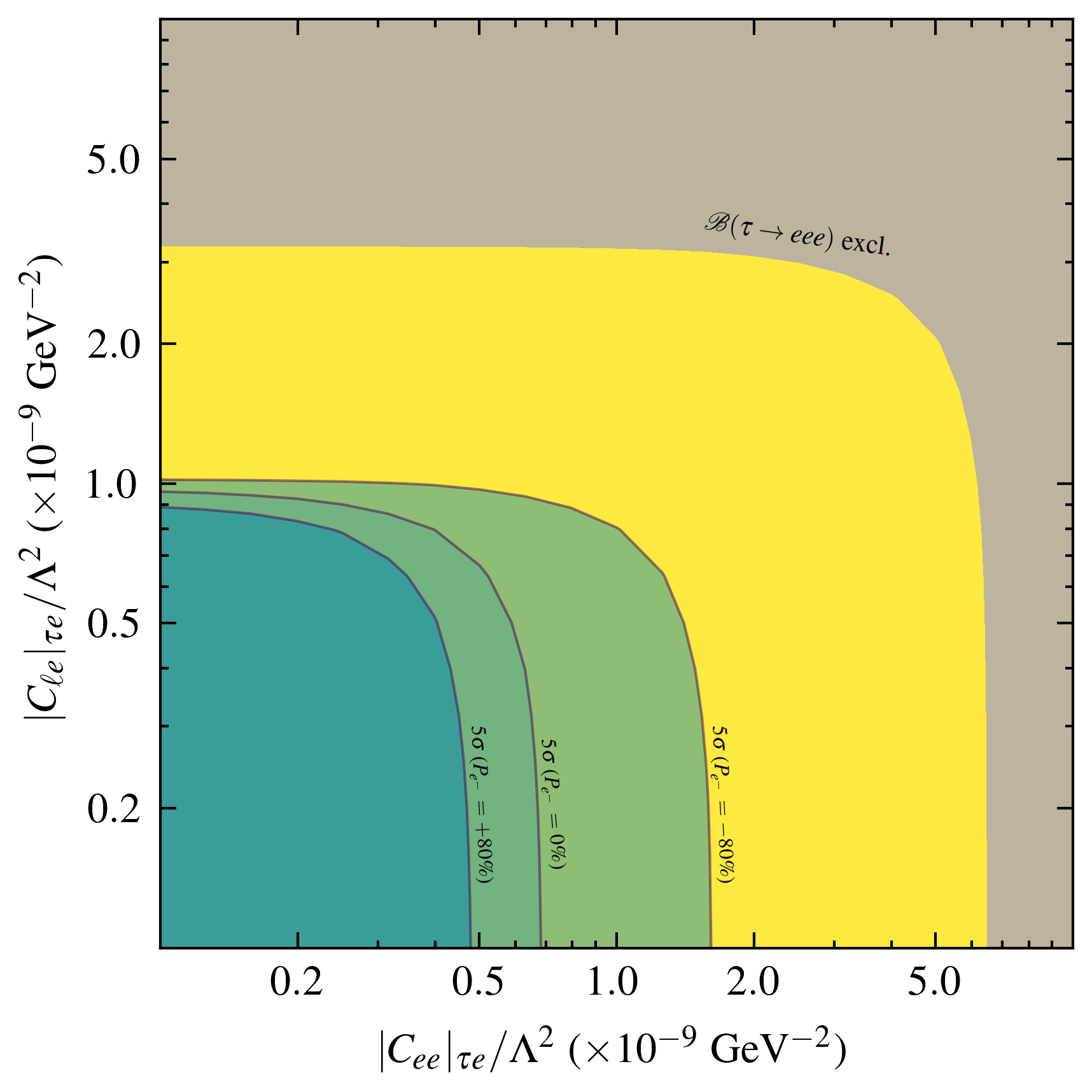} \\
	\includegraphics[width = 0.4 \textwidth, height = 0.37 \textwidth]{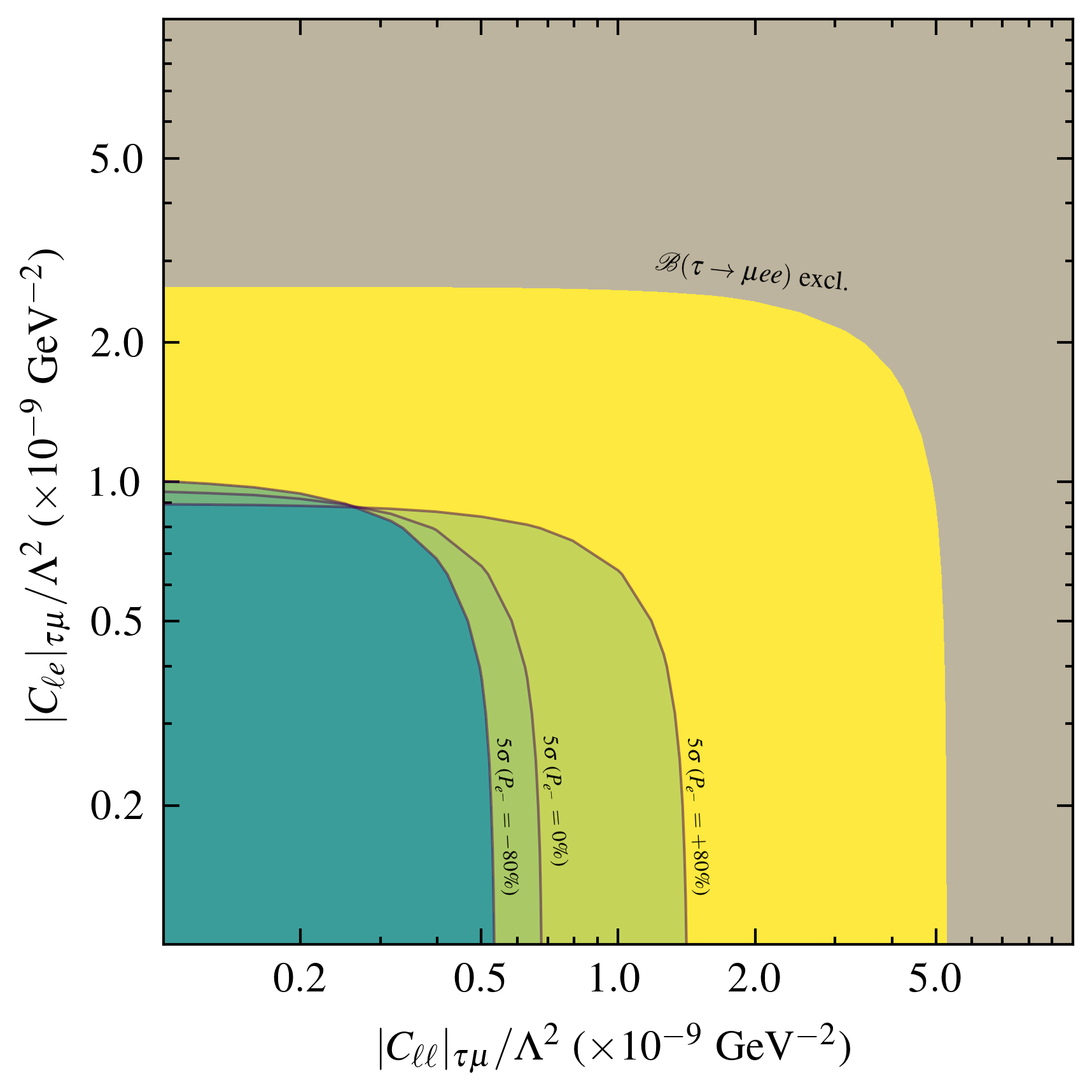}
	\includegraphics[width = 0.4 \textwidth, height = 0.37 \textwidth]{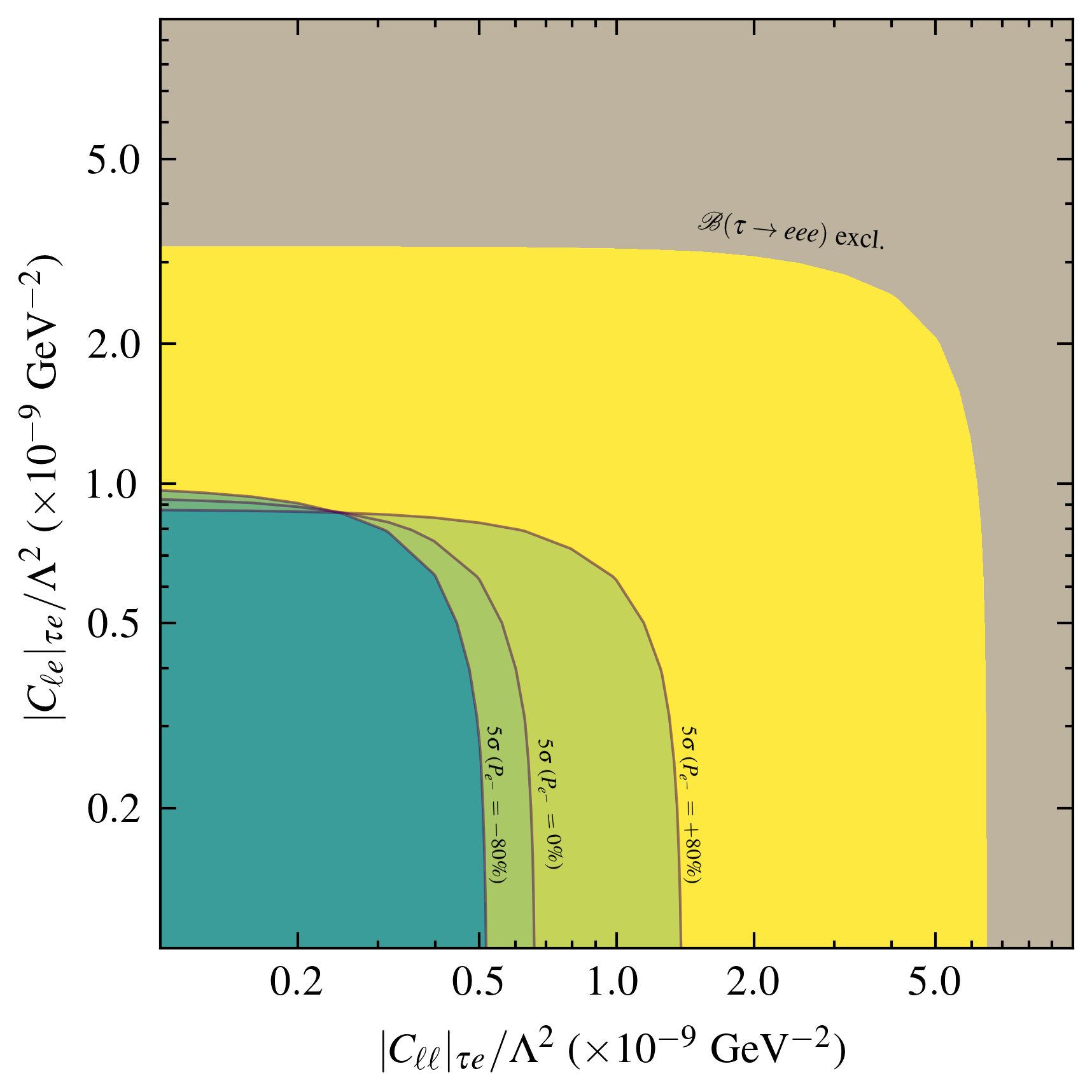} \\
	\caption{Significance plots corresponding to the process $e^{+} e^{-} \rightarrow \ell \tau_h$ at $\sqrt{s}$ = 3 TeV and $\mathfrak{L}_{\text{int}} = 1000$ $\rm{fb^{-1}}$. The solid lines refers to the $5\sigma$ signal significance for different polarization settings. The region excluded from $\tau$ branching ratios are also shown in the plots.}
	\label{fig:dist-s1}
\end{figure}

For signal significance we use the definition:
\begin{equation}
	\mathfrak{Z} = \dfrac{S}{\sqrt{B}} = \dfrac{\sigma_{S} \times \mathfrak{L}_{\text{int}}}{\sqrt{\sigma_{B} \times \mathfrak{L}_{\text{int}}}} = \dfrac{\sigma_{S} \times \sqrt{\mathfrak{L}_{\text{int}}}}{\sqrt{\sigma_{B}}}
\end{equation}
Here, $S$, $B$, $\sigma_{S}$, $\sigma_{B}$ and $\mathfrak{L}_{\text{int}}$ are the number of signal and background events, the signal and background cross-sections, and the integrated luminosity, respectively. $\mathfrak{Z}$ gives the number of sigmas, by which the NP signal, $S$, out-stands over the uncertainty in the SM background, $\sqrt{B}$. The significance plots (at 5$\sigma$ level), based on the cut based analysis in the previous section, on the parameter spaces of the EFT coefficients are shown in Fig. \ref{fig:dist-s1} in two-parameter plane along with the exclusion from the flavor-violating three body decays of $\tau$. It is observed that due to the chiral structure of the operators, polarization plays an important role in probing the operators.

\section{Optimal Observable Technique}
\label{sec:oot}
The Optimal Observable Technique (OOT) is a convenient tool for determining the statistical sensitivity of any NP coupling in an optimal way. Here, we provide a brief overview of the mathematical framework of the OOT, which has already been explained in detail in previous studies \cite{Diehl:1993br,Gunion:1996vv,Bhattacharya:2021ltd}. Any observable such as the differential cross-section that receives contributions from both the SM and  BSM can be expressed as
\begin{equation}
	\mathcal{O}(\phi)=\frac{d\sigma}{d\phi}=g_i f_i(\phi),
\end{equation}
where $g_i$'s are the function of NP coefficients and $f_i(\phi)$'s are the function of phase space variable $\phi$. As our analysis is based on the process $e^+e^- \to \mu \tau$, the cosine of the emerging angle of the outgoing particle ($\cos \theta$) is the phase-space variable of our interest. Alternative variables may be chosen instead of $\cos \theta$ depending on the specific observable/process being studied.

Our goal is to determine $g_i$. This can be achieved by utilizing an appropriate weighting function ($w_i(\phi)$):
\begin{equation}
	g_i=\int w_i(\phi) \mathcal{O}(\phi) d\phi,
\end{equation}
In principle, various options for $w_i(\phi)$ are feasible, but there exists a particular selection for which the covariance matrix ($V_{ij}$) is optimal. This choice minimizes statistical uncertainties in NP couplings. For this specific selection, $V_{ij}$ follows:
\begin{equation}
	V_{ij} \propto \int w_i(\phi) w_j(\phi) \mathcal{O}(\phi) d\phi,
\end{equation}
Hence, the weighting functions that fulfill the optimal condition $\delta V_{ij} = 0$ are
\begin{equation}
	w_i(\phi)=\frac{M^{-1}_{ij}f_j(\phi)}{\mathcal{O}(\phi)},
\end{equation}
where
\begin{equation}
	M_{ij}=\int \frac{f_i(\phi) f_j(\phi)}{\mathcal{O}(\phi)}d\phi.
\end{equation}
Next, the optimal covariance matrix takes shape as follows:
\begin{equation}
	V_{ij}=\frac{M^{-1}_{ij}}{\mathfrak{L}_{\tt int}}.
\end{equation}
Here, $\sigma_T = \int \mathcal{O}(\phi)d\phi$, and $N$ represents the total number of events ($N = \sigma_T \mathfrak{L}_{\tt int}$). $\mathfrak{L}_{\tt int}$ denotes the integrated luminosity.

The function $\chi^2$, which dictates the optimal constraint on NP couplings, is defined as
\begin{equation}
	\chi^2=\sum_{ij}(g_i-g_i^0)(g_j-g_j^0)V^{-1}_{ij},
	\label{eq:oot.chi.sqr}
\end{equation}
where  $g_0$'s are `seed values' that are dependent on the particular NP scenario. The limit set by $\chi^2 \leq n^2$ corresponds to $n\sigma$ standard deviations from these seed values ($g_0$), establishing the optimal limit for NP couplings while assuming the covariance matrix ($V_{ij}$) is minimized. Using the $\chi^2$ function definition in Eq.~\eqref{eq:oot.chi.sqr}, the optimal constraints on NP couplings have been investigated in subsequent sections.

\subsection{Optimal sensitivity of effective couplings}
\begin{figure}[htb!]
	\centering
	\includegraphics[height=6cm, width=5.4cm]{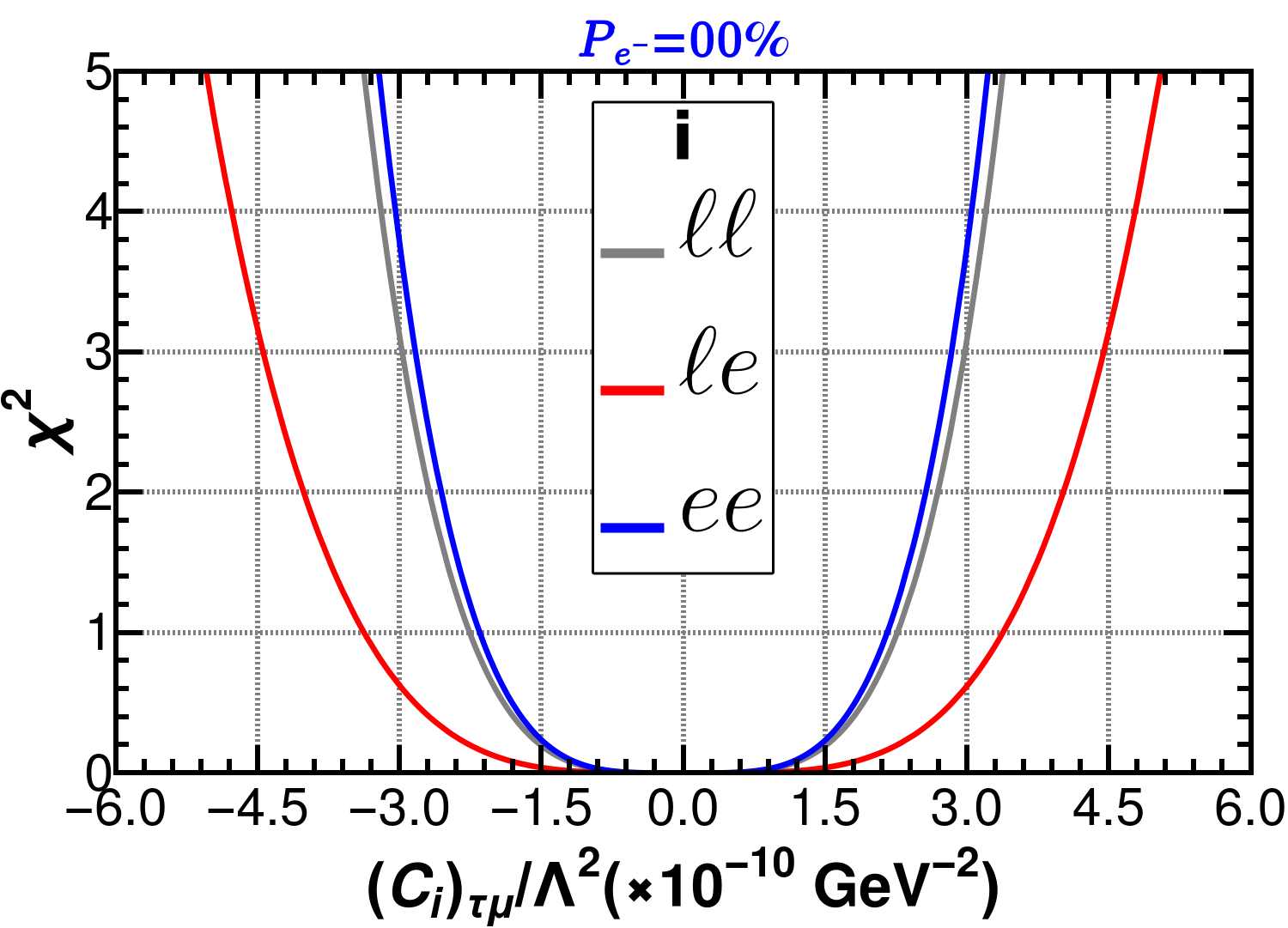}~
	\includegraphics[height=6cm, width=5.4cm]{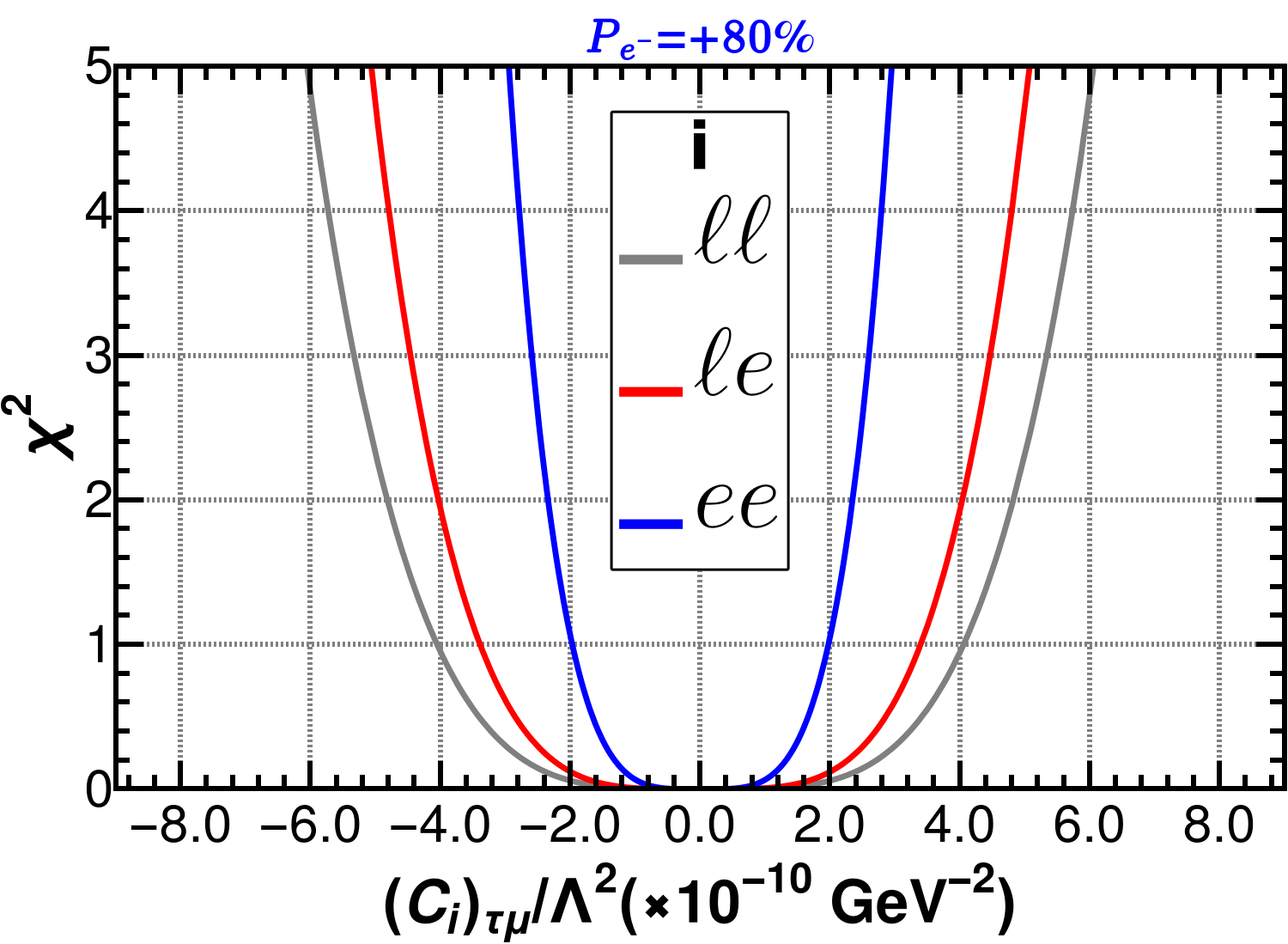}~
	\includegraphics[height=6cm, width=5.4cm]{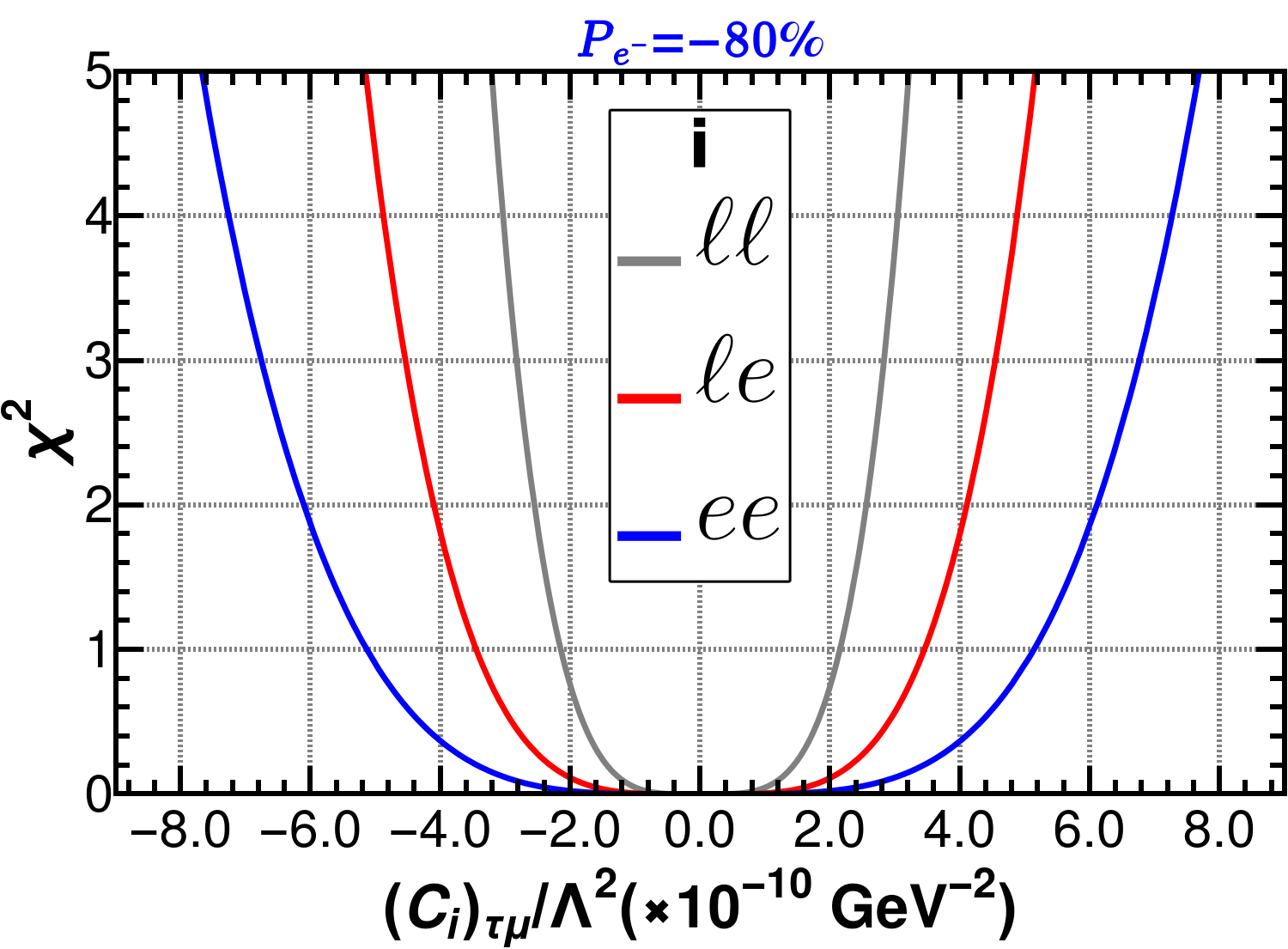}
	\caption{Optimal $\chi^2$ variations for four-Fermi effective couplings with different choices of beam polarization with $\sqrt{s}$ = 3 TeV and $\mathfrak{L}_{\text{int}}=1000~ \rm{fb}^{-1}$. Left: unpolarized beam, middle: $\{P_{e^-}:P_{e^+}=+80\%:00\%\}$, right: $\{P_{e^-}:P_{e^+}=-80\%:00\%\}$.}
	\label{fig:oot.1d}
\end{figure}

In this section, we explore the optimal sensitivity of dimension-6 flavor-violating effective couplings via $\tau \mu$ production at the $e^+e^-$ collider with $\sqrt{s}$ = 3 TeV and $\mathfrak{L}_{\text{int}}=1000~\rm{fb^{-1}}$. Using Eq.~\eqref{eq:oot.chi.sqr}, optimal $\chi^2$ variation with different NP couplings (one operator scenario) are shown in Fig.~\ref{fig:oot.1d} for several choices of beam polarization and optimal limits (95\% C.L.) are presented in Table~\ref{tab:95cl}. Given the CM energy and luminosity of a specific collider, the sensitivity of a particular flavor violating NP coupling depends on the relative contribution to the $\tau \mu$ production, efficiency factor for a final state, and beam polarization.  For unpolarized beam, the sensitivity of $(C_{\ell \ell})_{\tau \mu}/\Lambda^2$ and $(C_{\ell e})_{\tau \mu}/\Lambda^2$ appear to be similar as their contributions to the $\tau \mu$ production are equal. However, there is a slight betterment for $(C_{\ell \ell})_{\tau \mu}/\Lambda^2$ as the efficiency factor is relatively better for this coupling. Due to the spinor structure of $\mathcal{O}_{\ell \ell}$, $(C_{\ell \ell})_{\tau \mu}/\Lambda^2$ provides the maximum cross-section for $P_{e^-}=-80\%$ choice, hence provides best sensitivity among these three polarization combination. Whereas, due to the similar reason, best sensitivity is achieved for $P_{e^-}=+80\%$ choice in case of $(C_{e e})_{\tau \mu}/\Lambda^2$. For $(C_{\ell e})_{\tau \mu}/\Lambda^2$, all polarization combinations have the same cross-section, therefore, the sensitivity of this coupling are expected to be similar for all polarization combinations. However, due to the SM background reduction, $P_{e^-}=+80\%$ provides a delicate enhancement for this coupling. It is noteworthy to mention that, at $\sqrt{s}$ = 3 TeV, the CLIC is expected to surpass the flavor sensitivity of NP couplings (from three body decays of $\tau$ lepton) with $\mathfrak{L}_{\text{int}}=1~\rm{fb^{-1}}$. At $\mathfrak{L}_{\text{int}}=1000~\rm{fb^{-1}}$, the sensitivity of these NP couplings could enhanced by one order compared to flavor violating tau decays.

\begin{table}[htb!]
	\centering
	\begin{tabular}{| c | c |  c | c |  c |  } 
		\hline
		\multicolumn{1}{|c}{Couplings} &
		\multicolumn{3}{|c|}{Sensitivity (95\% C.L.) $\times 10^{-10}$} \\
		\cline{2-4}
		\multicolumn{1}{|c}{$\rm (GeV^{-2})$}&
		\multicolumn{1}{|c|}{$P_{e^-} = 00\%$} &
		\multicolumn{1}{c|}{$P_{e^-} = -80\%$}&
		\multicolumn{1}{c|}{$P_{e^-} = +80\%$}\\
		\hline
		$(C_{\ell \ell})_{\mu \tau}/\Lambda^2$& $\pm3.18$ & $\pm3.04$ & $\pm5.69$   \\
		\hline
		$(C_{\ell e})_{\mu \tau}/\Lambda^2$& $\pm4.75$ & $\pm4.82$ & $\pm4.70$ \\
		\hline
		$(C_{ee})_{\mu \tau}/\Lambda^2$& $\pm3.03$ & $\pm7.21$ & $\pm2.80$ \\
		\hline
	\end{tabular}
	\caption{Optimal sensitivity at 95\% C.L. on dimension-6 flavor-violating effective couplings at the CLIC with $\sqrt{s}=3$ TeV CM energy and $\mathfrak{L}_{\text{int}}=1000~\rm{fb}^{-1}$ luminosity.}
	\label{tab:95cl}
\end{table}

\begin{figure}[htb!]
	\centering
	\includegraphics[height=6cm, width=5.4cm]{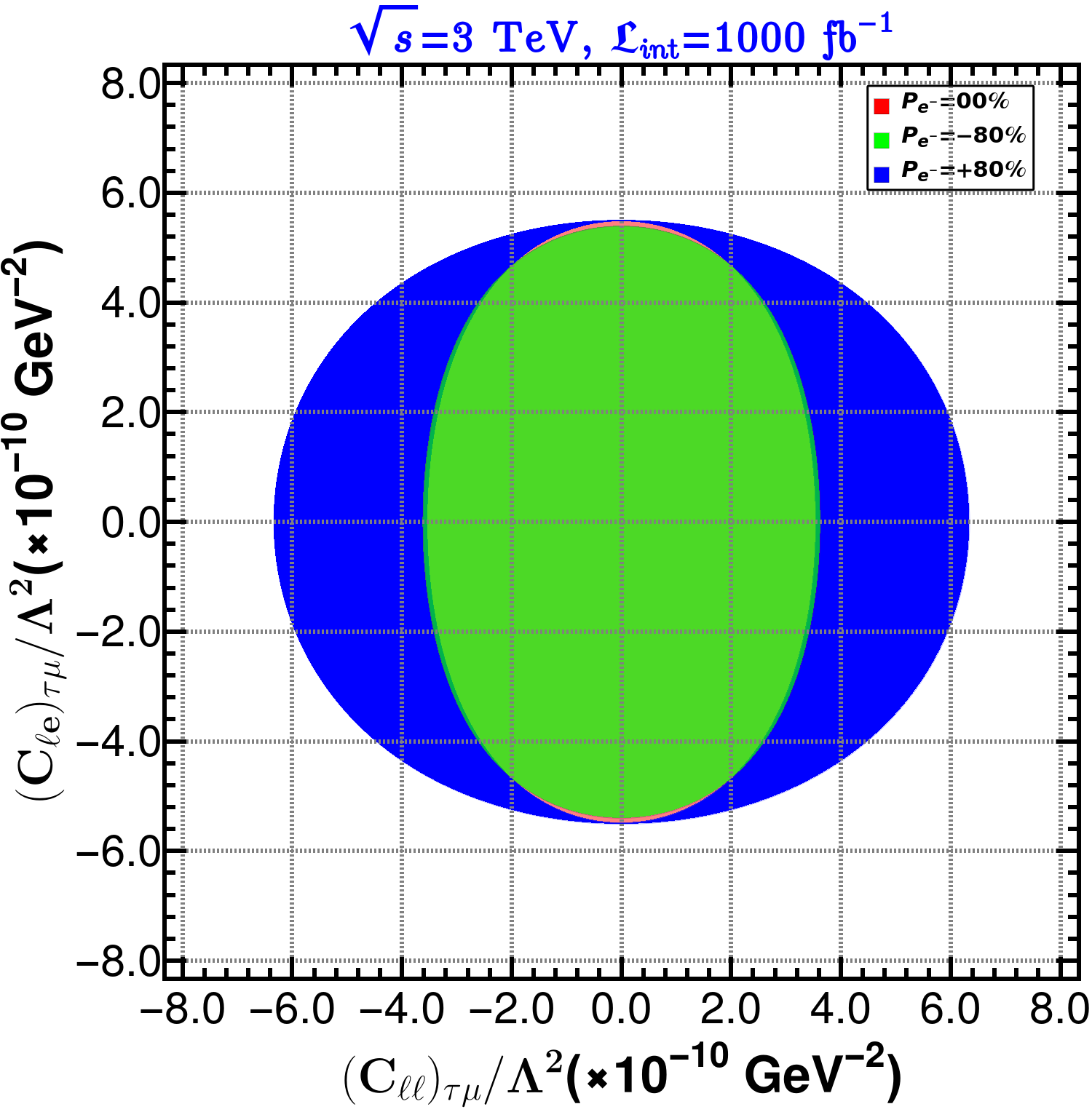}~
	\includegraphics[height=6cm, width=5.4cm]{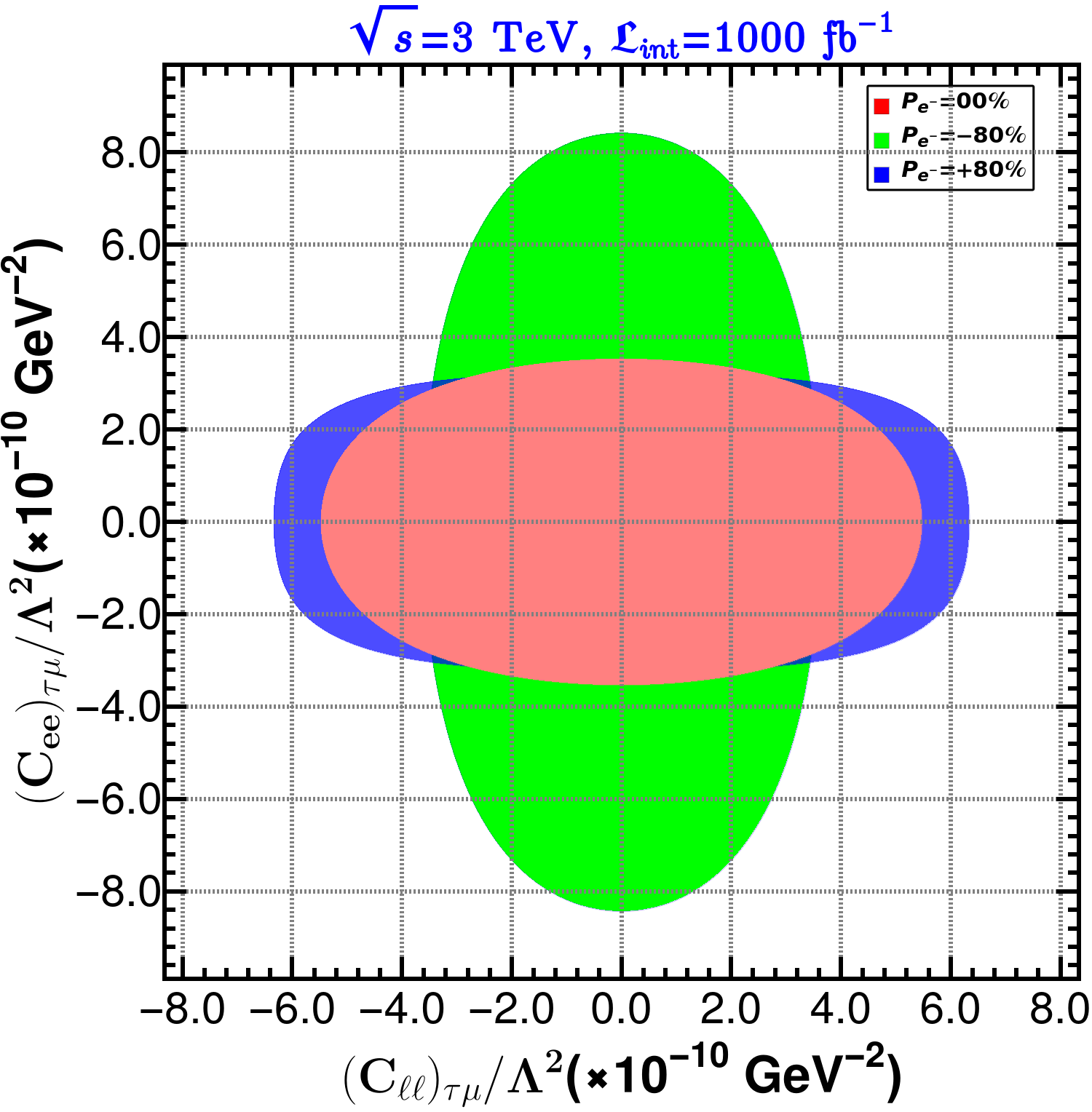}~
	\includegraphics[height=6cm, width=5.4cm]{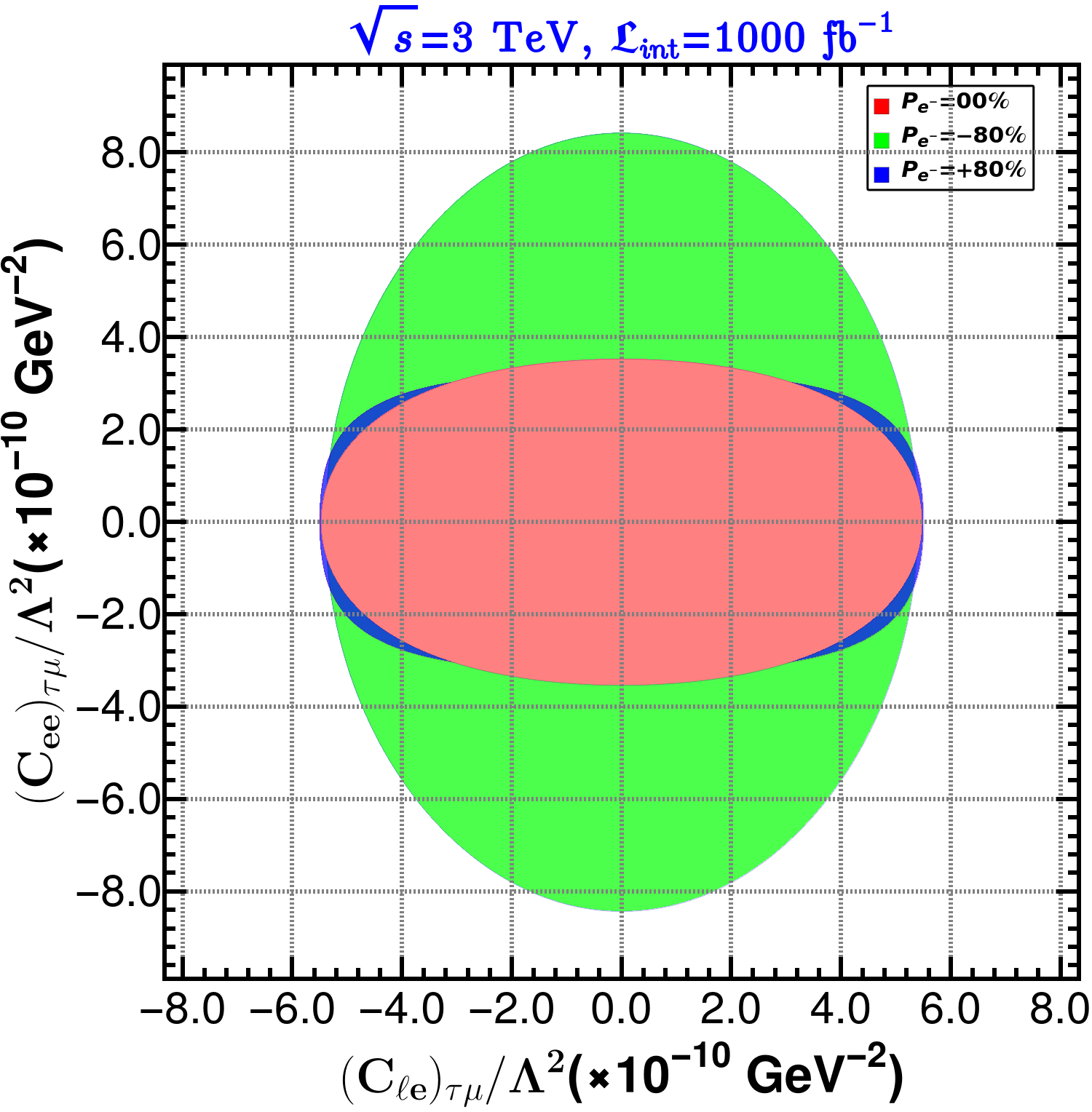}
	\caption{Optimal 95\% C.L. region between two different dimension-6 effective couplings at $\sqrt{s}$ = 3 TeV and $\mathfrak{L}_{\text{int}}=1000~ \rm{fb}^{-1}$. Left: $(C_{\ell \ell})_{\tau \mu}/\Lambda^2-(C_{\ell e})_{\tau \mu}/\Lambda^2$ plane, middle: $(C_{\ell \ell})_{\tau \mu}/\Lambda^2-(C_{ee})_{\tau \mu}/\Lambda^2$ plane, right: $(C_{\ell e})_{\tau \mu}/\Lambda^2-(C_{e e})_{\tau \mu}/\Lambda^2$ plane.}
	\label{fig:oot.2d}
\end{figure}

\begin{figure}[htb!]
	\centering
	\includegraphics[height=6cm, width=6.4cm]{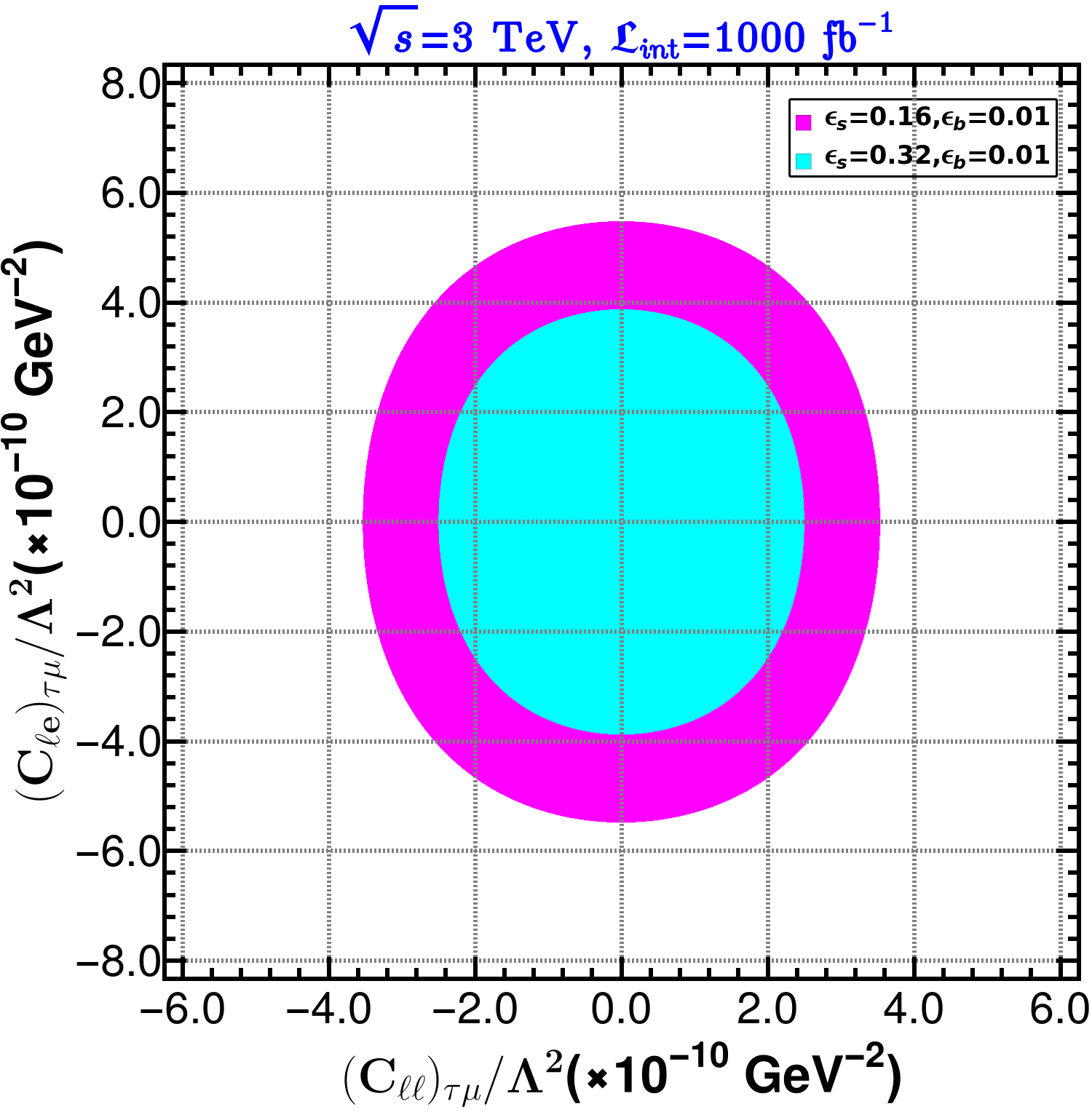}~
	\includegraphics[height=6cm, width=6.4cm]{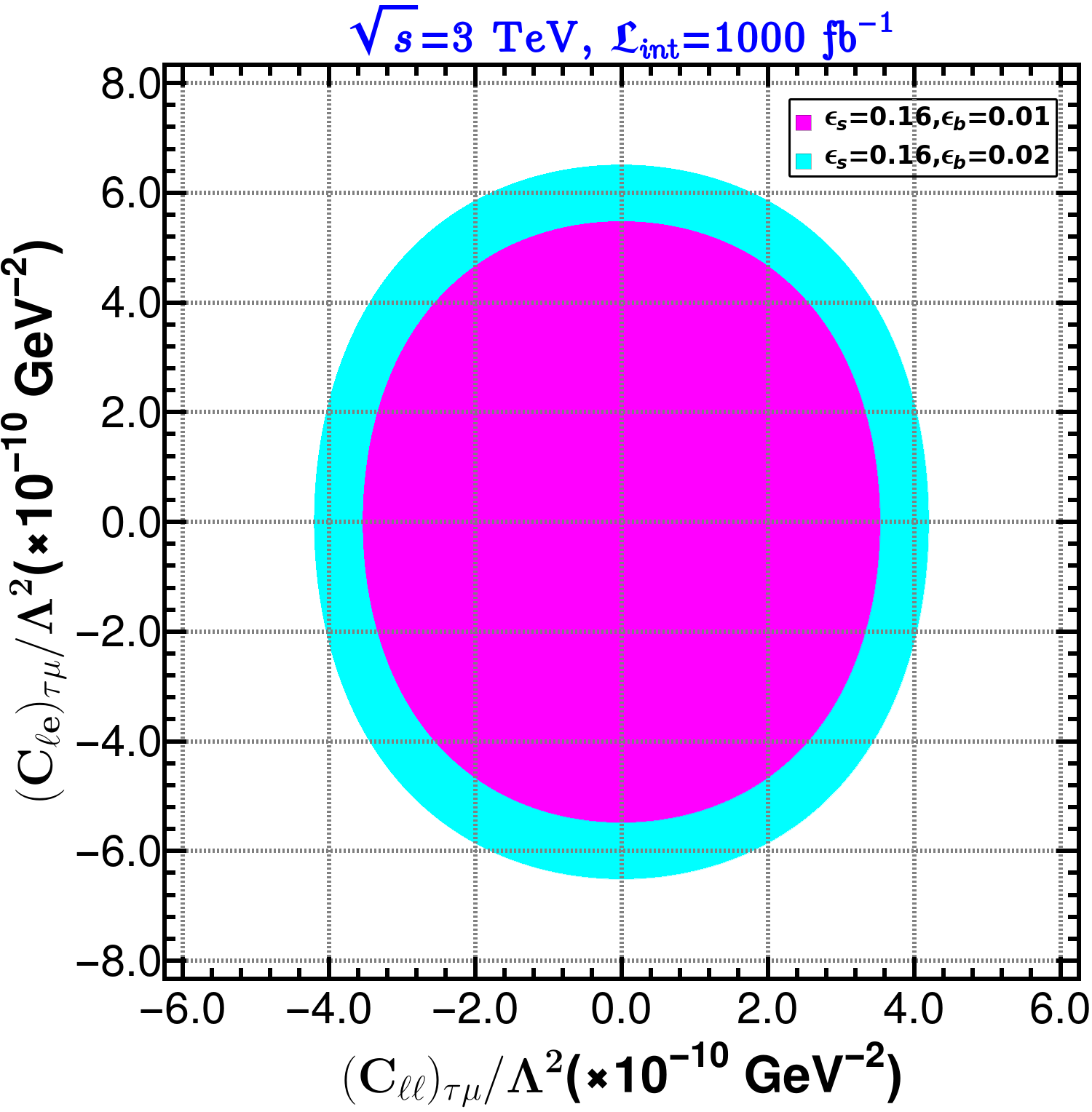}
	\caption{Variation of optimal 95\% C.L. region with the change in the efficiency factors. Left: $\epsilon_s$ is enhanced by a factor of 2 by keeping $\epsilon_b$ fixed, right: $\epsilon_b$ is enhanced by a factor of 2 keeping $\epsilon_s$ fixed.}
	\label{fig:2d.eps}
\end{figure}

Now, we turn to discuss the effect of signal and background efficiency to estimate the sensitivity of NP couplings. If we increase\footnote{The enhancement (reduction) of $\epsilon_s~(\epsilon_b)$ can be achieved by using multivariate analysis \cite{TMVA:2007ngy,Coadou:2013lca}.} the $\epsilon_s$ by a factor of 2 by keeping $\epsilon_b$ fixed, the sensitivity of a particular NP coupling improves by 30\% as shown in the left of Fig.~\ref{fig:2d.eps}. On the other hand, if we decrease $\epsilon_b$ by a factor of 2 by keeping $\epsilon_s$ constant, the sensitivity of the NP couplings enhances by 16\% (right plot of Fig.~\ref{fig:2d.eps}). Therefore, we conclude that increasing $\epsilon_s$ is more economical than decreasing $\epsilon_b$. In our analysis, $\epsilon_s$ is 15 times larger than $\epsilon_b$, making the signal contribution five times greater than the background in the chosen final state. As a result, our analysis is signal-dominated, which is why changing $\epsilon_s$ is more effective than changing $\epsilon_b$ in estimating the sensitivity of the NP couplings. It is worthwhile to mention that the change in sensitivity of the NP couplings depends on the relative contribution of signal and background to the final state which means if there is a scenario where background dominates (unlike our scenario) compare to signal for a particular final state then change in $\epsilon_b$ will affect the change in sensitivity of the NP couplings compare to the change in $\epsilon_s$. In Fig.~\ref{fig:oot.2d}, we show the 95\% C.L. allowed region in 2D parameter space for different choice of beam polarizations. We would like to highlight that, under the signal-only hypothesis ($\epsilon_b \to 0$), $\Delta g$  is inversely proportional to the CM energy and inversely proportional to the square root of the luminosity (for the case of contact interaction). This suggests that a high-energy lepton collider would be advantageous for estimating these type of couplings compared to high-luminosity lepton colliders.

\section{Conclusion}
\label{sec:con}

Lepton flavor violation (LFV), evidenced by neutrino oscillations yet absent in the SM, is crucial in particle physics for uncovering fundamental interactions beyond the SM. In this paper, we have discussed the estimation of dimension-6 flavor-violating effective couplings through $\ell \tau~(\ell=\mu,e)$ production at the future electron-positron colliders. After evaluating the upper limits on NP couplings from flavor-violating tau decays, we have performed cut-based analysis using $\ell \tau_{h}$ as our final state signal. Invariant dilepton mass and $H_T$ are the collider kinematical variables that play the crucial role to estimate the signal background estimation. 

After performing cut-based analysis, we have espoused the optimal observable technique to determine the optimal sensitivity of flavor violating effective couplings at the $e^+e^-$ colliders. At 3 TeV CM energy, the CLIC could surpass the upper bound on effective couplings obtained from flavor-violating tau decays at 1 $\text{fb}^{-1}$ integrated luminosity. If we further increase the luminosity up to 1000 $\text{fb}^{-1}$, then the upper bound the NP couplings can be tighter by one order of magnitude compared to the flavor bound. The signal and background efficiencies play a very important role to estimate the optimal sensitivity of NP couplings. In our scenario, as we are able to reduce the non-interfering SM backgrounds maximally after employing prudent kinematical cuts therefore enhancing the signal efficiency is more beneficial to achieve better optimal precision of NP couplings. Judicious choice of beam polarization is advantageous for assessing the sensitivity of the NP couplings. For instance, left (right)-polarized electron beam improves sensitivity by approximately (4\%) 8\% for $(C_{\ell \ell})_{\tau \mu}/\Lambda^2$ ($(C_{ee})_{\tau \mu}/\Lambda^2$) compared to the unpolarized beam. On a contrary, there is minuscule improvement of the sensitivity estimation in case of $(C_{\ell e})_{\tau \mu}/\Lambda^2$ for $P_{e^-}=+80\%$ choice. The interplay between the signal and background plays a very important role to estimate the sensitivity of NP couplings. As far as four-Fermi flavor violating operators of our concern,the optimal sensitivity is inversely proportional (for signal-only hypothesis) to CM energy, therefore, the high energy muon collider ($\sqrt{s}$ = 10, 14, and 30 TeV) is expected to provide better estimation of these type of couplings. Although, our analysis focuses on $\tau \mu$ production, but a similar approach can be applied to $\tau e$ production. While the effect of polarization will remain same, the sensitivity in estimating the NP couplings will have slight variations because of different efficiency factors.

\section*{Acknowledgements}
The authors would like to thank Subhaditya Bhattacharya and Jose Wudka for useful discussions.  SJ acknowledges financial support from CRG/2021/005080 project under the SERB, Govt. of India.

\appendix

\section{Others kinematical distributions}
\label{sec:ca}

\begin{figure}[htbp!]
	\centering
	\includegraphics[width = 0.4 \textwidth]{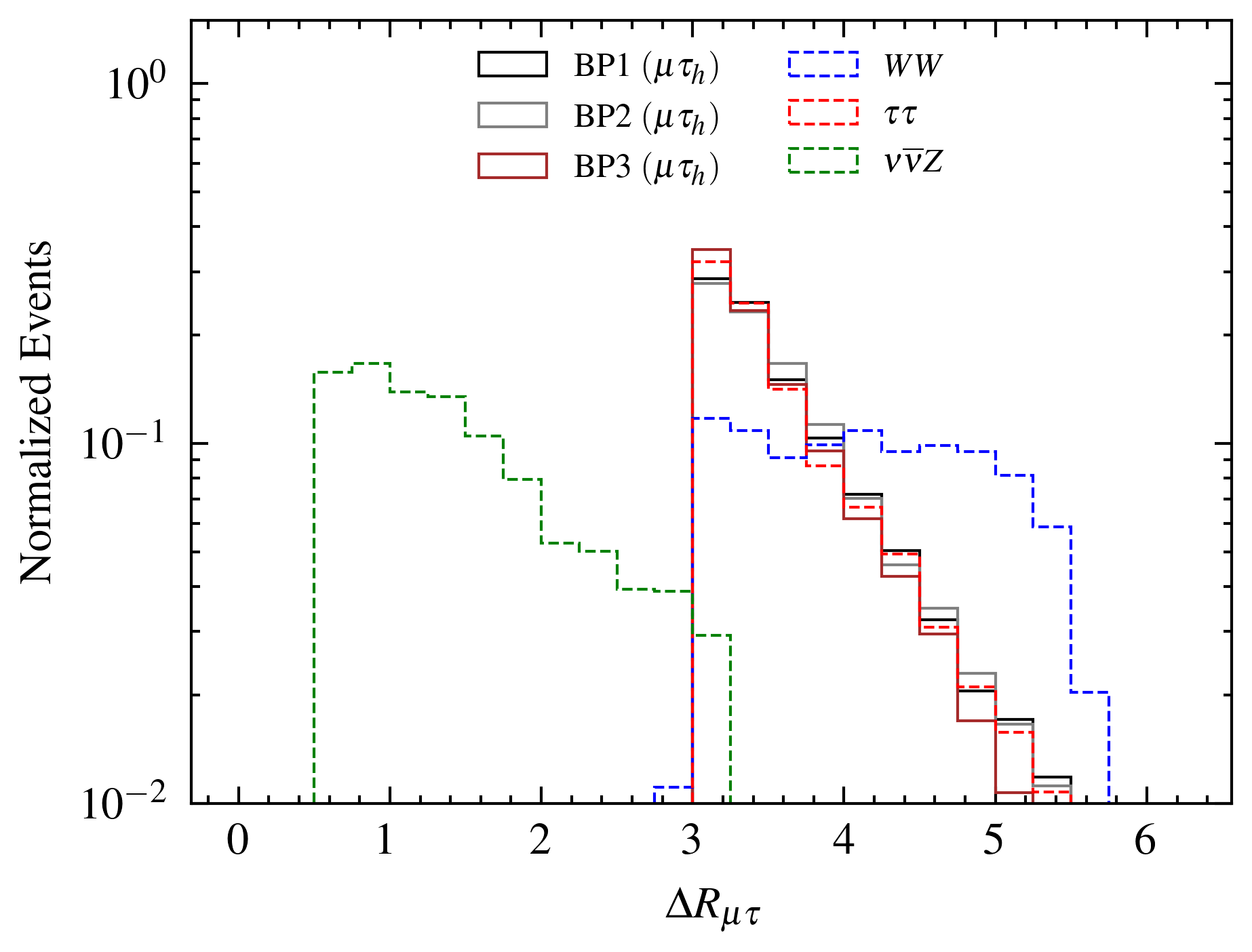}
	\includegraphics[width = 0.4 \textwidth]{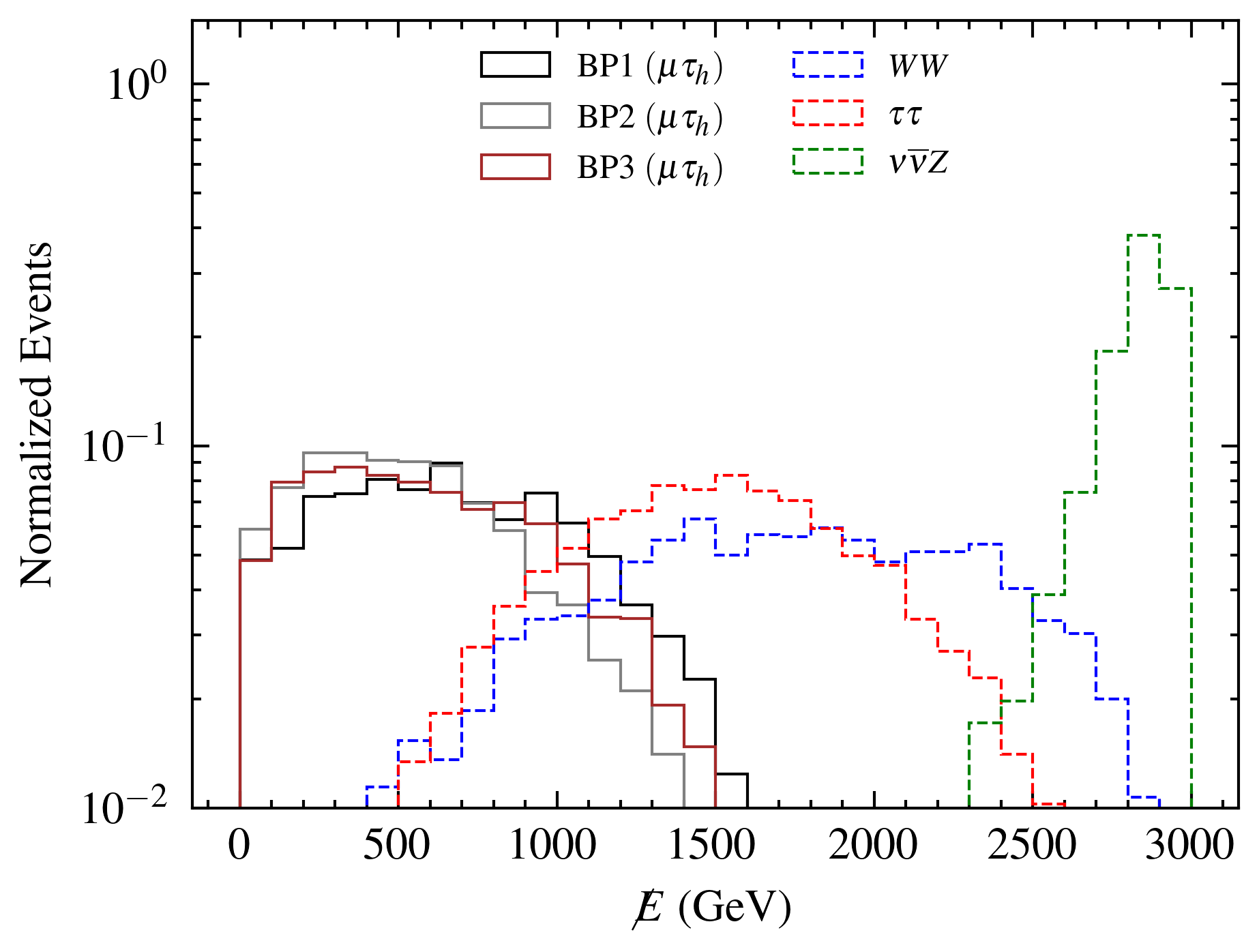}
	\includegraphics[width = 0.4 \textwidth]{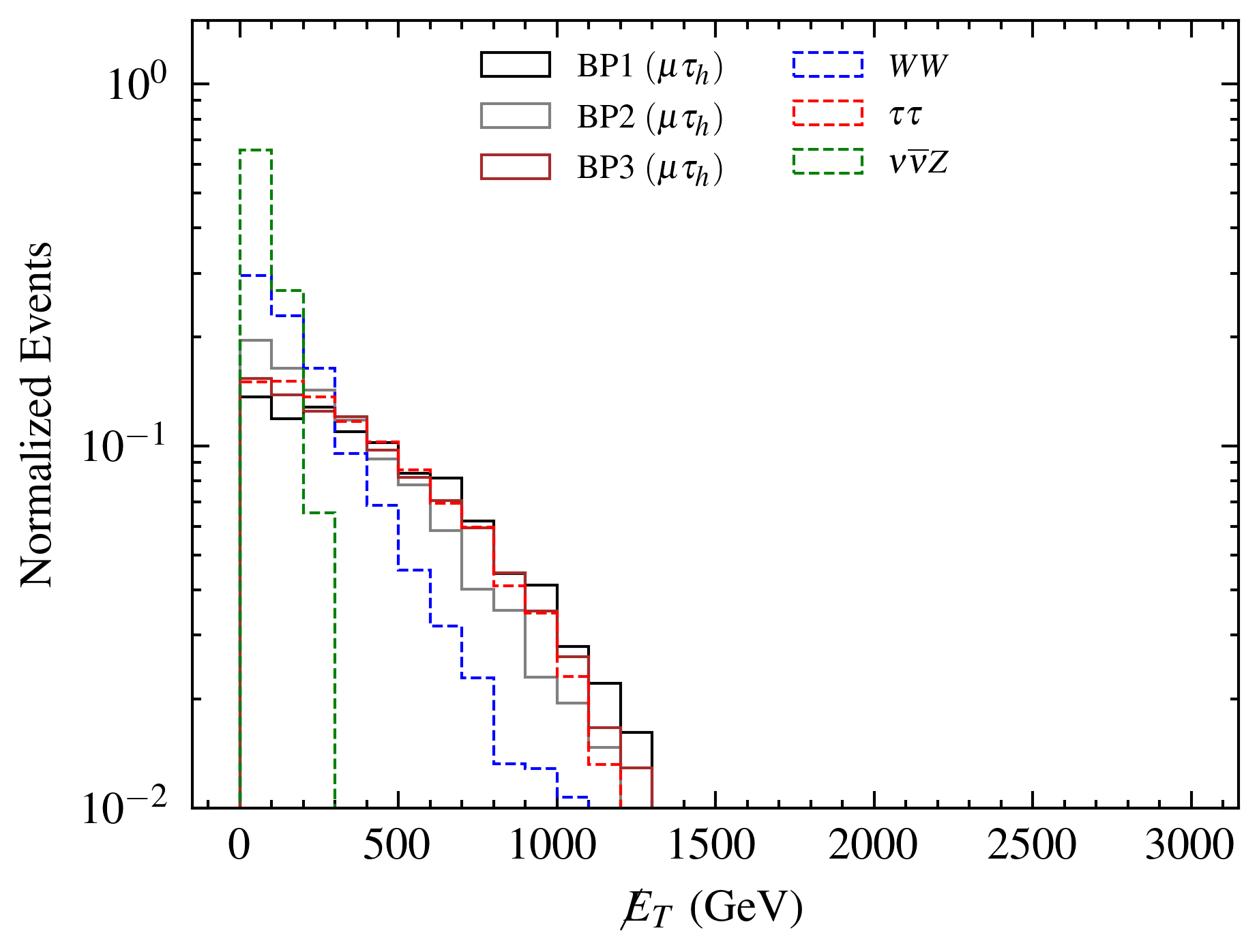}
	\includegraphics[width = 0.4 \textwidth]{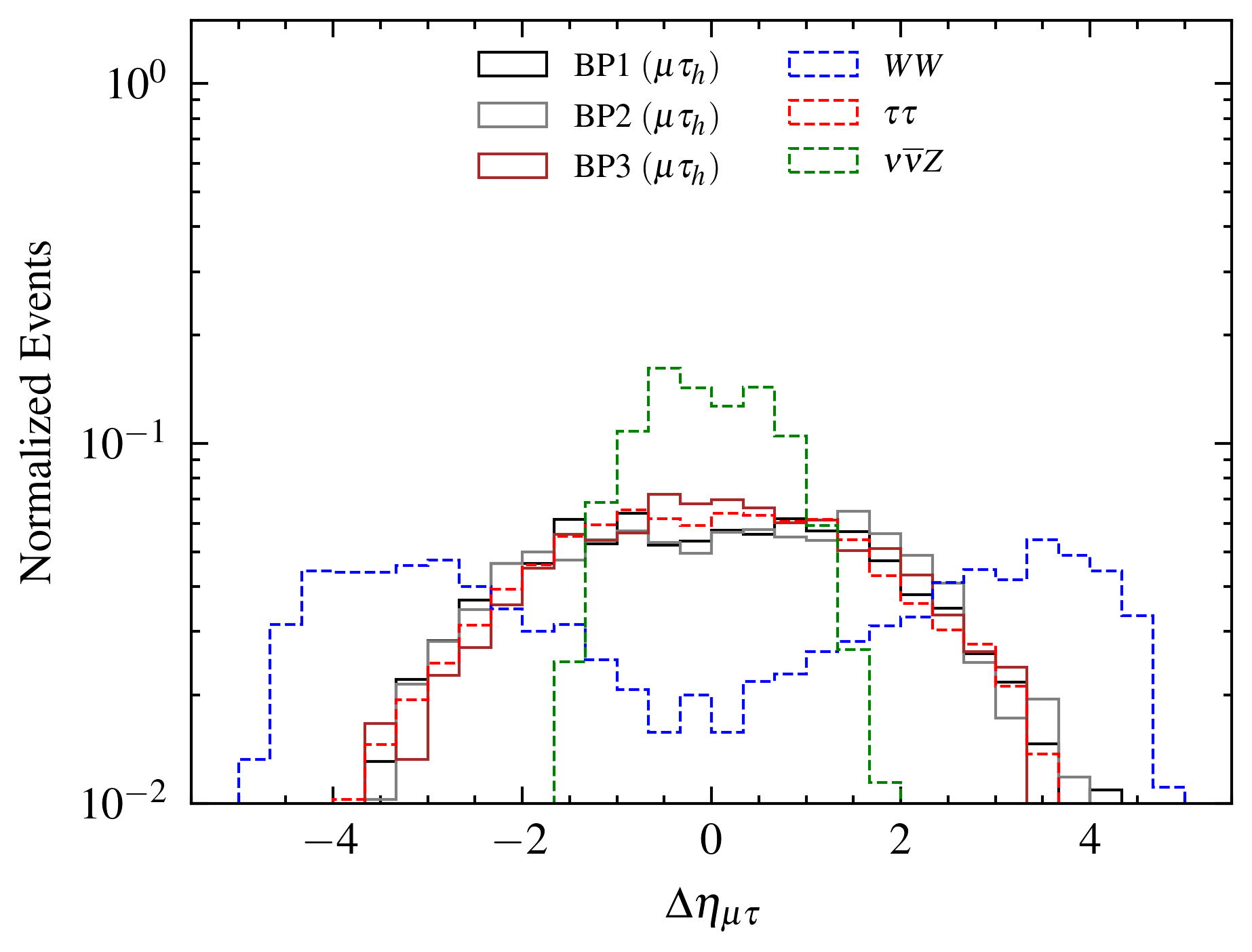}
	\caption{Additional kinematic distributions corresponding to signal and main background processes for $e^{+} e^{-} \rightarrow \mu \tau_h$ production at CLIC 3 TeV.}
	\label{fig:dist-mt1}
\end{figure}

\begin{figure}[htbp!]
	\centering
	\includegraphics[width = 0.4 \textwidth]{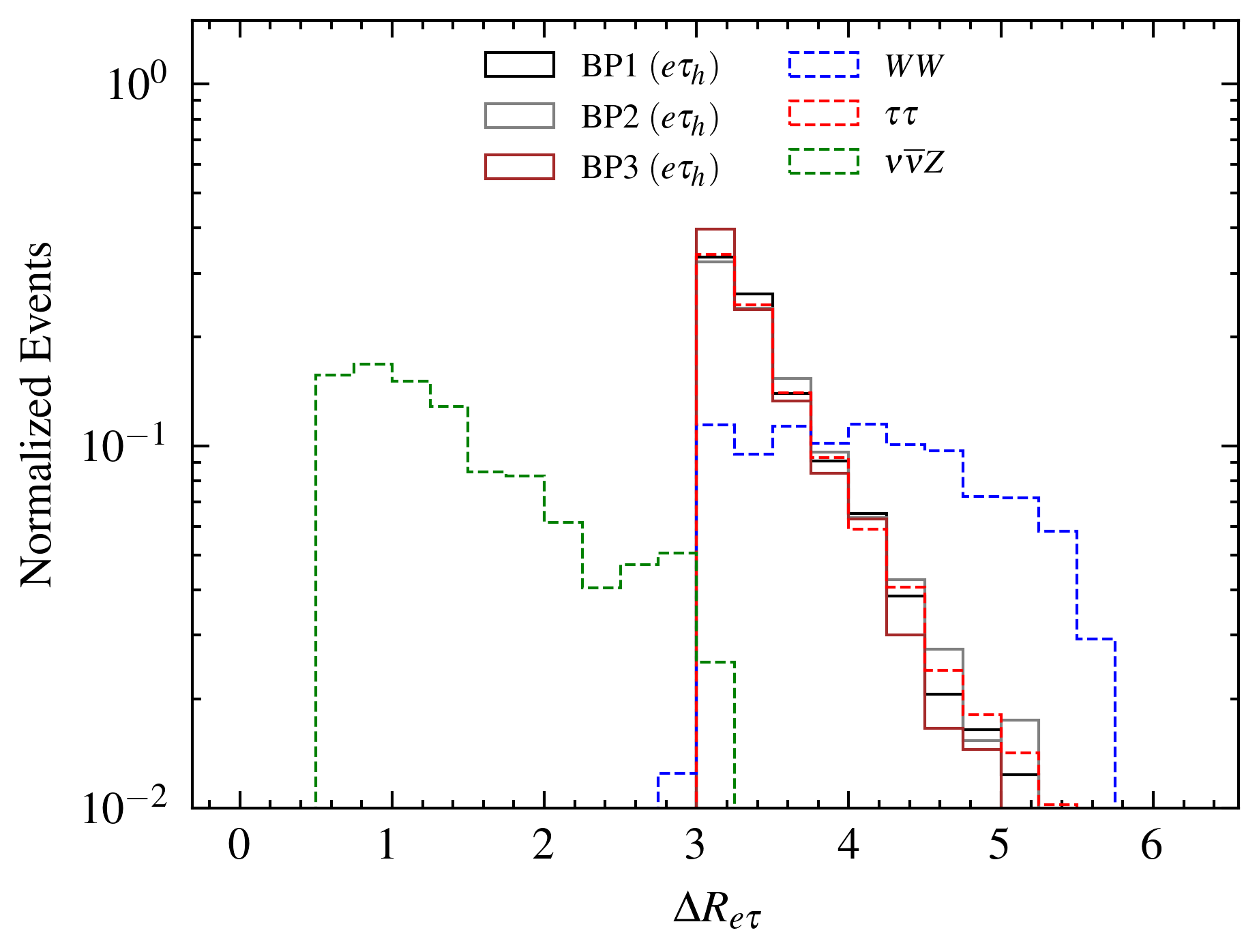}
	\includegraphics[width = 0.4 \textwidth]{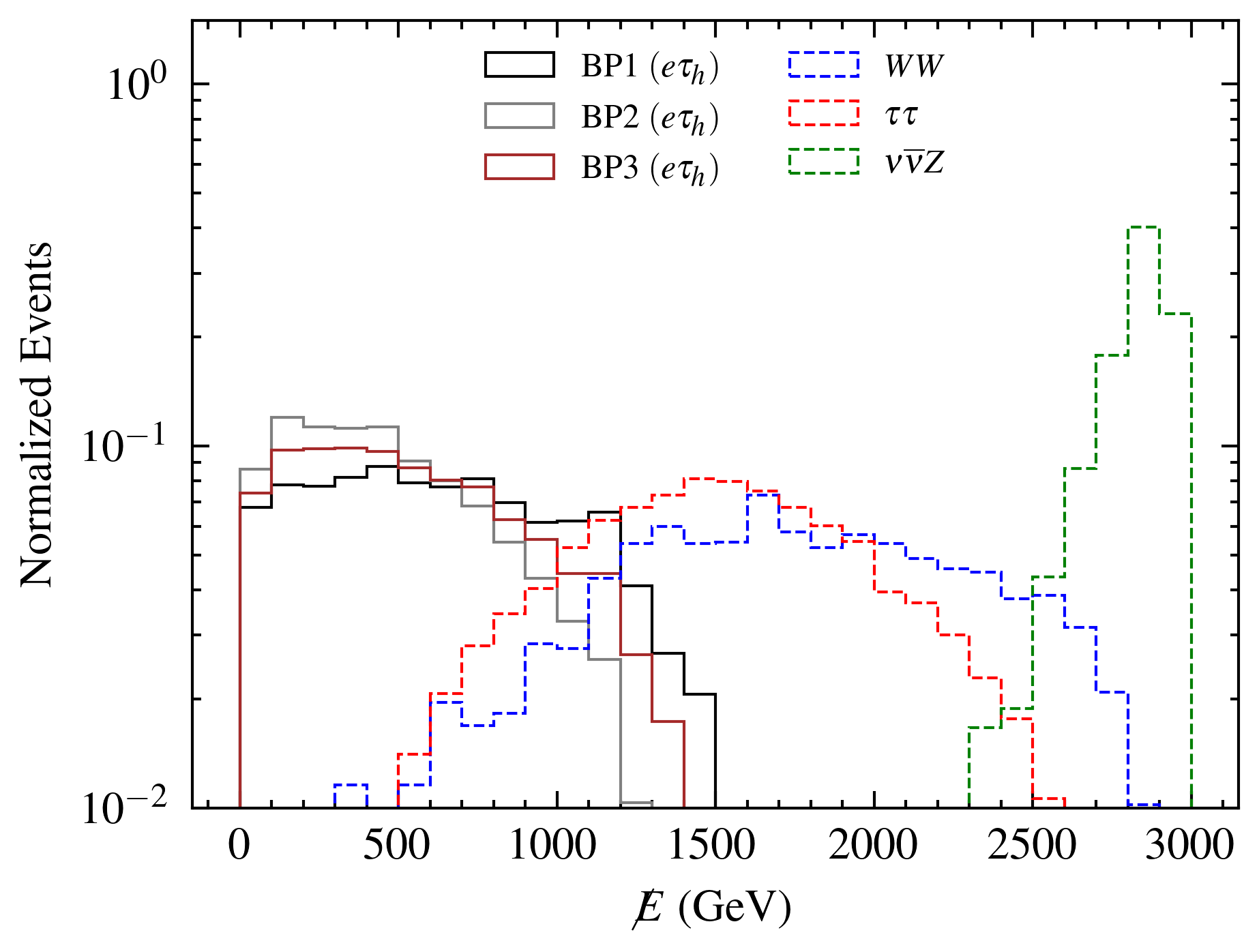}
	\includegraphics[width = 0.4 \textwidth]{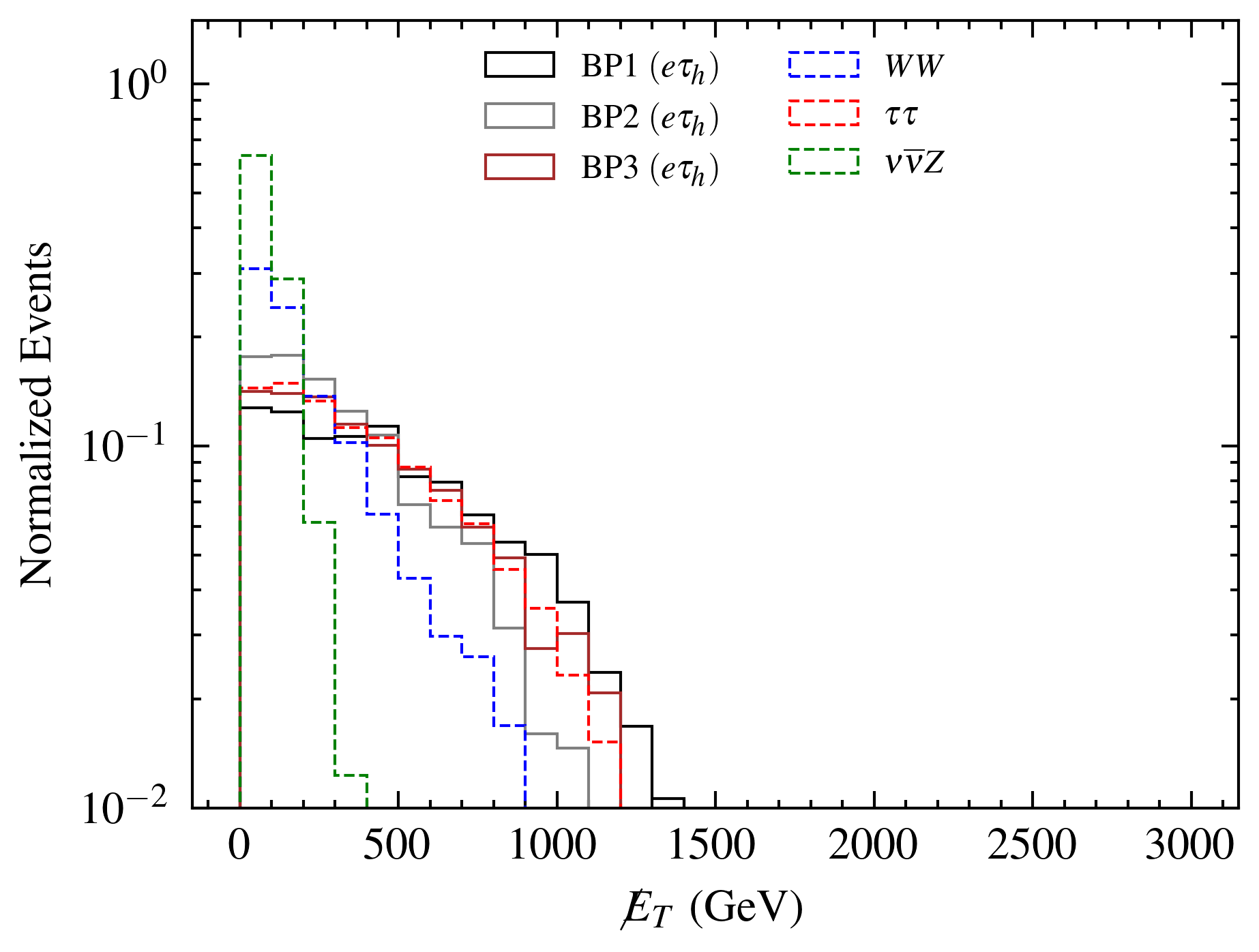}
	\includegraphics[width = 0.4 \textwidth]{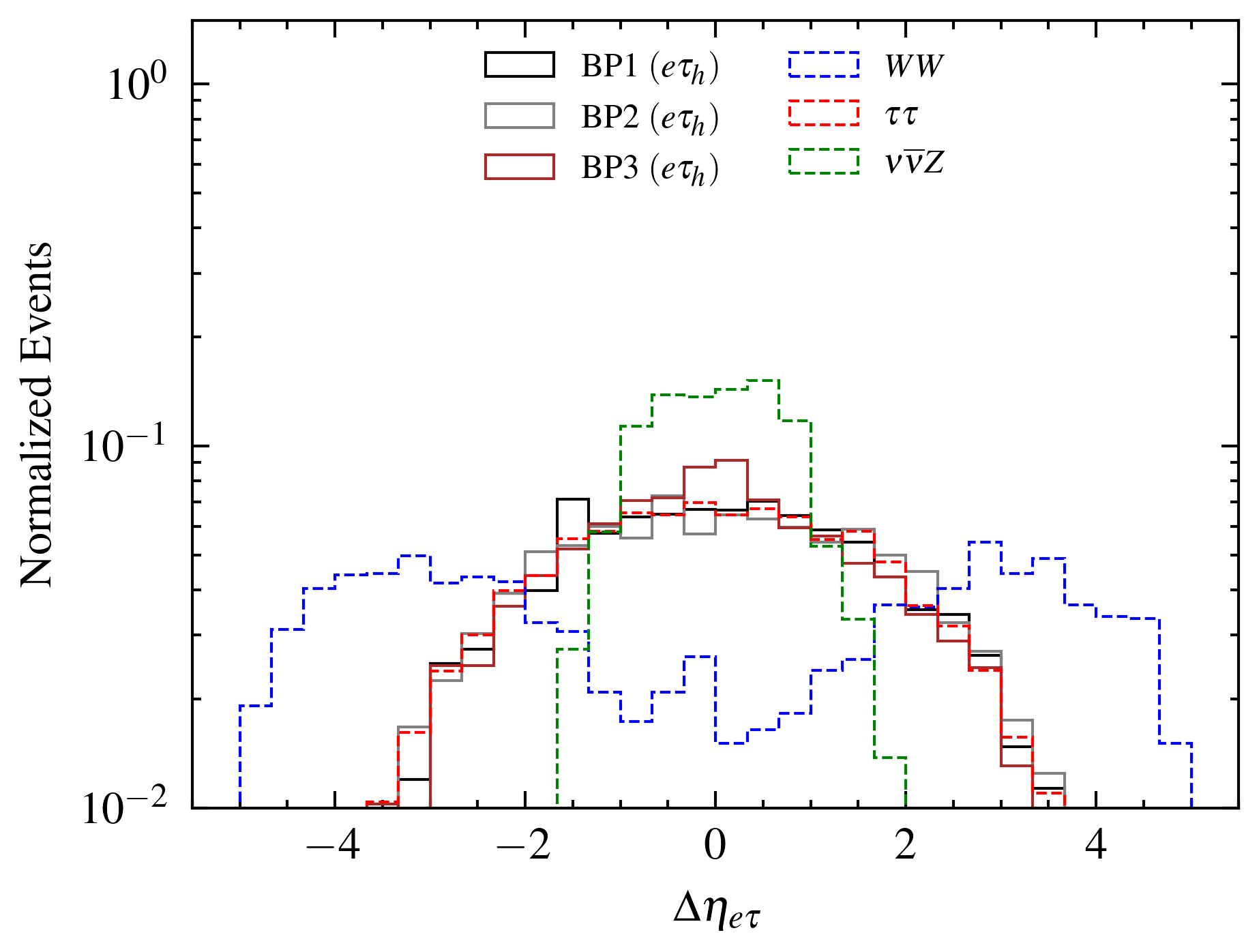}
	\caption{Additional kinematic distributions corresponding to signal and main background processes for $e^{+} e^{-} \rightarrow e \tau_h$ production at CLIC 3 TeV.}
	\label{fig:dist-et1}
\end{figure}

The additional kinematic distributions of the signal and background processes are shown in Figs.~\ref{fig:dist-mt1} and \ref{fig:dist-et1}. The detailed cutflow for the SM backgrounds is detailed in Table~\ref{tab:cutflow_b}.

\begin{table}[h!]
	\centering
	{\renewcommand{\arraystretch}{1.2}
		\begin{tabular}{|c|c|c|c|c|c|c|c|c|c|}
			\hline
			\multirow{2}*{Processes} & \multicolumn{3}{c|}{$\mathcal{C}_{0}:$ Basic cuts} & \multicolumn{3}{c|}{$\mathcal{C}_{1}: M_{\mu \tau} > 2$ TeV} & \multicolumn{3}{c|}{$\mathcal{C}_{2}: H_{T} > 1.5$ TeV} \\
			\cline{2-10}
			& $P_{0}$ & $P_{+}$ & $P_{-}$ & $P_{0}$ & $P_{+}$ & $P_{-}$ & $P_{0}$ & $P_{+}$ & $P_{-}$ \\
			\hline
			$W^{+} W^{-} (\mu \tau_h)$ & 1.270 & 0.257 & 2.279 & 0.138 & 0.028 & 0.248 & 0.043 & 0.009 & 0.077 \\
			$\tau^{+} \tau^{-} (\mu \tau_h)$ & 1.420 & 1.358 & 1.481 & 0.174 & 0.166 & 0.182 & 0.119 & 0.114 & 0.124 \\
			$\nu \overline{\nu} Z (\mu \tau_h)$ & 3.289 & 0.660 & 5.903 & 0.000 & 0.000 & 0.000 & 0.000 & 0.000 & 0.000 \\
			\hline
	\end{tabular}}
	\vspace{0.25cm}
	
	{\renewcommand{\arraystretch}{1.2}
		\begin{tabular}{|c|c|c|c|c|c|c|c|c|c|}
			\hline
			\multirow{2}*{Processes} & \multicolumn{3}{c|}{$\mathcal{C}_{0}:$ Basic cuts} & \multicolumn{3}{c|}{$\mathcal{C}_{1}: M_{e \tau} > 2$ TeV} & \multicolumn{3}{c|}{$\mathcal{C}_{2}: H_{T} > 1.5$ TeV} \\
			\cline{2-10}
			& $P_{0}$ & $P_{+}$ & $P_{-}$ & $P_{0}$ & $P_{+}$ & $P_{-}$ & $P_{0}$ & $P_{+}$ & $P_{-}$ \\
			\hline
			$W^{+} W^{-} (e \tau_h)$ & 1.021 & 0.207 & 1.832 & 0.106 & 0.021 & 0.190 & 0.031 & 0.006 & 0.056 \\
			$\tau^{+} \tau^{-} (e \tau_h)$ & 1.223 & 1.170 & 1.276 & 0.146 & 0.140 & 0.152 & 0.102 & 0.098 & 0.106 \\
			$\nu \overline{\nu} Z (e \tau_h)$ & 2.890 & 0.580 & 5.187 & 0.000 & 0.000 & 0.000 & 0.000 & 0.000 & 0.000 \\
			\hline
	\end{tabular}}
	\caption{Cutflow cross-sections (in fb) corresponding to SM backgrounds ($\ell \tau_{h}$ + $\sla{E}$) for different beam polarization choices at $\sqrt{s} = 3$ TeV. Here, $P_{0}: P_{e^{-}}= 0\%$, $P_{+}: P_{e^{-}}= +80\%$ and $P_{-}: P_{e^{-}}= -80\%$.}
	\label{tab:cutflow_b}
\end{table}

\section{Possible new physics connections} \label{sec:npc}
In this section, we qualitatively map out some of the tree level NP possibilities arising from the four-Fermi operators. At tree level, only possible mediators of NP turns out to be a gauge boson (say, $Z^{'}$) or a scalar boson (say, $H^{'}$), arising from extended BSM sectors. Tables~\ref{tab:np-eft} and \ref{tab:np-eft2} show the NP connections for $\mu \tau$ production, same applies for $e \tau$ production as well.

\begin{table}[htb!]
	\centering
	{\renewcommand{\arraystretch}{1.2}
		\begin{tabular}{|c|c|c|}
			\hline
			SMEFT Operators & LEFT Operators & NP Processes \\ \hline
			&  & \multirow{5}{*}{
				\begin{tikzpicture}[scale = 0.1]
					\begin{feynman}
						\vertex (a);
						\vertex [above left=1cm of a] (b) {$e^{-}_{L}$};
						\vertex [below left=1cm of a] (c) {$e^{+}_{L}$};
						\vertex[ right=1.2cm of a] (d);
						\vertex [above right=1cm of d] (e) {$\tau^{\pm}_{L}$};
						\vertex [below right=1cm of d] (f) {$\mu^{\mp}_{L}$};
						\diagram{
							(b) -- [fermion, thick, arrow size = 0.9 pt] (a) -- [fermion, thick, arrow size = 0.9 pt] (c);
							(a) -- [ very thick, boson, edge label={$Z^{'}$}] (d);
							(f) -- [fermion, thick, arrow size = 0.9 pt] (d) -- [fermion, thick, arrow size = 0.9 pt] (e);
						};
					\end{feynman}
			\end{tikzpicture}}
			\\
			$\left(\mathcal{O}_{\ell \ell}\right)_{ee \mu \tau}$ & $\left(\bar e_L \gamma^{\alpha} e_L\right) \left( \bar \tau_L \gamma_{\alpha} \mu_L \right)$ &  \\
			$\left(\mathcal{O}_{\ell \ell}\right)_{\mu \tau ee}$ & $ \left( \bar \tau_L \gamma_{\alpha} \mu_L \right) \left(\bar e_L \gamma^{\alpha} e_L\right)$ &  \\
			&  &  \\
			&  &  \\
			\hline
			&  & \multirow{5}{*}{
				\begin{tikzpicture}[scale = 0.1]
					\begin{feynman}
						\vertex (a);
						\vertex[below=1.2cm of a] (d);
						\vertex [above left=0.0cm and 1cm of a] (b) {$e^{-}_{L}$};
						\vertex [below left=0.0cm and 1cm of d] (c) {$e^{+}_{L}$};
						\vertex [above right=0.0cm and 1cm of a] (e) {$\tau^{\pm}_{L}$};
						\vertex [below right=0.0cm and 1cm of d] (f) {$\mu^{\mp}_{L}$};
						\diagram{
							(b) -- [fermion, thick, arrow size = 0.9 pt] (a) -- [fermion, thick, arrow size = 0.9 pt] (e);
							(a) -- [ very thick, boson, edge label={$Z^{'}$}] (d);
							(f) -- [fermion, thick, arrow size = 0.9 pt] (d) -- [fermion, thick, arrow size = 0.9 pt] (c);
						};
					\end{feynman}
			\end{tikzpicture}}
			\\ 
			$\left(\mathcal{O}_{\ell \ell}\right)_{\mu ee \tau}$ & $\left(\bar \tau_L \gamma^{\alpha} e_L\right) \left( \bar e_L \gamma_{\alpha} \mu_L \right)$ &  \\
			$\left(\mathcal{O}_{\ell \ell}\right)_{e \mu \tau e}$ & $\left( \bar e_L \gamma_{\alpha} \mu_L \right) \left(\bar \tau_L \gamma^{\alpha} e_L\right)$ &  \\
			&  &  \\
			&  &  \\
			\hline
			&  & \multirow{5}{*}{
				\begin{tikzpicture}[scale = 0.1]
					\begin{feynman}
						\vertex (a);
						\vertex [above left=1cm of a] (b) {$e^{-}_{R}$};
						\vertex [below left=1cm of a] (c) {$e^{+}_{R}$};
						\vertex[ right=1.2cm of a] (d);
						\vertex [above right=1cm of d] (e) {$\tau^{\pm}_{R}$};
						\vertex [below right=1cm of d] (f) {$\mu^{\mp}_{R}$};
						\diagram{
							(b) -- [fermion, thick, arrow size = 0.9 pt] (a) -- [fermion, thick, arrow size = 0.9 pt] (c);
							(a) -- [ very thick, boson, edge label={$Z^{'}$}] (d);
							(f) -- [fermion, thick, arrow size = 0.9 pt] (d) -- [fermion, thick, arrow size = 0.9 pt] (e);
						};
					\end{feynman}
			\end{tikzpicture}}
			\\ 
			$\left(\mathcal{O}_{ee}\right)_{ee \mu \tau}$ & $\left(\bar e_R \gamma^{\alpha} e_R\right) \left( \bar \tau_R \gamma_{\alpha} \mu_R \right)$ &  \\
			$\left(\mathcal{O}_{ee}\right)_{\mu \tau ee}$ & $ \left( \bar \tau_R \gamma_{\alpha} \mu_R \right) \left(\bar e_R \gamma^{\alpha} e_R\right)$ &  \\
			&  &  \\
			&  &  \\
			\hline
			&  & \multirow{5}{*}{
				\begin{tikzpicture}[scale = 0.1]
					\begin{feynman}
						\vertex (a);
						\vertex[below=1.2cm of a] (d);
						\vertex [above left=0.0cm and 1cm of a] (b) {$e^{-}_{R}$};
						\vertex [below left=0.0cm and 1cm of d] (c) {$e^{+}_{R}$};
						\vertex [above right=0.0cm and 1cm of a] (e) {$\tau^{\pm}_{R}$};
						\vertex [below right=0.0cm and 1cm of d] (f) {$\mu^{\mp}_{R}$};
						\diagram{
							(b) -- [fermion, thick, arrow size = 0.9 pt] (a) -- [fermion, thick, arrow size = 0.9 pt] (e);
							(a) -- [ very thick, boson, edge label={$Z^{'}$}] (d);
							(f) -- [fermion, thick, arrow size = 0.9 pt] (d) -- [fermion, thick, arrow size = 0.9 pt] (c);
						};
					\end{feynman}
			\end{tikzpicture}}
			\\ 
			$\left(\mathcal{O}_{ee}\right)_{\mu ee \tau}$ & $\left(\bar \tau_R \gamma^{\alpha} e_R\right) \left( \bar e_R \gamma_{\alpha} \mu_R \right)$ &  \\
			$\left(\mathcal{O}_{ee}\right)_{e \mu \tau e}$ & $\left( \bar e_R \gamma_{\alpha} \mu_R \right) \left(\bar \tau_R \gamma^{\alpha} e_R\right)$ &  \\
			&  &  \\
			&  &  \\
			\hline
			&  & \multirow{5}{*}{
				\begin{tikzpicture}[scale = 0.1]
					\begin{feynman}
						\vertex (a);
						\vertex [above left=1cm of a] (b) {$e^{-}_{L/R}$};
						\vertex [below left=1cm of a] (c) {$e^{+}_{L/R}$};
						\vertex[ right=1.2cm of a] (d);
						\vertex [above right=1cm of d] (e) {$\tau^{\pm}_{R/L}$};
						\vertex [below right=1cm of d] (f) {$\mu^{\mp}_{R/L}$};
						\diagram{
							(b) -- [fermion, thick, arrow size = 0.9 pt] (a) -- [fermion, thick, arrow size = 0.9 pt] (c);
							(a) -- [ very thick, boson, edge label={$Z^{'}$}] (d);
							(f) -- [fermion, thick, arrow size = 0.9 pt] (d) -- [fermion, thick, arrow size = 0.9 pt] (e);
						};
					\end{feynman}
			\end{tikzpicture}}
			\\ 
			$\left(\mathcal{O}_{\ell e}\right)_{ee \mu \tau}$ & $\left(\bar e_L \gamma^{\alpha} e_L\right) \left( \bar \tau_R \gamma_{\alpha} \mu_R \right)$ &  \\
			$\left(\mathcal{O}_{\ell e}\right)_{\mu \tau ee}$ & $ \left( \bar \tau_L \gamma_{\alpha} \mu_L \right) \left(\bar e_R \gamma^{\alpha} e_R\right)$ &  \\
			&  &  \\
			&  &  \\
			\hline
			&  & \multirow{5}{*}{
				\begin{tikzpicture}[scale = 0.1]
					\begin{feynman}
						\vertex (a);
						\vertex[below=1.2cm of a] (d);
						\vertex [above left=0.0cm and 1cm of a] (b) {$e^{-}_{L/R}$};
						\vertex [below left=0.0cm and 1cm of d] (c) {$e^{+}_{R/L}$};
						\vertex [above right=0.0cm and 1cm of a] (e) {$\tau^{\pm}_{L/R}$};
						\vertex [below right=0.0cm and 1cm of d] (f) {$\mu^{\mp}_{R/L}$};
						\diagram{
							(b) -- [fermion, thick, arrow size = 0.9 pt] (a) -- [fermion, thick, arrow size = 0.9 pt] (e);
							(a) -- [ very thick, boson, edge label={$Z^{'}$}] (d);
							(f) -- [fermion, thick, arrow size = 0.9 pt] (d) -- [fermion, thick, arrow size = 0.9 pt] (c);
						};
					\end{feynman}
			\end{tikzpicture}}
			\\ 
			$\left(\mathcal{O}_{\ell e}\right)_{\mu ee \tau}$ & $\left(\bar \tau_L \gamma^{\alpha} e_L\right) \left( \bar e_R \gamma_{\alpha} \mu_R \right)$ &  \\
			$\left(\mathcal{O}_{\ell e}\right)_{e \mu \tau e}$ & $\left( \bar e_L \gamma_{\alpha} \mu_R \right) \left(\bar \tau_R \gamma^{\alpha} e_R\right)$ &  \\
			&  &  \\
			&  &  \\
			\hline
	\end{tabular}}
	\caption{$Z^{'}$ mediated NP connections to SMEFT and Low-energy EFT (LEFT) operators.}
	\label{tab:np-eft}
\end{table}

\begin{table}[htb!]
	\centering
	{\renewcommand{\arraystretch}{1.2}
		\begin{tabular}{|c|c|c|}
			\hline
			SMEFT Operators & LEFT Operators & NP Processes \\ \hline
			&  & \multirow{5}{*}{
				\begin{tikzpicture}[scale = 0.1]
					\begin{feynman}
						\vertex (a);
						\vertex [above left=1cm of a] (b) {$e^{-}_{L/R}$};
						\vertex [below left=1cm of a] (c) {$e^{+}_{R/L}$};
						\vertex[ right=1.2cm of a] (d);
						\vertex [above right=1cm of d] (e) {$\tau^{\pm}_{R/L}$};
						\vertex [below right=1cm of d] (f) {$\mu^{\mp}_{L/R}$};
						\diagram{
							(b) -- [fermion, thick, arrow size = 0.9 pt] (a) -- [fermion, thick, arrow size = 0.9 pt] (c);
							(a) -- [ very thick, scalar, edge label={$H^{'}$}] (d);
							(f) -- [fermion, thick, arrow size = 0.9 pt] (d) -- [fermion, thick, arrow size = 0.9 pt] (e);
						};
					\end{feynman}
			\end{tikzpicture}}
			\\ 
			$\left(\mathcal{O}_{\ell e}\right)_{\mu ee \tau}$ & $\left(\bar e_R  \mu_L\right) \left( \bar \tau_R e_L \right)$ &  \\
			$\left(\mathcal{O}_{\ell e}\right)_{e \mu \tau e}$ & $ \left( \bar \tau_L e_R \right) \left(\bar e_L \mu_R\right)$ &  \\
			&  &  \\
			&  &  \\
			\hline
			&  & \multirow{5}{*}{
				\begin{tikzpicture}[scale = 0.1]
					\begin{feynman}
						\vertex (a);
						\vertex[below=1.2cm of a] (d);
						\vertex [above left=0.0cm and 1cm of a] (b) {$e^{-}_{L/R}$};
						\vertex [below left=0.0cm and 1cm of d] (c) {$e^{+}_{L/R}$};
						\vertex [above right=0.0cm and 1cm of a] (e) {$\tau^{\pm}_{R/L}$};
						\vertex [below right=0.0cm and 1cm of d] (f) {$\mu^{\mp}_{R/L}$};
						\diagram{
							(b) -- [fermion, thick, arrow size = 0.9 pt] (a) -- [fermion, thick, arrow size = 0.9 pt] (e);
							(a) -- [ very thick, scalar, edge label={$H^{'}$}] (d);
							(f) -- [fermion, thick, arrow size = 0.9 pt] (d) -- [fermion, thick, arrow size = 0.9 pt] (c);
						};
					\end{feynman}
			\end{tikzpicture}}
			\\ 
			$\left(\mathcal{O}_{\ell e}\right)_{ee \mu \tau}$ & $\left(\bar \tau_R e_L\right) \left( \bar e_L \mu_R \right)$ &  \\
			$\left(\mathcal{O}_{\ell e}\right)_{\mu \tau ee}$ & $\left( \bar e_R \mu_L \right) \left(\bar \tau_L e_R\right)$ &  \\
			&  &  \\
			&  &  \\
			\hline
	\end{tabular}}
	\caption{$H^{'}$ mediated NP connections to SMEFT and Low-energy EFT (LEFT) operators.}
	\label{tab:np-eft2}
\end{table}

\section{Renormalization group equations}
\label{sec:rge}
\begin{figure}[htb!]
	\centering
	\includegraphics[width = 0.4 \textwidth]{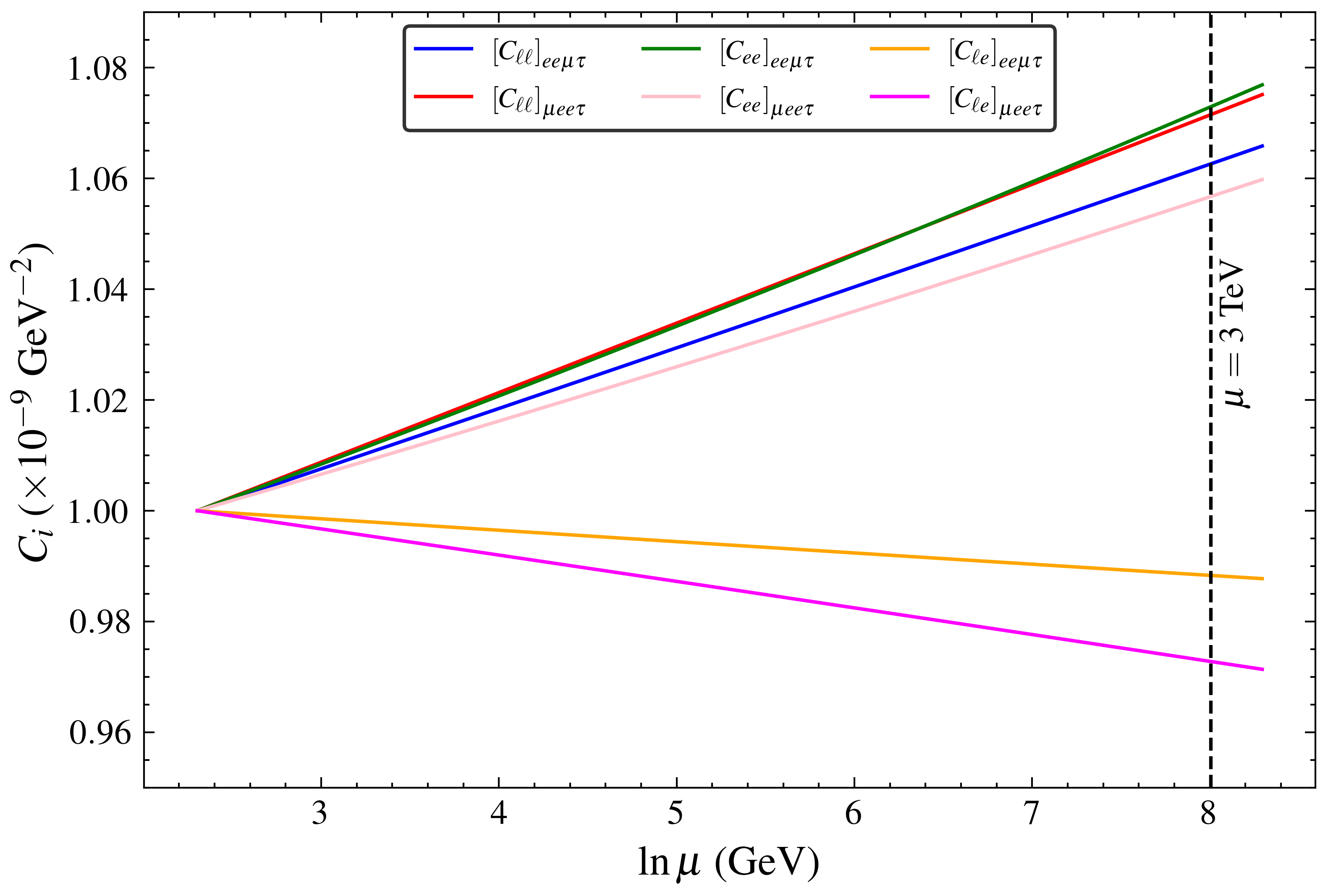}
	\includegraphics[width = 0.4 \textwidth]{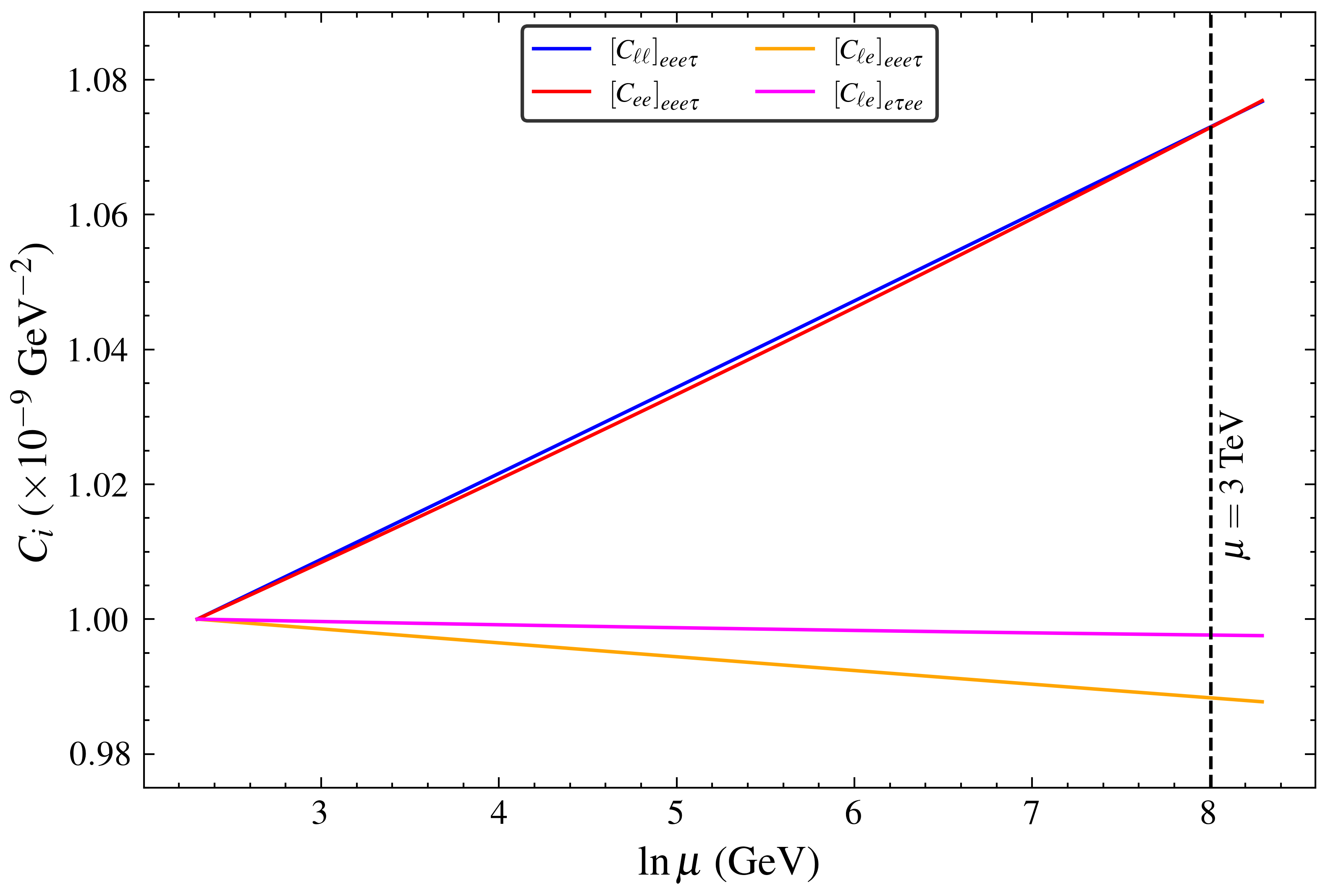}
	\caption{RGE flow for some of the LFV operators (\textit{Left:} $\mu \tau$ operators, \textit{Right:} $e \tau$ operators). The black dashed line corresponds to the energy scale of the CLIC collider.}
	\label{fig:rge}
\end{figure}

The RG equations corresponding to the SMEFT operators are represented as follows:
\begin{equation}
	\frac{d[\mathcal{C}_{i}]}{d\log{\mu}} = \frac{1}{16 \pi^{2}} \beta_{i}.
\end{equation}
Here, $[\mathcal{C}_{i}] = C_{i}/\Lambda^{2}$ are the operator coefficients, $\mu$ is the renormalization scale, and $\beta_{i}$ is the RGE $\beta$ function corresponding to the operator $\mathcal{O}_{i}$. The RG equations for the SM gauge couplings are detailed in Eq.~\eqref{eq:rgex}.
\begin{equation} \label{eq:rgex}
	\begin{split}
		\frac{dg}{d\log{\mu}} &= \frac{1}{16 \pi^{2}} \left(-\frac{19}{6} g^{3} \right), \\
		\frac{dg'}{d\log{\mu}} &= \frac{1}{16 \pi^{2}} \left(\frac{41}{6} g'^{3}\right), \\
		\frac{dg_{s}}{d\log{\mu}} &= \frac{1}{16 \pi^{2}} \left( -7 g_{s}^{3} \right). \\
	\end{split}
\end{equation}
The RG equation for the Top Yukawa coupling is shown in Eq.~\eqref{eq:rgey}. All Yukawa couplings other than the one of the Top quark are considered as negligible.
\begin{equation} \label{eq:rgey}
	\begin{split}
		\frac{dy_{t}}{d\log{\mu}} &= \frac{y_{t}}{16 \pi^{2}} \left( \frac{9}{2} y^{2}_{t} - 8 g^{2}_{s} - \frac{9}{4} g^{2} - \frac{17}{12} g'^{2} \right).
	\end{split}
\end{equation}
The general form of the $\beta$ function corresponding to the LFV four-Fermi operators are detailed in Eqs.~\eqref{eq:rge1}-\eqref{eq:rge3}. Here, $\left[C_{ij}\right]_{prst} = \left(C_{ij}\right)_{prst}/\Lambda^{2}$.
\begin{equation} \label{eq:rge1}
	\begin{split}
		\left[\beta_{\ell \ell} \right]_{prst} &= \frac{1}{3} g'^{2} \left(2 \left[C_{\ell \ell}\right]_{prww} + \left[C_{\ell \ell}\right]_{pwwr} \right) \delta_{st} + \frac{1}{3} g'^{2} \left(2 \left[C_{\ell \ell}\right]_{wwst} + \left[C_{\ell \ell}\right]_{wtsw} \right) \delta_{pr} \\
		&- \frac{1}{3} g^{2} \left[C_{\ell \ell}\right]_{pwwr} \delta_{st} + \frac{2}{3} g^{2} \left[C_{\ell \ell}\right]_{swwr} \delta_{pt} - \frac{1}{3} g^{2} \left[C_{\ell \ell}\right]_{wtsw} \delta_{pr} + \frac{2}{3} g^{2} \left[C_{\ell \ell}\right]_{wtpw} \delta_{sr} \\
		&+ \frac{1}{3} g'^{2} \left[C_{\ell e}\right]_{prww} \delta_{st} + \frac{1}{3} g'^{2} \left[C_{\ell e}\right]_{wwst} \delta_{pr} + 6 g^{2} \left[C_{\ell \ell}\right]_{ptsr} + 3 \left(g'^{2} - g^{2}\right) \left[C_{\ell \ell}\right]_{prst}, \\
	\end{split}
\end{equation}
\begin{equation} \label{eq:rge2}
	\begin{split}
		\left[\beta_{ee} \right]_{prst} &= \frac{2}{3} g'^{2} \left( \left[C_{\ell e}\right]_{wwpr} + 4 \left[C_{ee}\right]_{prww} \right) \delta_{st} + \frac{2}{3} g'^{2} \left( \left[C_{\ell e}\right]_{stww} + 4 \left[C_{ee}\right]_{wwst} \right) \delta_{pr} \\
		&+ 6 g'^{2} \left[C_{ee}\right]_{prst} + 6 g'^{2} \left[C_{ee}\right]_{stpr}, \\
	\end{split}
\end{equation}
\begin{equation} \label{eq:rge3}
	\begin{split}
		\left[\beta_{\ell e} \right]_{prst} &= \frac{8}{3} g'^{2} \left[C_{\ell \ell}\right]_{prww} \delta_{st} + \frac{4}{3} g'^{2} \left[C_{\ell \ell}\right]_{pwwr} \delta_{st} + \frac{4}{3} g'^{2} \left[C_{\ell e}\right]_{prww} \delta_{st} + \frac{2}{3} g'^{2} \left[C_{\ell e}\right]_{wwst} \delta_{pr} \\
		&+ \frac{8}{3} g'^{2} \left[C_{ee}\right]_{wwst} \delta_{pr} - 6 g'^{2} \left[C_{\ell e}\right]_{prst}.
	\end{split}
\end{equation}
Using the general form, we construct $\beta$ function for each of the different operators. The equations are listed in Eqs.~\eqref{eq:rg1}-\eqref{eq:rg6}.
\begin{itemize}
	\item \textbf{Operator Class:} $\left(\mathcal{O}_{\ell \ell}\right)_{\tau \mu}$
	\begin{equation} \label{eq:rg1}
		\begin{split}
			\left[\beta_{\ell \ell} \right]_{ee \mu \tau} &= \frac{1}{3} \left(11 g'^{2} - 9 g^{2} \right) \left[C_{\ell \ell} \right]_{ee \mu \tau} + \frac{1}{3} \left(g'^{2} + 17 g^{2} \right) \left[C_{\ell \ell} \right]_{e \tau \mu e} + \frac{1}{3} g'^{2} \left[C_{\ell e} \right]_{ee \mu \tau}, \\
			\left[\beta_{\ell \ell} \right]_{\mu \tau ee} &= \frac{1}{3} \left(11 g'^{2} - 9 g^{2} \right) \left[C_{\ell \ell} \right]_{\mu \tau ee} + \frac{1}{3} \left(g'^{2} + 17 g^{2} \right) \left[C_{\ell \ell} \right]_{\mu ee \tau} + \frac{1}{3} g'^{2} \left[C_{\ell e} \right]_{\mu \tau ee}, \\
			\left[\beta_{\ell \ell} \right]_{\mu ee \tau} &= \frac{2}{3} g^{2} \left[C_{\ell \ell} \right]_{e \tau \mu e} + 6 g^{2} \left[C_{\ell \ell} \right]_{\mu \tau ee} + 3 \left(g'^{2} - g^{2} \right) \left[C_{\ell \ell} \right]_{\mu ee \tau}, \\
			\left[\beta_{\ell \ell} \right]_{e \tau \mu e} &= \frac{2}{3} g^{2} \left[C_{\ell \ell} \right]_{\mu ee \tau} + 6 g^{2} \left[C_{\ell \ell} \right]_{ee \mu \tau} + 3 \left(g'^{2} - g^{2} \right) \left[C_{\ell \ell} \right]_{e \tau \mu e}. \\
		\end{split}
	\end{equation}
	
	\item \textbf{Operator Class:} $\left(\mathcal{O}_{\ell \ell}\right)_{\tau e}$
	\begin{equation} \label{eq:rg2}
		\begin{split}
			\left[\beta_{\ell \ell} \right]_{eee \tau} &= \frac{1}{3} \left(11 g'^{2} - 9 g^{2} \right) \left[C_{\ell \ell} \right]_{eee \tau} + \frac{1}{3} \left(g'^{2} + 19 g^{2} \right) \left[C_{\ell \ell} \right]_{e \tau ee} + \frac{1}{3} g'^{2} \left[C_{\ell e} \right]_{eee \tau}, \\
			\left[\beta_{\ell \ell} \right]_{e \tau ee} &= \frac{1}{3} \left(11 g'^{2} - 9 g^{2} \right) \left[C_{\ell \ell} \right]_{e \tau ee} + \frac{1}{3} \left(g'^{2} + 19 g^{2} \right) \left[C_{\ell \ell} \right]_{eee \tau} + \frac{1}{3} g'^{2} \left[C_{\ell e} \right]_{e \tau ee}. \\
		\end{split}
	\end{equation}
	
	\item \textbf{Operator Class:} $\left(\mathcal{O}_{ee}\right)_{\tau \mu}$
	\begin{equation} \label{eq:rg3}
		\begin{split}
			\left[\beta_{ee} \right]_{ee \mu \tau} &= \frac{26}{3} g'^{2} \left[C_{ee} \right]_{ee \mu \tau} + 6 g'^{2} \left[C_{ee} \right]_{\mu \tau ee} + \frac{2}{3} g'^{2} \left[C_{\ell e} \right]_{\mu \tau ee}, \\
			\left[\beta_{ee} \right]_{\mu \tau ee} &= \frac{26}{3} g'^{2} \left[C_{ee} \right]_{\mu \tau ee} + 6 g'^{2} \left[C_{ee} \right]_{ee \mu \tau} + \frac{2}{3} g'^{2} \left[C_{\ell e} \right]_{ee \mu \tau}, \\
			\left[\beta_{ee} \right]_{\mu ee \tau} &= 6 g'^{2} \left[C_{ee} \right]_{\mu ee \tau} + 6 g'^{2} \left[C_{ee} \right]_{e \tau \mu e}, \\
			\left[\beta_{ee} \right]_{e \tau \mu e} &= 6 g'^{2} \left[C_{ee} \right]_{e \tau \mu e} + 6 g'^{2} \left[C_{ee} \right]_{\mu ee \tau}. \\
		\end{split}
	\end{equation}
	
	\item \textbf{Operator Class:} $\left(\mathcal{O}_{ee}\right)_{\tau e}$
	\begin{equation} \label{eq:rg4}
		\begin{split}
			\left[\beta_{ee} \right]_{eee \tau} &= \frac{26}{3} g'^{2} \left[C_{ee} \right]_{eee \tau} + 6 g'^{2} \left[C_{ee} \right]_{e \tau ee} + \frac{2}{3} g'^{2} \left[C_{\ell e} \right]_{e \tau ee}, \\
			\left[\beta_{ee} \right]_{e \tau ee} &= \frac{26}{3} g'^{2} \left[C_{ee} \right]_{e \tau ee} + 6 g'^{2} \left[C_{ee} \right]_{eee \tau} + \frac{2}{3} g'^{2} \left[C_{\ell e} \right]_{eee \tau}. \\
		\end{split}
	\end{equation}
	
	\item \textbf{Operator Class:} $\left(\mathcal{O}_{\ell e}\right)_{\tau \mu}$
	\begin{equation} \label{eq:rg5}
		\begin{split}
			\left[\beta_{\ell e} \right]_{ee \mu \tau} &= -\frac{16}{3} g'^{2} \left[C_{\ell e} \right]_{ee \mu \tau} + \frac{8}{3} g'^{2} \left[C_{ee} \right]_{ee \mu \tau}, \\
			\left[\beta_{\ell e} \right]_{\mu \tau ee} &= - \frac{14}{3} g'^{2} \left[C_{\ell e} \right]_{\mu \tau ee} + \frac{8}{3} g'^{2} \left[C_{\ell \ell} \right]_{\mu \tau ee} + \frac{4}{3} g'^{2} \left[C_{\ell \ell} \right]_{\mu ee \tau}, \\
			\left[\beta_{\ell e} \right]_{\mu ee \tau} &= -6 g'^{2} \left[C_{\ell e} \right]_{\mu ee \tau}, \\
			\left[\beta_{\ell e} \right]_{e \tau \mu e} &= -6 g'^{2} \left[C_{\ell e} \right]_{e \tau \mu e}. \\
		\end{split}
	\end{equation}
	
	\item \textbf{Operator Class:} $\left(\mathcal{O}_{\ell e}\right)_{\tau e}$
	\begin{equation} \label{eq:rg6}
		\begin{split}
			\left[\beta_{\ell e} \right]_{eee \tau} &= -\frac{16}{3} g'^{2} \left[C_{\ell e} \right]_{eee \tau} + \frac{8}{3} g'^{2} \left[C_{ee} \right]_{eee \tau}, \\
			\left[\beta_{\ell e} \right]_{e \tau ee} &= - \frac{14}{3} g'^{2} \left[C_{\ell e} \right]_{e \tau ee} + \frac{8}{3} g'^{2} \left[C_{\ell \ell} \right]_{e \tau ee} + \frac{4}{3} g'^{2} \left[C_{\ell \ell} \right]_{eee \tau}. \\
		\end{split}
	\end{equation}
\end{itemize}
Fig. \ref{fig:rge} shows the RGE flow of relevant operators obtained by solving Eqs.~\eqref{eq:rg1}-\eqref{eq:rg6}. The operators show very little variation to change in renormalization scale, hence, bounds obtained at lower energies will still be relevant even at higher energy scales. 

\bibliographystyle{JHEP}
\bibliography{lfv.bib}
\end{document}